\documentclass[letterarticle,12pt,peerreviewca,draftcls]{IEEEtran}
\usepackage{csm16}
\usepackage[margin=1in]{geometry}
\usepackage{amsmath}
\usepackage{amssymb}
\usepackage{dsfont}
\usepackage[labelformat=simple]{subcaption}

\usepackage{wrapfig}

\usepackage{url}
\usepackage{graphicx,xcolor}
\usepackage{verbatim}
\makeatletter
\newcommand{\verbatimfont}[1]{\def\verbatim@font{#1}}%
\makeatother
\verbatimfont{\ttfamily\small}

\newcommand{\bi}{\begin{itemize}}\newcommand{\ei}{\end{itemize}}
\newcommand{\be}{\begin{equation}}\newcommand{\ee}{\end{equation}}
\newcommand{\bee}{\begin{enumerate}}\newcommand{\eee}{\end{enumerate}}
\newcommand{\bea}{\begin{eqnarray}}\newcommand{\eea}{\end{eqnarray}}

\newcommand{\bc}{\begin{center}}\newcommand{\ec}{\end{center}}
\usepackage[left,pagewise]{lineno} 
\usepackage[english]{babel} 
\usepackage{blindtext}
\usepackage{BernsteinStyle}
\usepackage{tcolorbox}

\usepackage{nomencl}
\usepackage[linesnumbered,ruled,vlined]{algorithm2e}

\SetCommentSty{mycommfont}
\makenomenclature

\usepackage[maxbibnames=50,style=ieee]{biblatex}
\newcounter{example}
\newenvironment{example}[1][]{\refstepcounter{example}\par\medskip
   \noindent \textbf{\noindent Example~\theexample. #1} \rmfamily}{\medskip}

\title{ Output-Feedback  Model Predictive Control\\
with Online  Identification\\[1ex]
\Large  A Numerical Investigation of\\ Persistency, Consistency, and Exigency}

\author{Tam W. Nguyen, Syed Aseem Ul Islam,\\ Dennis S. Bernstein, and Ilya V. Kolmanovsky\\
	POC: Ilya V. Kolmanovsky (ilya@umich.edu)\\ \today }

\newif\ifPDF \ifx\pdfoutput\undefined\PDFfalse \else\ifnum\pdfoutput > 0\PDFtrue \else\PDFfalse \fi \fi
\ifPDF 
\fi

\usepackage{tikz}
\usetikzlibrary{shapes,arrows,calc}

\begin{document}
\maketitle
\CSMsetup

Among the multitude of modern control methods, model predictive control (MPC) is among the most successful \cite{cairano2018MPC,eren2017model,industrysamad}.
As noted in ``Summary,'' this success is largely due to the ability of MPC to respect constraints  on controls and enforce constraints on outputs, both of which are difficult to handle with linear control methods, such as LQR and LQG, and nonlinear control methods, such as feedback linearization and sliding mode control.
Although MPC is computationally intensive, it is more broadly applicable than Hamilton-Jacobi-Bellman-based control, and more suitable for feedback control than the minimum principle.
In many cases, the constrained optimization problem for MPC is convex, which facilitates computational efficiency \cite{boyd2004convex}.

In a nutshell, MPC uses a model of the system with constrained receding-horizon optimization to compute a sequence of future control inputs.
The first element of the optimized sequence is implemented, and the remaining elements are discarded.
At a first glance, this strategy seems ad hoc and wasteful.
However, MPC is known to be asymptotically stabilizing for a sufficiently long prediction horizon or with terminal constraints, and, if the initial state is feasible and the model matches the plant, it is recursively feasible with respect to constraints  \cite{keerthi1988optimal,camacho2013model,rawlings2017model,rakovic2019mpc}.

MPC handles convex control and output constraints efficiently compared to conventional feedback control methods that treat control constraints as input nonlinearities.
In particular, linear dynamics with control magnitude and move-size (rate) saturation comprise a Hammerstein system \cite{fruzzetti1997nolinear,bloemen2001model,zhang2018nonlinear}, whose importance is reflected by the enormous literature on bounded control and anti-windup strategies \cite{hu2001control,tarbouriech2011stability,bernstein1995sat}.
Within the context of MPC, however, magnitude and move-size control constraints are treated as optimization constraints rather than explicit nonlinearities, which simplifies the treatment of this problem.

From the perspective of classical control and modern optimal state-space control, MPC is a paradigm shift.
For example, conventional feedback methods use integral action to asymptotically follow step commands and reject step disturbances, and it is common practice to embed integrators within LQR and LQG controllers.
MPC, however, invokes numerical optimization at each step, and thus no integrator per se appears.
This raises the question of whether or not MPC can provide integral action \cite{rawlings2017model} and whether any such ``integral action'' is degraded by integrator windup.

Since MPC uses model-based prediction to determine future control inputs, it follows that these predictions require knowledge of future commands and disturbances.
Future commands are sometimes known;  this is the aim of preview control and trajectory-tracking methods \cite{limon2008mpc,goodwin2011preview}.
On the other hand, except for disturbance-feedforward control architectures \cite{kuomorgan}, disturbances are rarely known, especially in the future.
The lack of future knowledge of commands and disturbances thus represents a potential obstacle for MPC.

Although MPC is often based on linear models, this technique is also applicable to nonlinear systems \cite{grune2017nonlinear}.
An additional challenge is the case of output feedback, where not all of the plant states are measured.
In this case, state estimation can be used to provide estimates of unmeasured states, where the state estimates serve as ersatz states for the receding-horizon optimization  \cite{ricker1990model,Kolmanovsky2006MPCwithPE,ZHANG2020108974,ADETOLA2009320}.
Transient errors in the state estimates, however, can impede the ability to enforce constraints on unmeasured states.
Output-feedback MPC that does not rely on a state estimator is developed in \cite{clarke1987generalized,saraf2017fast,barlow2006direct,goulart_OFMPC}.

In addition to the lack of knowledge about future commands and disturbances, a potential weakness of MPC is the need for a sufficiently accurate model for effective predictive optimization.
This dependence is mitigated by robust MPC techniques \cite{Rakovic2013,cannon2016},
minimax techniques for minimizing the loss of a worst-case scenario, and robust tube-based MPC \cite{lofberg2003minimax,LANGSON2004125,fontes2017rigid}.

As an alternative to robust MPC, extensions of MPC to include online identification and learning are considered in \cite{shouche1998simultaneous,LORENZEN2019461,cairano2014dualMPC,allgower2019dual}.
These techniques can potentially overcome the worst-case considerations of robust MPC by allowing MPC to learn the true plant dynamics and disturbances.
For system identification, the commands, disturbances, and control inputs must provide sufficient {\it persistency} to facilitate identification.
The use of concurrent learning with MPC can be viewed as a form of indirect adaptive control \cite{ioannoufidan}.
Within this context, the role of persistency is a longstanding issue \cite{Chowdhary2014May}.

Beyond persistency, since online identification and learning occur during closed-loop operation, the control input is correlated with the measurements due to disturbances and sensor noise.
When RLS is used for closed-loop identification, as in the present article, this correlation may obstruct {\it consistency}, and thus lead to asymptotic bias in the parameter estimates \cite{forssell1999closed,aljanaideh2016closed,frantCLID}.
Alternative identification methods, such as instrumental variables, provide consistency despite signal correlation, albeit at higher computational cost \cite{IVCLID}.

Since MPC is an inherently discrete-time control technique, its application to systems with continuous-time dynamics entails a sampled-data system with analog-to-digital sampling and digital-to-analog input reconstruction.
As in any sampled-data controller implementation, the sampling rate must be chosen to minimize aliasing and folding effects  \cite{yuz2014sampled,astromwittenmark1997}.
Within this context, MPC provides direct digital control of discretized plants without the need to discretize continuous-time controllers.
The present article considers examples with sampled-data dynamics in order to assess the intersample response \cite{FONTES20179840,fontes_mpc_discontinuous,rolf_sampled_NMPC}.

The present article focuses on MPC for constrained linear systems with two nonstandard features.
First, only a limited number of states of the plant are assumed to be measured;  this is the case of output feedback rather than full-state feedback.
Instead of using an observer to provide estimates of unmeasured states, the present article takes advantage of the block observable canonical form \cite{polderman1989state}, which is a state-space realization whose state is a function of measured inputs and outputs.
The use of this realization thus removes the need to build an observer to estimate unmeasured states.
Within this framework, a key assumption is the availability of measurements of all constrained outputs.

The second feature of the present article is the use of online identification to construct and refine a model for the constrained receding-horizon optimization.
Concurrent identification is performed with recursive least squares (RLS) \cite{astrom,islam2019recursive} with the additional benefit of variable-rate forgetting \cite{adamVRF,ankit2020VDF}.

The present article describes output-feedback MPC with online identification (OFMPCOI), which incorporates the two features described above.
No attempt is made in this article to derive stability or performance guarantees for this approach.
Instead, the goal is to systematically investigate OFMPCOI through a collection of numerical examples.
These examples are chosen to highlight the stability and performance of OFMPCOI for a diverse collection of plants and scenarios.
These scenarios include the effect of model order, sensor noise, unknown step, harmonic, and broadband disturbances, stabilization, sampled-data effects, magnitude and move-size control constraints, output constraints, and abrupt and gradual changes in the plant.
The ability of OFMPCOI to emulate a nonlinear controller is investigated by considering a chain of integrators under magnitude saturation, which cannot be globally stabilized by saturating a linear stabilizing controller \cite{teel1992global}.
Unlike \cite{teel1992global}, which assumes continuous-time, full-state feedback control, OFMPCOI uses only a measurement of the output of the last integrator in the chain under sampled-data control.
The analogous scenario involving a chain of identical undamped oscillators is also considered.

The numerical examples in this article provide a venue for investigating the interplay between identification and control, a longstanding problem in control theory \cite{ID4control,GeversJoint,HjalmarssonFrom}.
This interplay is addressed by dual control, where the objective is to determine probing signals that enhance the speed and accuracy of the concurrent identification \cite{feldbaum1960dual,wittenmark1995adaptive,filatovdual2004}.
In contrast to the use of probing signals, OFMPCOI does not use a separate probing perturbation, such as the dither signal used to estimate gradients in extremum-seeking control \cite{krsticES}.
Instead, the present article investigates the phenomenon of {\it self-generated persistency}, where OFMPCOI automatically increases the persistency of the control signal in response to the closed-loop performance. 
Within the context of OFMPCOI, the interplay between identification and control is embodied by {\it exigency,} which refers to the ability of the identification to prioritize features of the identified model that impact closed-loop performance.
A key goal of the numerical examples is thus to explore manifestations of persistency, consistency, and exigency within OFMPCOI.
This article extends the preliminary investigation of OFMPCOI given in \cite{tam2020MPC,nima2020MPC} through refinements of the control algorithm, more extensive numerical examples, and investigation of persistency, consistency, and exigency.
Table \ref{tab:examples} lists the numerical examples that investigate these phenomena.
\begin{table}[!h]
\centering
    \begin{tabular}{|c|c|c|c|} \hline
         Example & Persistency & Consistency & Exigency  \\
         \hline
         1 & $\bullet$ & $\bullet$ & $\bullet$ \\ \hline
         2 & $\bullet$ & $\bullet$ & \\ \hline
         6 & & & $\bullet$ \\ \hline
         8 & & & $\bullet$ \\ \hline
         11 & & $\bullet$ & \\ \hline
         13 & & $\bullet$ & $\bullet$  \\ \hline
    \end{tabular}
    \caption{Numerical examples that investigate persistency, consistency, and exigency.}
    \label{tab:examples}
\end{table}

The next section defines the BOCF for an input-output  predictive model.
Next, the control objectives and control architecture of OFMPCOI are described.
Linear SISO examples 1--7 are then presented, followed by linear MIMO examples 8--13.
Conclusions and directions for future research are then presented.

\begin{tcolorbox}[
    standard jigsaw,
    title=Notation,
    opacityback=0,
]
The $n\times n$ identity matrix is denoted by $I_n,$ the $m\times n$ matrix of ones by $1_{m\times n},$ the $m\times n$ matrix of zeros by $0_{m\times n},$ and the Kronecker product by $\otimes.$
The infinity norm of a vector $x=[x_1 \ \cdots \ x_n]^\rmT\in\mathbb{R}^n$ is defined by $\|x\|_\infty \isdef \max(|x_1|,\ldots,|x_n|).$
\end{tcolorbox}

\section{Block Observable Canonical Form for Input-Output Models}
\label{sec:predictive_model}

To provide a framework for online system identification, we consider a MIMO input-output model of the form
\begin{align}
    {y}_{k} = -\sum_{i=1}^{n}{F}_{i} y_{k-i} + \sum_{i=0}^{n}{G}_{i} u_{k-i}, \label{eq:yhat}
\end{align}
where $k\ge0$ is the time step, $n\ge1$ is the data window, $u_k\in\mathbb{R}^{m}$
is the control, $y_k\in\mathbb{R}^{p}$ is the measurement, and ${F}_{i}\in\mathbb{R}^{p\times p}$ and ${G}_{i}\in\mathbb{R}^{p\times m}$ are model coefficients.
Next, we construct the block observable canonical form (BOCF) state-space realization of \eqref{eq:yhat} given by
\begin{align}
    x_{k+1} & = Ax_{k} + Bu_{k}, \label{eq:x1k} \\
    y_k & = Cx_k + D u_{k}, \label{eq:yk}
\end{align}
where the state $x_{k} \isdef [ x_{1,k}^\rmT \  \cdots \  x_{{n},k}^\rmT ]^\rmT \in \mathbb{R}^{{n}p}$ is constructed as
\begin{align}
    x_{1,k} & \isdef y_{k} - {G}_{0}u_{k}, \label{eq:xhat1} \\
    x_{i,k} & \isdef -\sum_{j=1}^{n-i+1}{F}_{i+j-1} y_{k-j} + \sum_{j=1}^{n-i+1}{G}_{i+j-1} u_{k-j}, \quad i = 2, \ldots,{n},\label{eq:xhati}
\end{align}
and ${A} \in\mathbb{R}^{{n}p\times {n}p},$ ${B} \in\mathbb{R}^{{n}p\times m},$ ${C}\in\mathbb{R}^{p\times {n}p},$ and ${D}\in\mathbb{R}^{p\times m}$ are defined by 
\begin{align}
    {A} & \isdef
    \left[
    \begin{matrix}
        -{F}_{1} & I_p & \cdots & \cdots & 0_{p\times p} \\
        \vdots & 0_{p\times p} & \ddots & & \vdots \\
        \vdots & \vdots & \ddots & \ddots & 0_{p\times p} \\
        \vdots & \vdots & & \ddots & I_p \\
        -{F}_{{n}} & 0_{p\times p} & \cdots & \cdots & 0_{p\times p}
    \end{matrix}
    \right], \quad
    {B} \isdef
    \left[
    \begin{matrix}
        {G}_{1} - {F}_{1}{G}_{0} \\
        {G}_{2} - {F}_{2}{G}_{0} \\
        \vdots \\
        {G}_{{n}} - {F}_{{n}}{G}_{0}
    \end{matrix}
    \right], \label{ABmatrices}\\
    {C} &\isdef [ I_p \ \ 0_{p\times p} \  \ \cdots \ \ 0_{p\times p} ], \quad {D} \isdef {G}_{0}.\label{CDmatrices}
\end{align}
The model structure \eqref{eq:yhat} and its realization \eqref{ABmatrices}, \eqref{CDmatrices}, which extends the SISO construction given in \cite{polderman1989state}, provides the framework for system identification and output prediction described below.

Based on the structure of \eqref{eq:yhat}, the 1-step predicted output is given by
\begin{align}
    {y}_{1|k} = -\sum_{i=1}^{\hat{n}}\hat{F}_{i,k} y_{k+1-i}
    + \hat{G}_{0,k} u_{1|k} + \sum_{i=1}^{\hat{n}}\hat{G}_{i,k} u_{k+1-i}, \label{eq:ypred}
\end{align}
where $u_{1|k} \in \mathbb{R}^{m}$ is the 1-step computed input, $\hat{F}_{i,k}\in\mathbb{R}^{p\times p}$ and $\hat{G}_{i,k}\in\mathbb{R}^{p\times m}$ are coefficient matrices estimated at step $k,$ and $\hat{n}\ge 1$ is the data window used for estimation.
Similarly, using the 1-step predicted output $y_{1|k}$ and the 2-step computed control $u_{2|k},$ the 2-step predicted output is given by
\begin{align}
    {y}_{2|k} &= - \hat{F}_{1,k} {y}_{1|k} -\sum_{i=2}^{\hat{n}}\hat{F}_{i,k} y_{k+2-i} +  \hat{G}_{0,k}u_{2|k} + \hat{G}_{1,k}u_{1|k} +\sum_{i=2}^{\hat{n}}\hat{G}_{i,k} u_{k+2-i}. \label{eq:yhat2}
\end{align}
In general, for all $j\ge1,$ the $j$-step predicted output is given by
\begin{align}
    y_{j|k} = - \sum_{i=1}^{j-1} \hat{F}_{i,k} y_{j-i|k} - \sum_{i=j}^{\hat{n}} \hat{F}_{i,k} y_{k+j-i} + \sum_{i=0}^{j-1}\hat{G}_{i,k}u_{j-i|k} + \sum_{i=j}^{\hat{n}} \hat{G}_{i,k} u_{k+j-i},
\end{align}
where the first term is zero for $j=1,$ and the second and fourth terms are zero for $j>\hat{n}.$

Next, based on the structure of \eqref{eq:x1k}, we define the 1-step predicted state $x_{1|k}$ by
\begin{align}
    x_{1|k} \isdef \hat{A}_k \hat{x}_k + \hat{B}_k u_k, \label{eq:xpred1}
\end{align}
where $\hat{x}_{k} \isdef [ \hat{x}_{1,k}^\rmT \  \cdots \  \hat{x}_{\hat{n},k}^\rmT ]^\rmT \in \mathbb{R}^{\hat{n}p},$ and
\begin{align}
    \hat{x}_{1,k} & \isdef y_{k} - \hat{G}_{0,k}u_{k}, \label{eq:xhat1k} \\
    \hat{x}_{i,k} & \isdef -\sum_{j=1}^{\hat{n}-i+1}\hat{F}_{i+j-1,k} y_{k-j} + \sum_{j=1}^{\hat{n}-i+1}\hat{G}_{i+j-1,k} u_{k-j}, \quad i = 2, \ldots,\hat{n}, \label{eq:xhatik} \\
    \hat{A}_k & \isdef
    \left[
    \begin{matrix}
        -\hat{F}_{1,k} & I_p & \cdots & \cdots & 0_{p\times p} \\
        \vdots & 0_{p\times p} & \ddots & & \vdots \\
        \vdots & \vdots & \ddots & \ddots & 0_{p\times p} \\
        \vdots & \vdots & & \ddots & I_p \\
        -\hat{F}_{\hat{n},k} & 0_{p\times p} & \cdots & \cdots & 0_{p\times p}
    \end{matrix}
    \right], \quad
    \hat{B}_k \isdef
    \left[
    \begin{matrix}
        \hat{G}_{1,k} - \hat{F}_{1,k}\hat{G}_{0,k} \\
        \hat{G}_{2,k} - \hat{F}_{2,k}\hat{G}_{0,k} \\
        \vdots \\
        \hat{G}_{\hat{n},k} - \hat{F}_{\hat{n},k}\hat{G}_{0,k}
    \end{matrix}
    \right].
\end{align}
Note that \eqref{eq:xhat1k} and \eqref{eq:xhatik} depend on past values of the measurement $y$ and control $u$ rather than the 1-step predicted output $y_{1|k}$  given by \eqref{eq:ypred} and the 1-step computed input $u_{1|k}$ given in the next section.
Hence, $x_{1|k}$ given by \eqref{eq:xpred1} is not necessarily equal to $\hat x_{k+1}.$

Using \eqref{eq:xpred1}, the 1-step predicted output given by 
\eqref{eq:ypred} can be rewritten as
\begin{align}
    y_{1|k} = Cx_{1|k} + \hat{D}_{k} u_{1|k}, \label{eq:BOCF2}
\end{align}
where $\hat{D}_k \isdef \hat{G}_{0,k}.$
Similarly, defining the $i$-step predicted state
\begin{align}
    x_{i|k} \isdef \hat{A}_k x_{i-1|k} + \hat{B}_k u_{i-1|k}, \quad i\ge 2,
\end{align}
the $i$-step predicted output is given by
\begin{align}
    y_{i|k} = C x_{i|k} + \hat{D}_k u_{i|k}, \quad i\ge 2.
\end{align}
Note that the $i$-step prediction at step $k$ uses the current estimates $\hat{A}_k,$ $\hat{B}_k,$ and $\hat{D}_k$ at each intermediate stage of the prediction horizon.

Next, defining
\begin{align}
Y_{1|k,\ell} \isdef  \left[\begin{matrix}{y}_{1|k} \\ \vdots \\ {y}_{\ell|k}\end{matrix}\right] \in \mathbb{R}^{\ell p},
\quad
{U}_{1|k,\ell} \isdef \left[\begin{matrix}{u}_{1|k} \\  \vdots \\ {u}_{\ell|k}\end{matrix}\right] \in \mathbb{R}^{\ell m}, \label{eq:YU}
\end{align}
it follows that
\begin{align}
    {Y}_{1|k,\ell} & = \hat{\Gamma}_{k,\ell} x_{1|k} + \hat{T}_{k,\ell} {U}_{1|k,\ell}, \label{eq:Yhat_pred}
\end{align}
where $\hat{\Gamma}_{k,\ell} \in \mathbb{R}^{\ell p \times \hat{n}p}$ and $\hat{T}_{k,\ell} \in \mathbb{R}^{\ell p\times \ell m}$ are the observability and block-Toeplitz matrices  
\begin{align}
    \hat{\Gamma}_{k,\ell} & \isdef
    \left[
    \begin{matrix}
        C \\
        C\hat{A}_{k} \\
        \vdots \\
        C\hat{A}_{k}^{\ell-1}
    \end{matrix}
    \right], \quad
    \hat{T}_{k,\ell} \isdef
    \left[
    \begin{matrix}
        \hat{D}_{k}   & 0_{p\times m} & \cdots & \cdots & \cdots  & \cdots &  0_{p\times m} \\
        \hat{H}_{k,1} & \hat{D}_{k} & \cdots & \cdots & \cdots  & \cdots & 0_{p\times m} \\
        \hat{H}_{k,2} & \hat{H}_{k,1} & \hat{D}_{k} & \cdots & \cdots  & \cdots & 0_{p\times m} \\
        \hat{H}_{k,3} & \hat{H}_{k,2} & \hat{H}_{k,1} & \hat{D}_{k} & \cdots & \cdots & 0_{p\times m} \\
        \hat{H}_{k,4} & \hat{H}_{k,3} & \hat{H}_{k,2} & \ddots & \ddots & & \vdots \\
        \vdots & \vdots & \vdots & \ddots & \ddots & \ddots & 0_{p\times m} \\
        \hat{H}_{k,\ell-1} & \hat{H}_{k,\ell-2} & \hat{H}_{k,\ell-3} & \cdots & \hat{H}_{k,2} & \hat{H}_{k,1} & \hat{D}_{k}
    \end{matrix}
    \right], \label{eq:toeplitz_pred}
\end{align}
where, for all $i=1,\ldots,\ell-1,$ $\hat{H}_{k,i}\in\mathbb{R}^{p \times m}$ is defined by
\begin{align}
    \hat{H}_{k,i} & \isdef C\hat{A}_{k}^{i-1}\hat{B}_{k}. \label{eq:hankel_pred}
\end{align}

\section{Tracking Output and Output Constraints}

Define the tracking output $y_{\rmt,k}\in \mathbb{R}^{p_\rmt}$ by
\begin{align}
y_{\rmt,k} \isdef C_\rmt y_k,  \label{eq:yt}
\end{align}
where $C_\rmt\in\mathbb{R}^{p_\rmt\times p}.$
The performance objective is to have $y_{\rmt,k}$ follow a command trajectory $r_k\in\mathbb{R}^{p_\rmt},$ whose future values may or may not be known, and thus command preview may or may not be available.

In addition to the performance objective, we define the constrained output  $y_{\rmc,k}\in\mathbb{R}^{p_\rmc}$ by
\begin{align}
y_{\rmc,k} \isdef C_\rmc y_k, \label{eq:yc}
\end{align}
where $C_\rmc\in\mathbb{R}^{p_\rmc\times p}.$
The objective is to enforce the inequality constraint 
\begin{align}
    \SC y_{\rmc,k} + \SD \leq 0_{n_\rmc\times 1}, \label{eq:constraint}
\end{align}
where $\SC \in\mathbb{R}^{n_\rmc\times p_\rmc}$ and $\SD \in\mathbb{R}^{n_\rmc}.$
Note that \eqref{eq:constraint}, where ``$\le$'' is interpreted component-wise, defines a convex set.

\section{Control Constraints}
 
In all physical systems, the control is constrained in both magnitude and rate.
The magnitude control constraint has the form
\begin{align}
    u_{\mathrm{min}} \le u_k \le u_{\mathrm{max}}, \label{eq:magsat}
\end{align}
where $u_{\mathrm{min}}\in\mathbb{R}^m$ is the vector of the minimum control magnitudes  and $u_{\mathrm{max}}\in\mathbb{R}^m$ is the vector of maximum control magnitudes.

In addition, the move-size control constraint has the form
\begin{align}
    \Delta u_{\mathrm{min}} \le u_{k+1}-u_k \le \Delta u_{\mathrm{max}}, \label{eq:movesat}
\end{align}
where $\Delta u_{\mathrm{min}}\in\mathbb{R}^m$ is the vector of minimum control move sizes  and $\Delta u_{\mathrm{max}}\in\mathbb{R}^m$ is the vector of maximum control move sizes.

\section{Control Architecture}

The control architecture combines output-feedback MPC with online identification.  In particular, these components are:
\begin{itemize}
    \item[{\it i})] \textbf{Identification}: The coefficients of an input-output model are estimated by RLS to update the identified model for use by MPC.
    \item[{\it ii})] \textbf{MPC}: The control input is updated by applying QP to the identified model.
\end{itemize}
Figure~\ref{fig:CEP_scheme} shows the control architecture of OFMPCOI.

\tikzstyle{block} = [draw, fill=gray!20, rectangle, 
    minimum height=3em, minimum width=6em]
\tikzstyle{input} = [coordinate]
\tikzstyle{output} = [coordinate]
\tikzstyle{pinstyle} = [pin edge={to-,thin,black}]

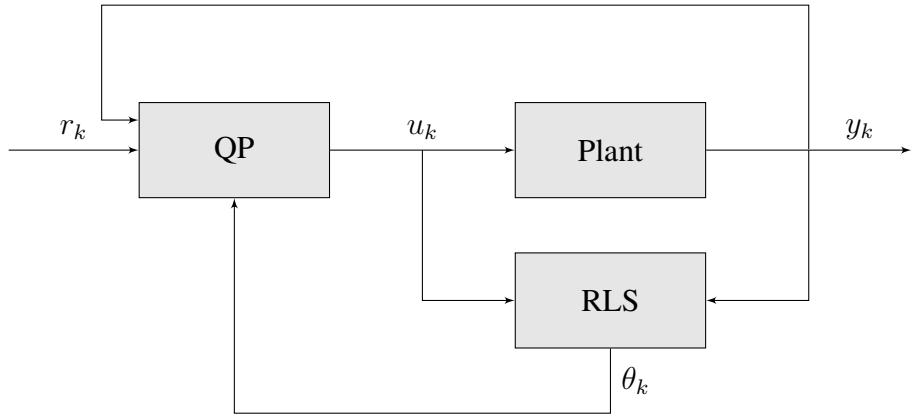
\begin{figure}[ht]
    \centering
    \begin{tikzpicture}[auto, node distance=4cm,>=latex']
    \node [input, name=input] {};
    \node [block, right of=input, node distance=3cm] (controller) {QP};
    \node [block, right of=controller,
            node distance=5cm] (system) {Plant};
    \draw [->] (controller) -- node[name=u] {$u_k$} (system);
    \node [output, right of=system] (output) {};
    \node [block, below of=system, align=center, draw, node distance=2cm] (ID) {RLS};
    \node [output, name=coefficient, below of=ID, node distance=1.5cm] {};

    \draw [draw,->] (input) -- node {$r_k$} (controller);
    \draw [->] (system) -- node [near end] {$y_k$}(output);
    \path (system) -- (output) node[midway] (y) {};
    \node [output, name=tmp, above of=y, node distance=1.5cm+8] {};
    \node [left of=tmp, node distance = 9.4cm] (ydown) {};
    \draw [->] (y) |- (ID);
    \draw [->] (u) |- (ID);
    \draw [->] (ID) -- node {$\theta_k$} (coefficient) -| node {} (controller);
    \draw [->,shorten <=-8] (y) -- (tmp) -- (ydown.center) |- ([yshift=0.4cm]controller.west);
    \end{tikzpicture}
    \caption{Output-feedback model predictive control with online identification.}
    \label{fig:CEP_scheme}
\end{figure}

\noindent {\it Online Identification}

For online identification, RLS is used to estimate the coefficients of an input-output model.
In particular, for all $k\ge0,$ RLS minimizes the cumulative cost
\begin{align}
    J_k(\hat{\theta}) = \sum_{i=0}^k \frac{\rho_i}{\rho_k}z_i^\rmT(\hat{\theta}) z_i(\hat{\theta}) + \frac{1}{\rho_k}(\hat{\theta}-\theta_0)^\rmT P_0^{-1}(\hat{\theta}-\theta_0), \label{eq:J}
\end{align}
where $\rho_k \isdef \prod_{j=0}^k \lambda_j^{-1} \in \mathbb{R},$ $\lambda_k\in(0,1]$ is the forgetting factor, $P_0\in\mathbb{R}^{[\hat{n}p(m+p)+mp]\times [\hat{n}p(m+p)+mp]}$ is positive definite, and
$\theta_0\in\mathbb{R}^{[\hat{n}p(m+p)+mp] \times 1}$ is the initial estimate of the coefficient vector.
The performance variable $z_k(\hat{\theta}) \in\mathbb{R}^{p}$ is defined by
\begin{align}
     z_{k}(\hat{\theta}) \isdef y_k + \sum_{i=1}^{\hat{n}} \hat{F}_{i} y_{k-i} - \sum_{i=0}^{\hat{n}}\hat{G}_{i} u_{k-i}, \label{eq:zid1}
\end{align}
where the vector of coefficients to be estimated $\hat{\theta}\in\mathbb{R}^{[\hat{n}p(m+p)+mp] \times 1}$ is defined by
\begin{align}
\hat{\theta}  \isdef \vek  \left[\begin{matrix} \hat{F}_{1} & \cdots & \hat{F}_{\hat{n}} & \hat{G}_{0} & \cdots & \hat{G}_{\hat{n}} \end{matrix}\right]^\rmT.
\end{align}
Defining the regressor matrix $\phi_{k}\in\mathbb{R}^{p\times [\hat{n}p(m+p)+mp]}$ by
\begin{align}
\phi_k \isdef I_p\otimes\left[\begin{matrix}-y_{k-1}^\rmT & \cdots & -y_{k-\hat{n}}^\rmT & u_{k}^\rmT & \cdots & u_{k-\hat{n}}^\rmT \end{matrix}\right], \label{eq:regressor}
\end{align}
it follows that the performance variable \eqref{eq:zid1} can be rewritten as
\begin{align}
    z_{k}(\hat{\theta}) = y_{k} - \phi_{k} \hat{\theta}. \label{eq:zid}
\end{align}
Note that, with \eqref{eq:zid}, the cost function \eqref{eq:J} is strictly convex and quadratic, and thus has a unique global minimizer.
The unique global minimizer $\theta_{k+1} \isdef \mathrm{argmin}_{\hat{\theta}} J_k(\hat{\theta})$ is computed by RLS as:
\begin{tcolorbox}[
    standard jigsaw,
    title=RLS Online Identification,
    opacityback=0,
]
\begin{align}
    L_{k} & = \lambda_{k}^{-1} P_{k}, \label{eq:RLS1} \\
    P_{k+1} & = L_{k} - L_{k}\phi_{k}^\rmT (I_{p} + \phi_{k} L_{k} \phi_{k}^\rmT)^{-1}\phi_{k} L_{k}, \label{eq:RLS2} \\
    \theta_{k+1} & = \theta_{k} + P_{k+1}\phi_{k}^\rmT(y_{k} - \phi_{k}\theta_{k}). \label{eq:RLS3}
\end{align}
\end{tcolorbox}

The step-dependent parameter $\lambda_k$ is the {\it forgetting factor.} 
In the case where $\lambda_k$ is constant, RLS uses {\it constant-rate forgetting} (CRF);  otherwise, RLS uses {\it variable-rate forgetting} (VRF) \cite{adamVRF}.
For VRF, $\lambda_k$ is given by 
\begin{align}
    \lambda_{k} &= \frac{1}{1 + \eta g( z_{k-\tau_\rmd}, \ldots , z_k) {\bf 1}[g( z_{k-\tau_\rmd}, \ldots , z_k)]   }, \label{eq:vrf}
\end{align}
where $\mathbf{1}\colon\mathbb{R}\to\{0,1\}$ is the unit step function, where $\mathbf{1}(x)=0$ for $x<0$ and $\mathbf{1}(x)=1$ for $x\ge0,$ and
\begin{align}
 g( z_{k-\tau_\rmd}, \ldots , z_k) \isdef
\frac{\sqrt{  \frac{1}{\tau_\rmn}   \sum_{i=k-\tau_\rmn}^k z_i^\rmT z_i  }   }{  \sqrt{ \frac{1}{\tau_\rmd}  \sum_{i=k-\tau_\rmd}^k  z_i^\rmT z_i } }-1. \label{eq:g}
\end{align}
In \eqref{eq:vrf} and \eqref{eq:g}, $\eta\ge0$ and $0<\tau_{\rmn} < \tau_{\rmd}$ are numerator and denominator window lengths, respectively. 
Define $g(0, \ldots , 0)\isdef 0.$  
If the sequence $z_{k-\tau_\rmd}, \ldots , z_k$ is zero-mean noise, then the numerator and denominator of \eqref{eq:g} approximate the average standard deviation of the noise over the intervals $[k-\tau_\rmn,k]$ and $[k-\tau_\rmd,k]$, respectively.
In particular, by choosing $\tau_\rmd>>\tau_\rmn$, it follows that the denominator of \eqref{eq:g} approximates the long-term-average standard deviation of $z_k,$ whereas the numerator of \eqref{eq:g} approximates the short-term-average standard deviation of $z_k.$
Consequently, the case $g(z_{k-\tau_\rmd}, \ldots , z_k)>1$ implies that the short-term-average standard deviation of $z_k$ is greater than the  long-term-average standard deviation of $z_k.$
The function $g(z_{k-\tau_\rmd}, \ldots , z_k)$ used in VRF suspends forgetting when the short-term-average standard deviation of $z_k$ drops below the long-term-average standard deviation of $z_k.$
This technique thus prevents forgetting in RLS-based online identification due to zero-mean sensor noise with constant standard deviation rather than due to the magnitude of the noise-free identification error.

\noindent {\it Control Update}

Let $\SR_{k,\ell}\isdef [r_{k+1}^\rmT \ \ldots r_{k+\ell}^\rmT ]^\rmT \in \mathbb{R}^{\ell p_\rmt}$ be the vector of $\ell$ future commands, 
define $C_{\rmt,\ell}\isdef I_\ell\otimes C_\rmt \in \mathbb{R}^{\ell p_\rmt \times \ell p},$ let $Y_{\rmt,1|k,\ell}\isdef C_{\rmt,\ell}Y_{1|k,\ell},$ where $Y_{1|k,\ell}$ is given by \eqref{eq:Yhat_pred}, be the $\ell$-step propagated tracking-output vector, 
define ${\SC}_{\ell} \isdef I_\ell \otimes (\SC C_\rmc) \in\mathbb{R}^{\ell n_\rmc \times \ell p}$ and $\SD_\ell \isdef 1_{\ell \times 1} \otimes \SD \in\mathbb{R}^{\ell n_\rmc},$
and define the sequence of differences of computed control inputs by
\begin{align}
    \Delta U_{1|k,\ell} \isdef [(u_{1|k}-u_k)^\rmT \ (u_{2|k}-u_{1|k})^\rmT \ \cdots \ (u_{\ell|k}-u_{\ell-1|k})^\rmT]^\rmT \in \mathbb{R}^{\ell m}.
\end{align}
With this notation, QP-based MPC is given by:
\begin{tcolorbox}[
    standard jigsaw,
    title=QP-based MPC,
    opacityback=0,
]
    \begin{align}
        \min_{U_{1|k,\ell}} (Y_{\rmt,1|k,\ell} - R_{k,\ell})^\rmT Q (Y_{\rmt,1|k,\ell}-R_{k,\ell}) + (\Delta U_{1|k,\ell})^\rmT R \Delta U_{1|k,\ell} \label{eq:QPnoslack}
    \end{align}
    subject to 
    \begin{align}
        \SC_{\ell}Y_{1|k,\ell} + \SD_{\ell} & \leq 0_{\ell n_\rmc}, \label{eq:Ycon} \\
        U_{\mathrm{min}} \leq U_{1|k,\ell} & \leq U_{\mathrm{max}}, \label{eq:Ucon} \\
        \Delta U_{\mathrm{min}} \leq \Delta U_{1|k,\ell} & \leq \Delta U_{\mathrm{max}}, \label{eq:DeltaUcon}
    \end{align}
\end{tcolorbox}
\noindent where $Q \isdef \left[\begin{smallmatrix}\bar{Q} & 0_{p_\rmt\times p_\rmt} \\ 0_{p_\rmt\times p_\rmt} & \bar{P} \end{smallmatrix}\right]\in\mathbb{R}^{\ell p_\rmt \times \ell p_\rmt}$ is the positive-definite output weight,
$\bar{Q}\in\mathbb{R}^{(\ell-1) p_\rmt \times (\ell-1) p_\rmt}$ is the positive-definite cost-to-go output weight,
$\bar{P}\in\mathbb{R}^{p_\rmt\times p_\rmt}$ is the positive-definite terminal output weight, $R\in\mathbb{R}^{\ell m \times \ell m}$ is the positive-definite control weight, $U_{\mathrm{min}}\isdef 1_{\ell\times 1} \otimes u_{\mathrm{min}}\in\mathbb{R}^{\ell m},$ $U_{\mathrm{max}}\isdef 1_{\ell\times 1} \otimes u_{\mathrm{max}}\in\mathbb{R}^{\ell m},$ $\Delta U_{\mathrm{min}} \isdef 1_{\ell\times1}\otimes \Delta u_{\mathrm{min}}\in\mathbb{R}^{\ell m},$ and $\Delta U_{\mathrm{max}}\isdef 1_{\ell\times1}\otimes \Delta u_{\mathrm{max}}\in\mathbb{R}^{\ell m}.$
Since $R$ is positive definite, QP-based MPC is a strictly convex optimization problem.


It may occur in practice that the constraint \eqref{eq:Ycon} on the predicted output cannot be satisfied for all values of the control input that satisfy \eqref{eq:magsat} and \eqref{eq:movesat}; in this case, QP-based MPC is {\it infeasible}.
To overcome this problem, a standard technique is to introduce a slack variable $\varepsilon \in\mathbb{R}^{\ell n_\rmc}$ in order to relax the constraint \eqref{eq:Ycon}.
With this modification, the QP-based MPC becomes:
\begin{tcolorbox}[
    standard jigsaw,
    title=QP-based MPC with Output-Constraint Relaxation,
    opacityback=0,
]
    \begin{align}
        \min_{U_{1|k,\ell},\varepsilon} (Y_{\rmt,1|k,\ell} - R_{k,\ell})^\rmT Q (Y_{\rmt,1|k,\ell}-R_{k,\ell}) + (\Delta U_{1|k,\ell})^\rmT R \Delta U_{1|k,\ell} + \varepsilon^\rmT S \varepsilon \label{eq:QPwithslack}
    \end{align}
    subject to
    \begin{align}
        \SC_{\ell}Y_{1|k,\ell} + \SD_{\ell} & \leq \varepsilon, \\
        U_{\mathrm{min}} \leq U_{1|k,\ell} & \leq U_{\mathrm{max}}, \label{eq:Ucon2} \\
        \Delta U_{\mathrm{min}} \leq \Delta U_{1|k,\ell} & \leq \Delta U_{\mathrm{max}}, \label{eq:DeltaUcon2} \\
        0_{\ell n_\rmc \times 1} & \leq \varepsilon, \label{eq:slackconstraint}
    \end{align}
\end{tcolorbox}
\noindent where $S\in\mathbb{R}^{\ell n_\rmc \times \ell n_\rmc}$ is the positive-definite relaxation weight.
\noindent Since $R$ and $S$ are positive definite,  \eqref{eq:QPwithslack}--\eqref{eq:slackconstraint} is a strictly convex optimization problem.

To solve \eqref{eq:QPnoslack}--\eqref{eq:DeltaUcon} and \eqref{eq:QPwithslack}--\eqref{eq:slackconstraint}, we use the accelerated dual gradient-projection algorithm, where the previously computed Lagrange multipliers are used as a warmstart for the next iteration. 
For real-time implementation,  the control computed between $k$ and $k+1$ is implemented at step $k+1.$

\clearpage

\section{Linear SISO Examples}
This section considers linear, time-invariant, single-input single-output (SISO) plants, where $y_{\rmt,k} = y_{\rmc,k} = y_k.$
None of the examples of this article consider command preview

In particular, for all command-following examples,   $R_{k,\ell}=1_{\ell\times 1}\otimes r_k,$
which implies that future commands over the prediction horizon are assumed to be equal to the current command.

We consider both discrete-time and sampled-data plants.
For all sampled-data examples, in this and later sections, the analog-to-digital conversion is an instantaneous sampler, and the digital-to-analog conversion is a zero-order-hold device.
OFMPCOI is run between samples, and the continuous-time dynamics are integrated by ode45.
All commands are discrete-time signals, all deterministic disturbances are discretized at the integration step size, and all stochastic disturbances are modeled as constant between sampling times.

\begin{figure*}[t!]
    \centering
    \begin{subfigure}[t]{0.49\textwidth}
        \centering
        \includegraphics[trim = 10 100 10 30, width=\textwidth]{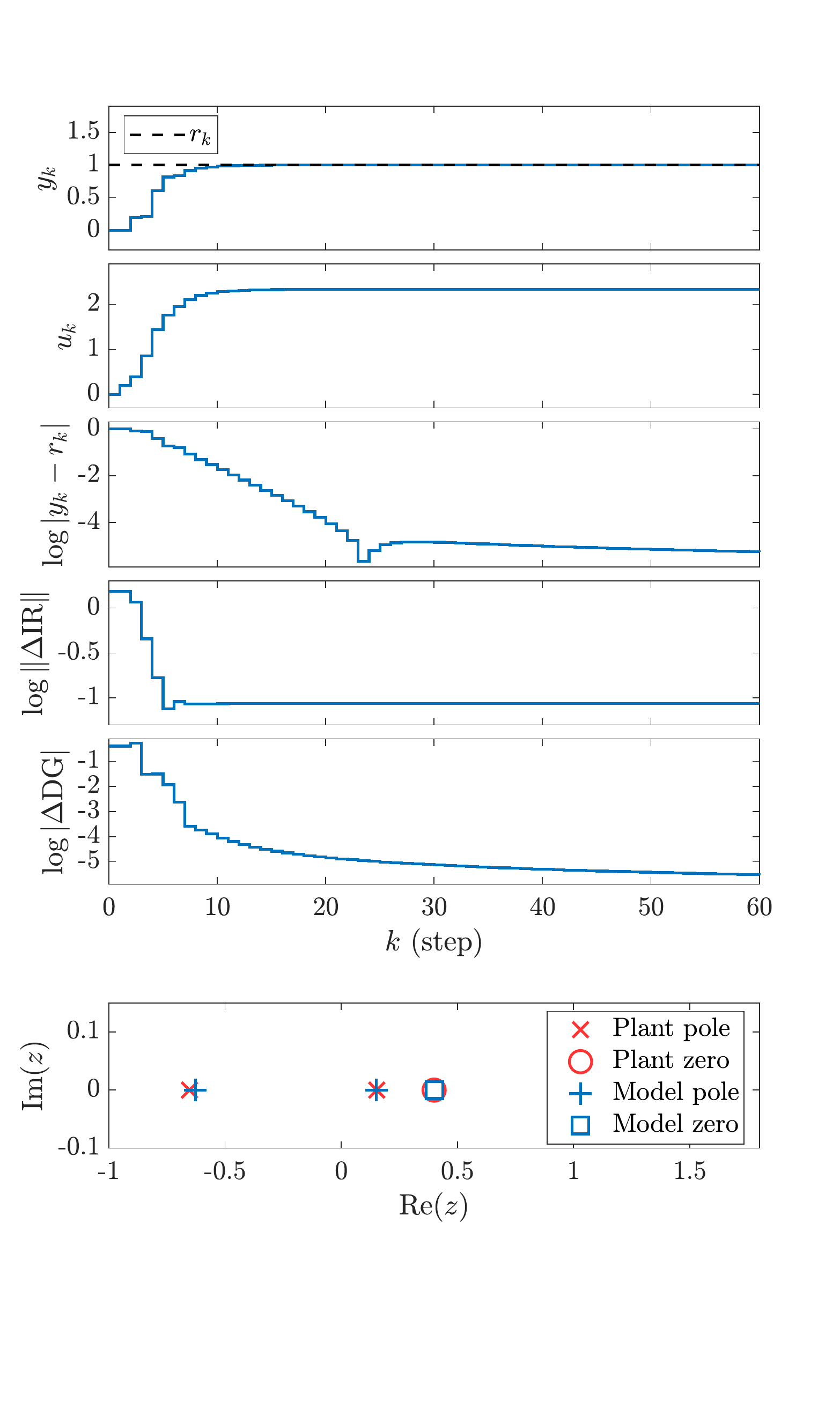}
        \caption{}
        \label{fig:ex1a}
    \end{subfigure}%
    ~ 
    \begin{subfigure}[t]{0.49\textwidth}
        \centering
        \includegraphics[trim = 10 100 10 30, width=\textwidth]{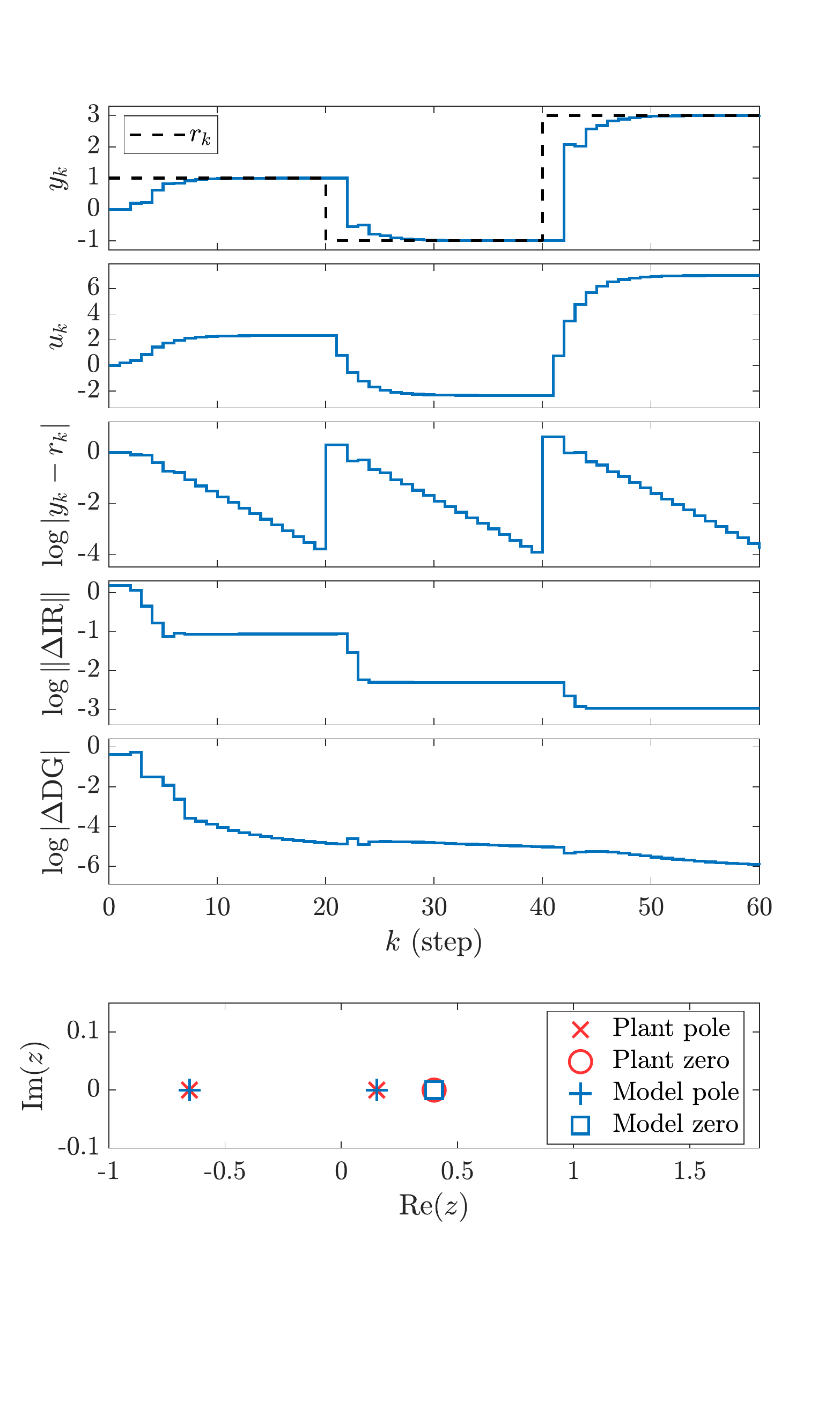}
        \caption{}
        \label{fig:ex1b}
    \end{subfigure}
    \caption{
    Example \ref{ex:ex1}.
    Command following for the SISO, asymptotically stable, discrete-time plant \eqref{eq:ex1} using $\hat{n}=n=2$ with a strictly proper model, that is, with $\hat G_0 = 0.$ 
    (a) Single-step command $r_k\equiv1.$
    The output $y_k$ approaches the step command with decreasing command-following error, which is $5.6e\mbox{-6}$ at $k=60.$ 
    Note that, although the impulse-response error
    $\Delta \mathrm{IR}$ stops decreasing after the initial transient, the estimate of the DC gain continues to be refined by RLS for the remainder of the simulation, which indicates exigency within the closed-loop identification; the identification of the DC gain is prioritized by RLS to decrease the command-following error. 
    The bottom-most plot compares the poles and zero of the identified model at $k=60$ to the poles and zero of the plant.
    (b) Multi-step command $r_k$ given by \eqref{eq:commandex1}. 
    For each step command, $y_k$ approaches the command with decreasing command-following error, which is $1.6e\mbox{-4}$ at $k=19,$ $1.2e\mbox{-4}$ at $k=39,$ and $1.6e\mbox{-4}$ at $k=60.$
    Note that, although $\Delta \mathrm{IR}$ stops decreasing after each transient, the estimate of the DC gain continues to be refined by RLS until the command changes.
    The bottom-most plot compares the poles and zero of the identified model at $k=60$ to the poles and zero of the plant.
    }
    \label{fig:ex1ab}
\end{figure*}

\begin{figure*}[t!]
    \centering
    \begin{subfigure}[t]{0.49\textwidth}
        \centering
        \includegraphics[trim = 10 100 10 30, width=\textwidth]{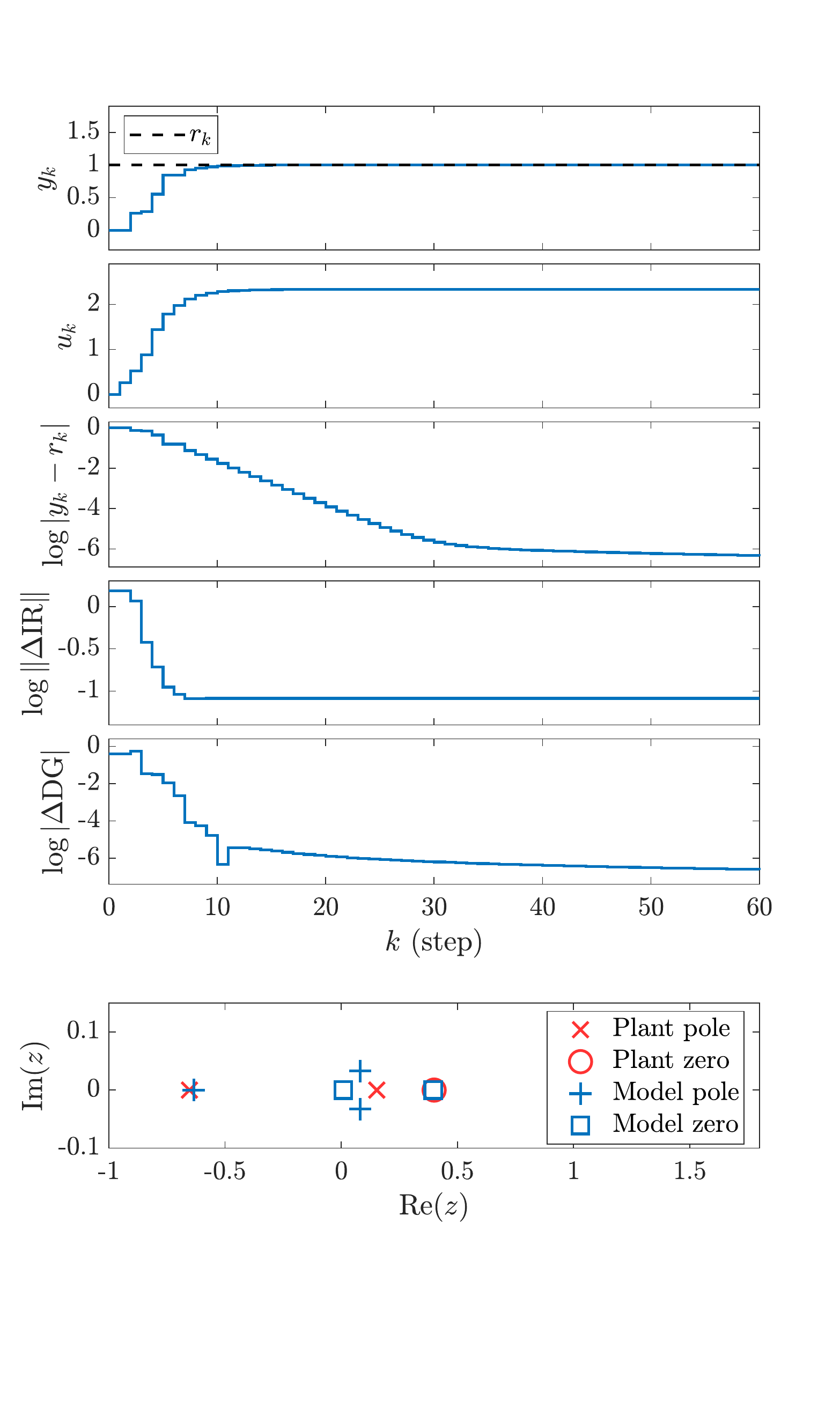}
        \caption{}
        \label{fig:ex1c}
    \end{subfigure}%
    ~ 
    \begin{subfigure}[t]{0.49\textwidth}
        \centering
        \includegraphics[trim = 10 100 10 30, width=\textwidth]{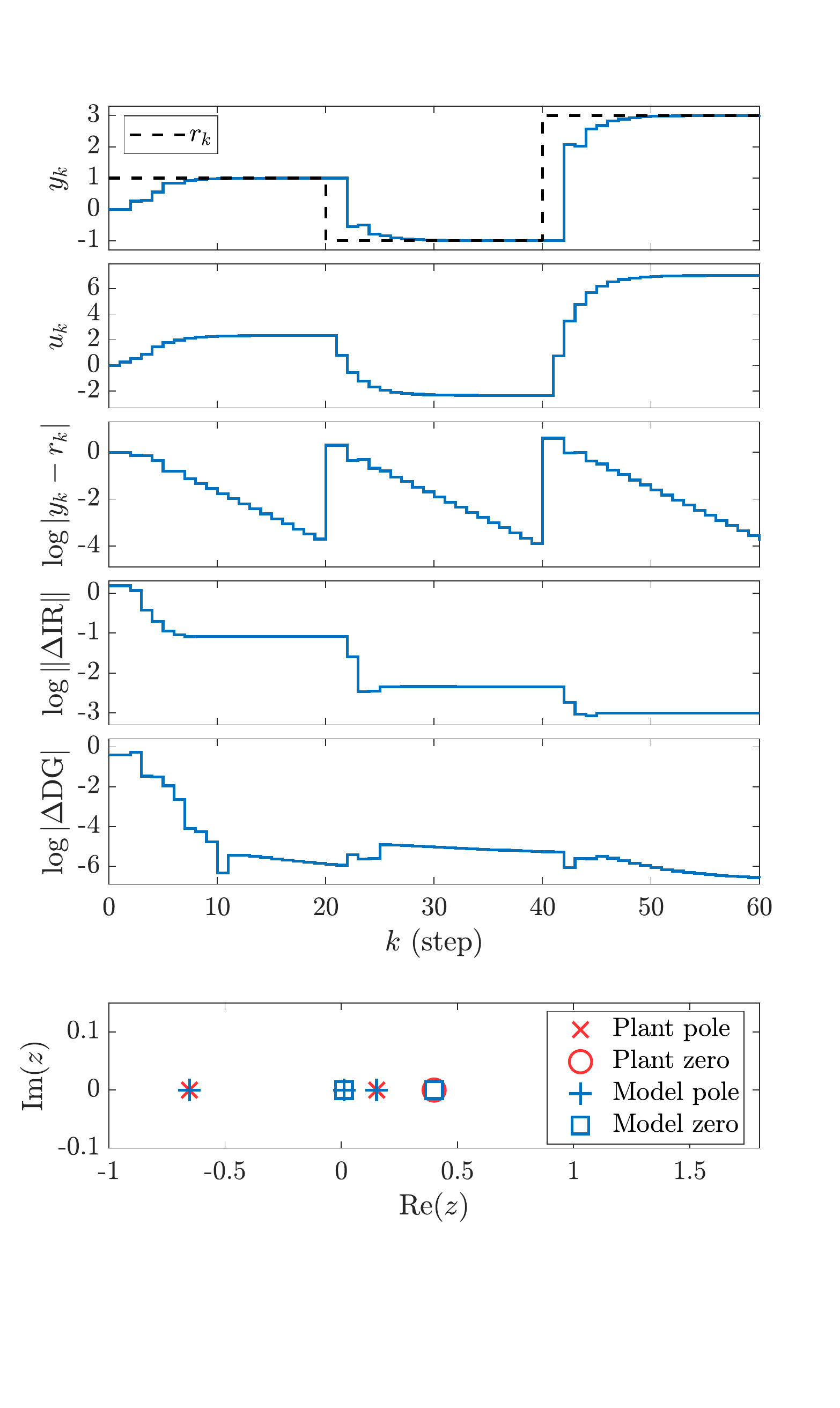}
        \caption{}
        \label{fig:ex1d}
    \end{subfigure}
    \caption{
    Example \ref{ex:ex1}.
    Command following for the SISO, asymptotically stable, discrete-time plant \eqref{eq:ex1} using $\hat{n}=3 > n = 2$  with a strictly proper model.
    (a) Single-step command $r_k\equiv1.$
    The output $y_k$ approaches the step command with decreasing command-following error, which is $ 4.7e\mbox{-7}$ at $k=60.$ 
    (b) Multi-step command $r_k$ given by \eqref{eq:commandex1}.
    For each step command, $y_k$ approaches the command with decreasing command-following error, which is $2e\mbox{-4}$ at $k=19,$ $1.3e\mbox{-4}$ at $k=39,$ and $1.7e\mbox{-4}$ at $k=60.$
    Note that, in (b), the estimates of the poles and zeros, including the pole-zero cancellation, are more accurate than in (a) due to the persistency arising from the changing step command.
    }
    \label{fig:ex1cd}
\end{figure*}

\begin{figure*}[t!]
    \centering
    \begin{subfigure}[t]{0.49\textwidth}
        \centering
        \includegraphics[trim = 10 100 10 30, width=\textwidth]{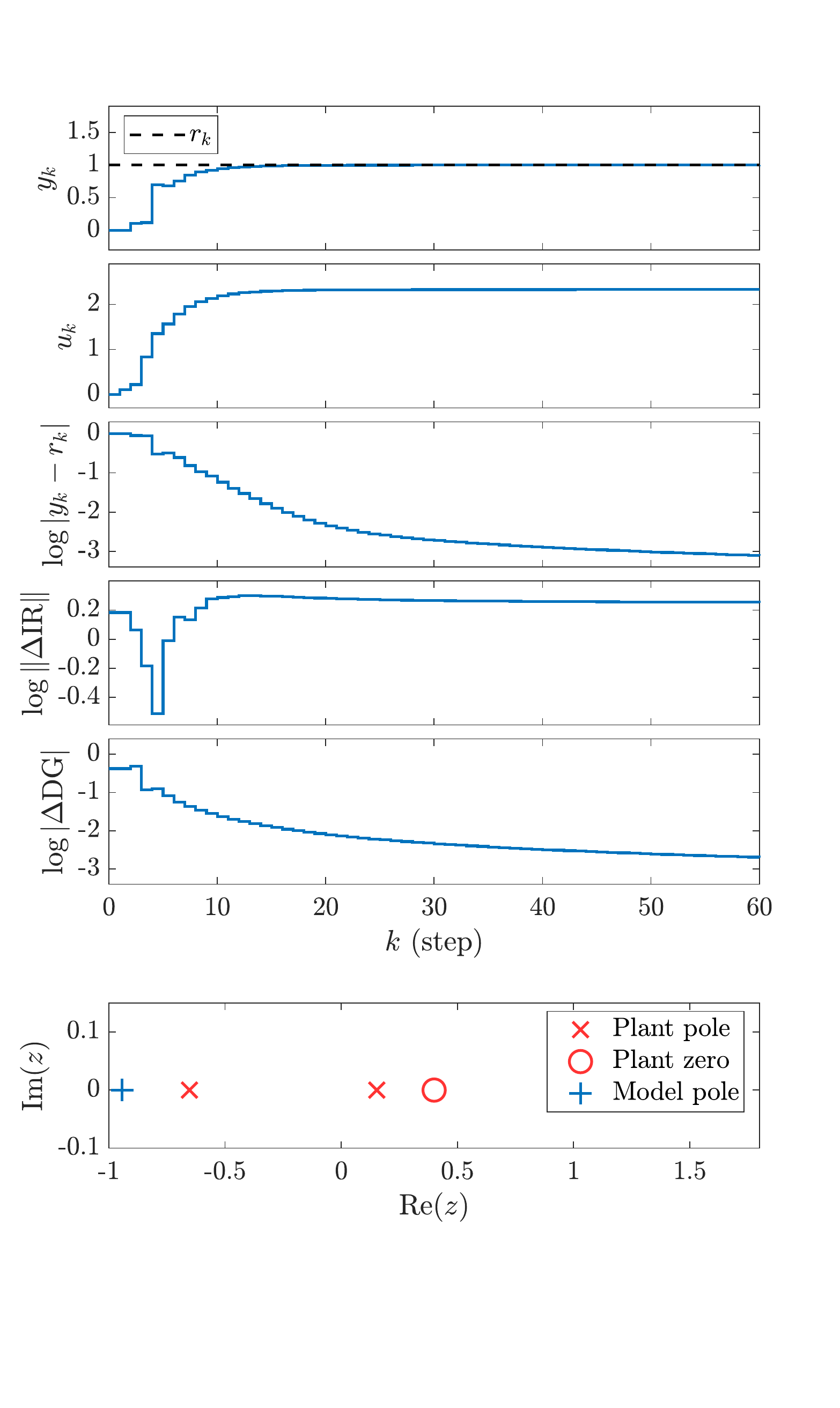}
        \caption{}
        \label{fig:ex1e}
    \end{subfigure}%
    ~ 
    \begin{subfigure}[t]{0.49\textwidth}
        \centering
        \includegraphics[trim = 10 100 10 30, width=\textwidth]{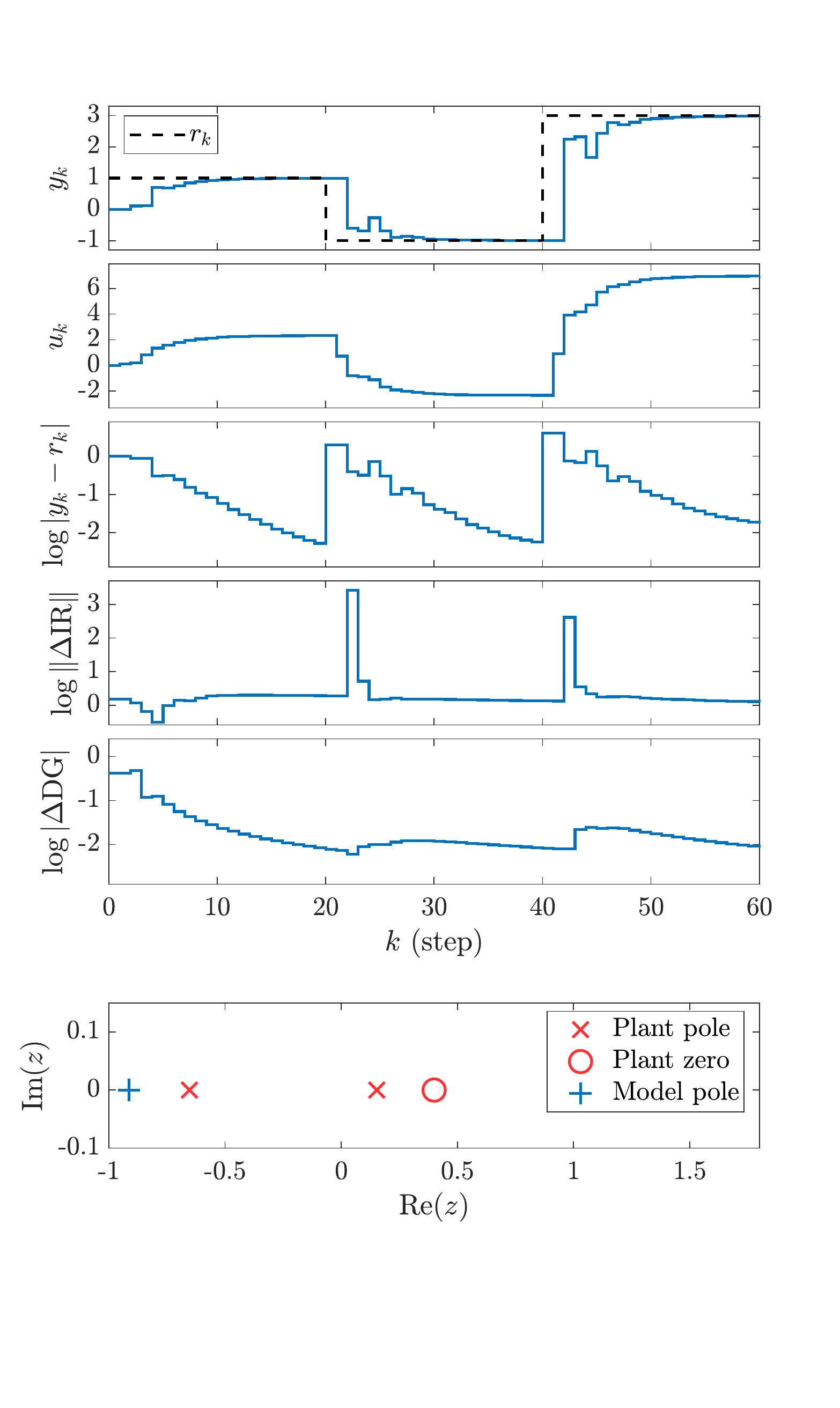}
        \caption{}
        \label{fig:ex1f}
    \end{subfigure}
    \caption{
    Example \ref{ex:ex1}.
    Command following for the SISO, asymptotically stable, discrete-time plant \eqref{eq:ex1} using $\hat{n}=1 < n = 2$  with a strictly proper model.
    (a) Single-step command $r_k\equiv1.$
    The output $y_k$ approaches the step command with decreasing command-following error, which is $ 7.8e\mbox{-4}$ at $k=60.$ 
    (b) Multi-step command $r_k$ given by \eqref{eq:commandex1}.
    For each step command, $y_k$ approaches the command with decreasing command-following error, which is $5.3e\mbox{-3}$ at $k=19,$ $5.7e\mbox{-3}$ at $k=39,$ and $1.8e\mbox{-2}$ at $k=60.$
    Note that the accuracy of the impulse response and the poles and zeros of the identified model in both (a) and (b) is poor due to the fact that $\hat{n} < n.$  Nevertheless, in both (a) and (b), the estimate of the DC gain is sufficiently accurate to allow OFMPCOI to approach the step command.
    This reflects the fact that RLS is exigent with the accuracy of the DC gain so that the output approaches the step commands.
    }
    \label{fig:ex1ef}
\end{figure*}

\begin{figure*}[t!]
    \centering
    \begin{subfigure}[t]{0.49\textwidth}
        \centering
        \includegraphics[trim = 10 100 10 30, width=\textwidth]{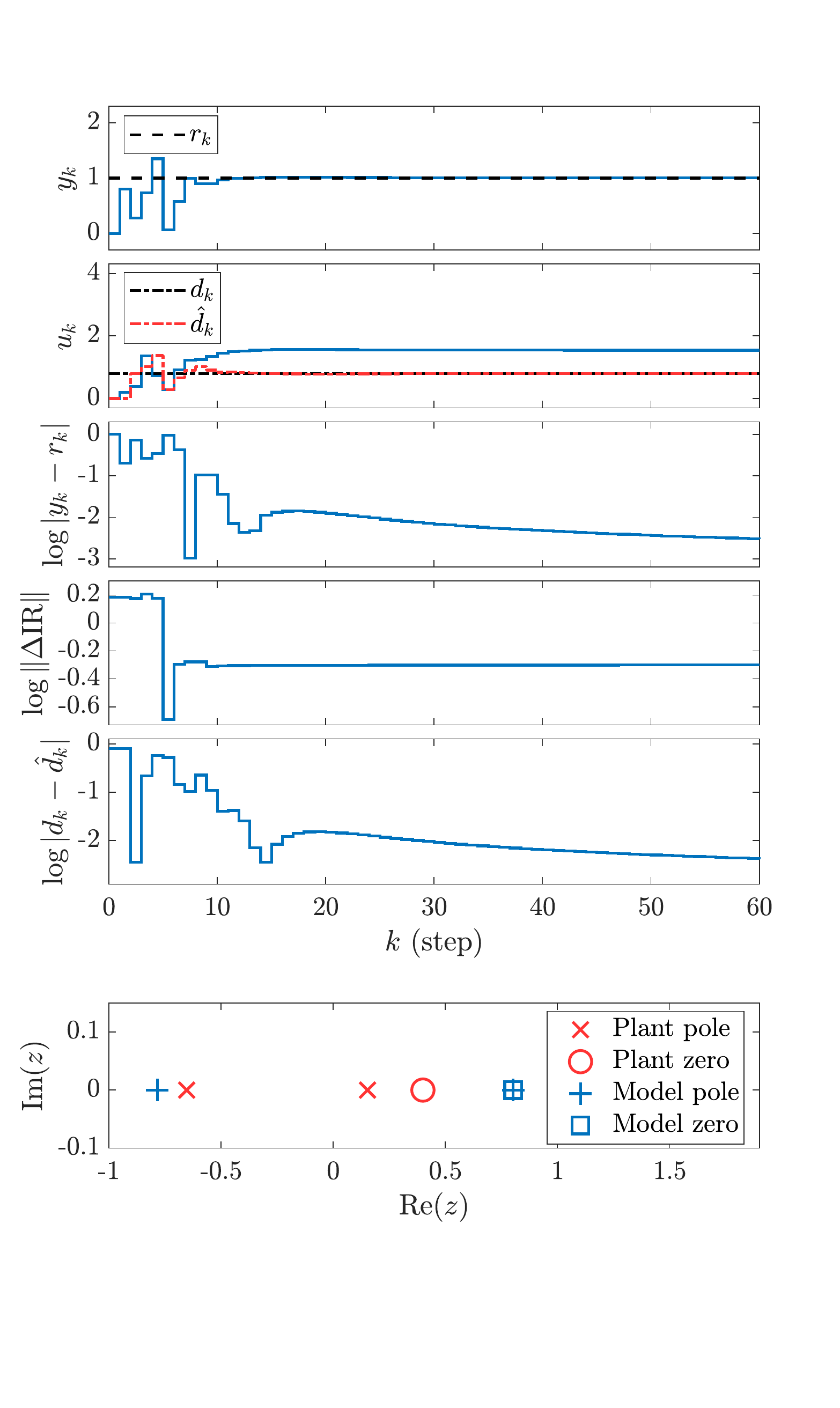}
        \caption{}
        \label{fig:ex1g}
    \end{subfigure}%
    ~ 
    \begin{subfigure}[t]{0.49\textwidth}
        \centering
        \includegraphics[trim = 10 100 10 30, width=\textwidth]{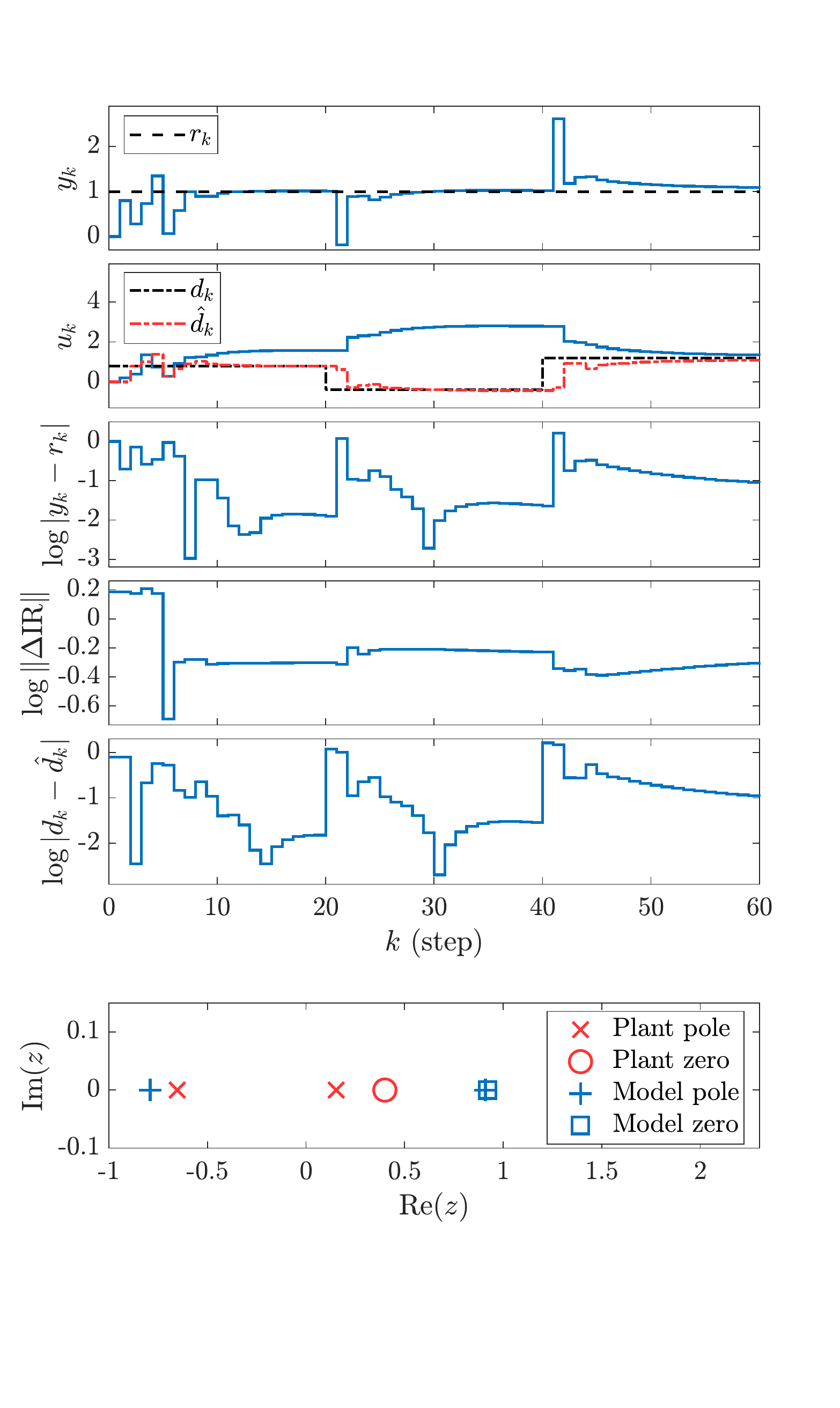}
        \caption{}
        \label{fig:ex1h}
    \end{subfigure}
    \caption{Example \ref{ex:ex1}. Command following and disturbance rejection for the SISO, asymptotically stable, discrete-time plant \eqref{eq:ex1} using $r_k\equiv1$ and $\hat{n}=n=2$ with a strictly proper model.
    (a) Single-step disturbance $d_k\equiv0.8$.
    The output $y_k$ approaches the step command with decreasing command-following error, which is $3 e\mbox{-3}$ at $k=60.$ 
    Note that, although $\Delta \mathrm{IR}$ stops decreasing after the initial transient, RLS continues to refine the estimate $\hat{d}_k$ of the disturbance for the remainder of the simulation.
    This refinement indicates exigency within the closed-loop identification, where the estimation of $\hat{d}_k$ is prioritized to decrease the command-following error.
    The bottom-most plot shows a pole-zero cancellation in the identified model at $k=60.$
    (b) Multi-step disturbance $d_k$ given by \eqref{eq:distex1}.
    For each step disturbance, $y_k$ approaches the command with decreasing command-following error, which is $1.3e\mbox{-2}$ at $k=19,$ $2.4e\mbox{-2}$ at $k=39,$ and $8.8e\mbox{-2}$ at $k=60.$
    Note that, although $\Delta \mathrm{IR}$ stops decreasing within a few steps after each change in the disturbance, 
    RLS continues to refine the estimate $\hat{d}_k$ of the disturbance until the command changes.
    Note that, in both (a) and (b), the step disturbances create a control offset in the model, which produces enough persistency for the output to approach the step commands. 
    The bottom-most plot shows a pole-zero cancellation in the identified model at $k=60.$
    }
    \label{fig:ex1gh}
\end{figure*}

\begin{figure*}[t!]
    \centering
    \begin{subfigure}[t]{0.49\textwidth}
        \centering
        \includegraphics[trim = 10 100 10 30, width=\textwidth]{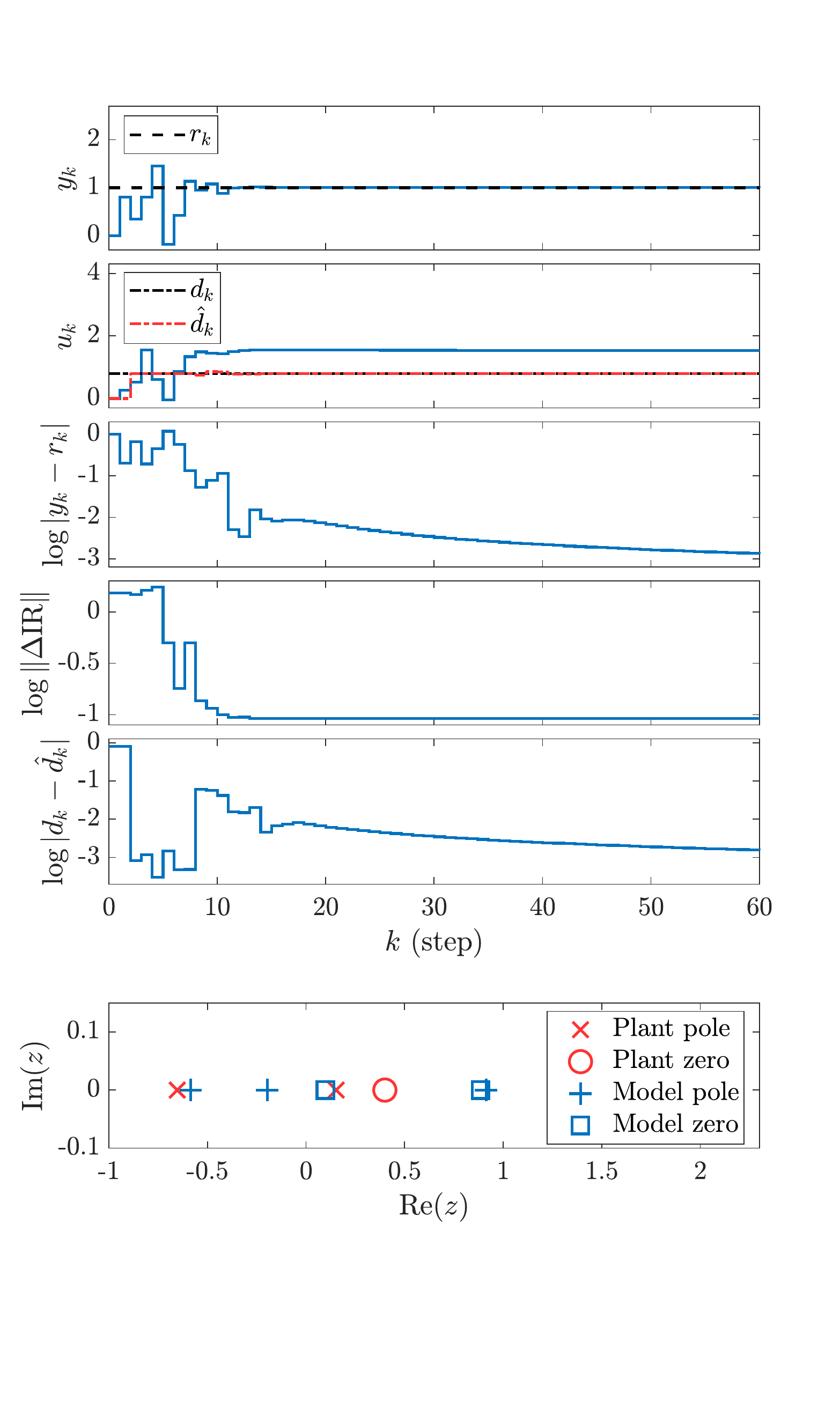}
        \caption{}
        \label{fig:ex1i}
    \end{subfigure}%
    ~ 
    \begin{subfigure}[t]{0.49\textwidth}
        \centering
        \includegraphics[trim = 10 100 10 30, width=\textwidth]{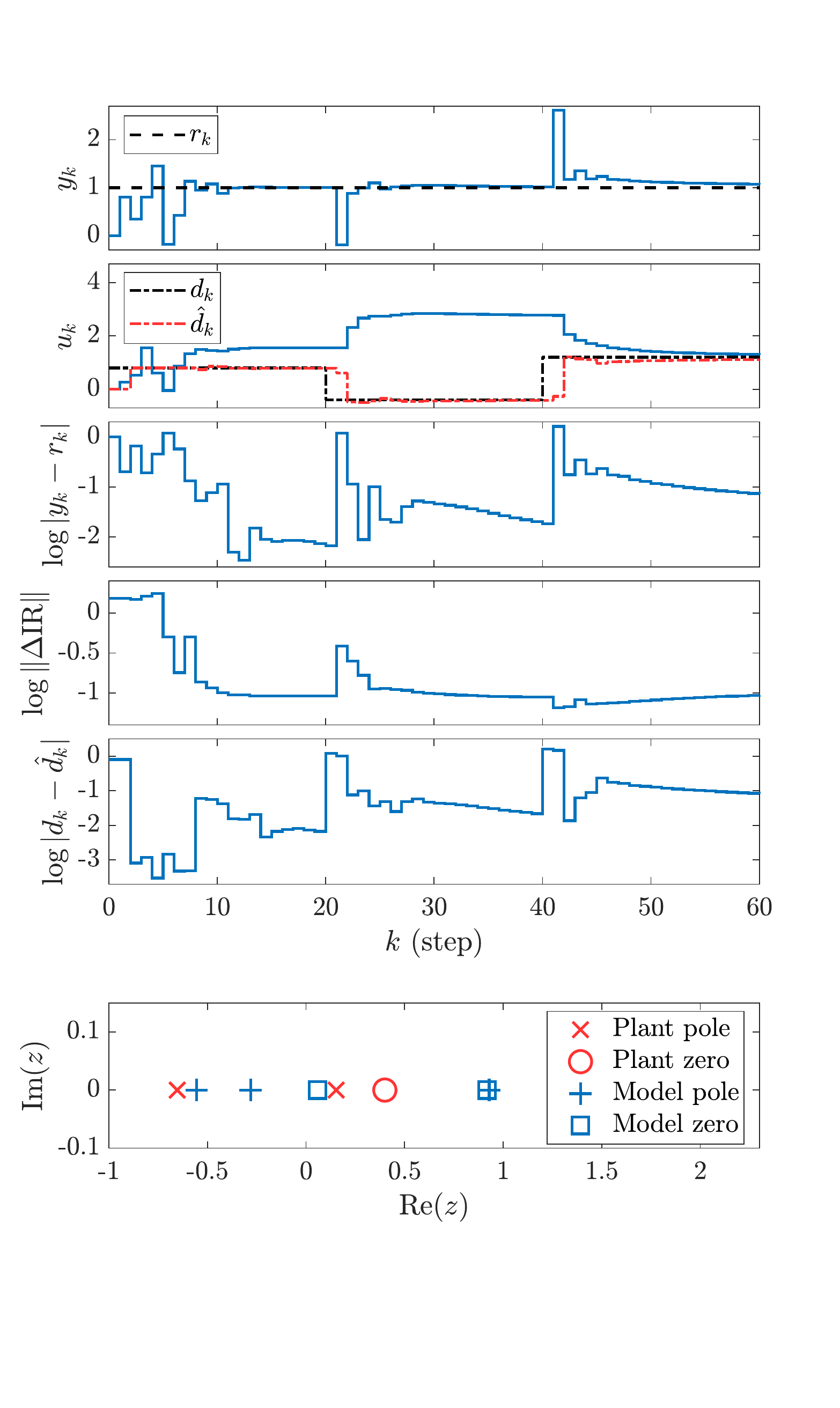}
        \caption{}
        \label{fig:ex1j}
    \end{subfigure}
    \caption{Example \ref{ex:ex1}. Command following and disturbance rejection for the SISO, asymptotically stable, discrete-time plant \eqref{eq:ex1} using $r_k\equiv1$ and $\hat{n}=3 > n = 2$ with a strictly proper model.
    (a) Single-step disturbance $d_k\equiv0.8$.
    The output $y_k$ approaches the step command with decreasing command-following error, which is $1.3 e\mbox{-3}$ at $k=60.$
    (b) Multi-step disturbance $d_k$ given by \eqref{eq:distex1}.
    For each step disturbance, $y_k$ approaches the command with decreasing command-following error, which is $7.4e\mbox{-3}$ at $k=19,$ $2e\mbox{-2}$ at $k=39,$ and $7.1e\mbox{-2}$ at $k=60.$
    Note that, although the accuracy of the poles and zeros of the identified model in both (a) and (b) is poor, the estimate of the disturbance is sufficiently accurate to allow OFMPCOI to approach the command and reject the disturbance.
    }
    \label{fig:ex1ij}
\end{figure*}

\begin{figure*}[t!]
    \centering
    \begin{subfigure}[t]{0.49\textwidth}
        \centering
        \includegraphics[trim = 10 100 10 30, width=\textwidth]{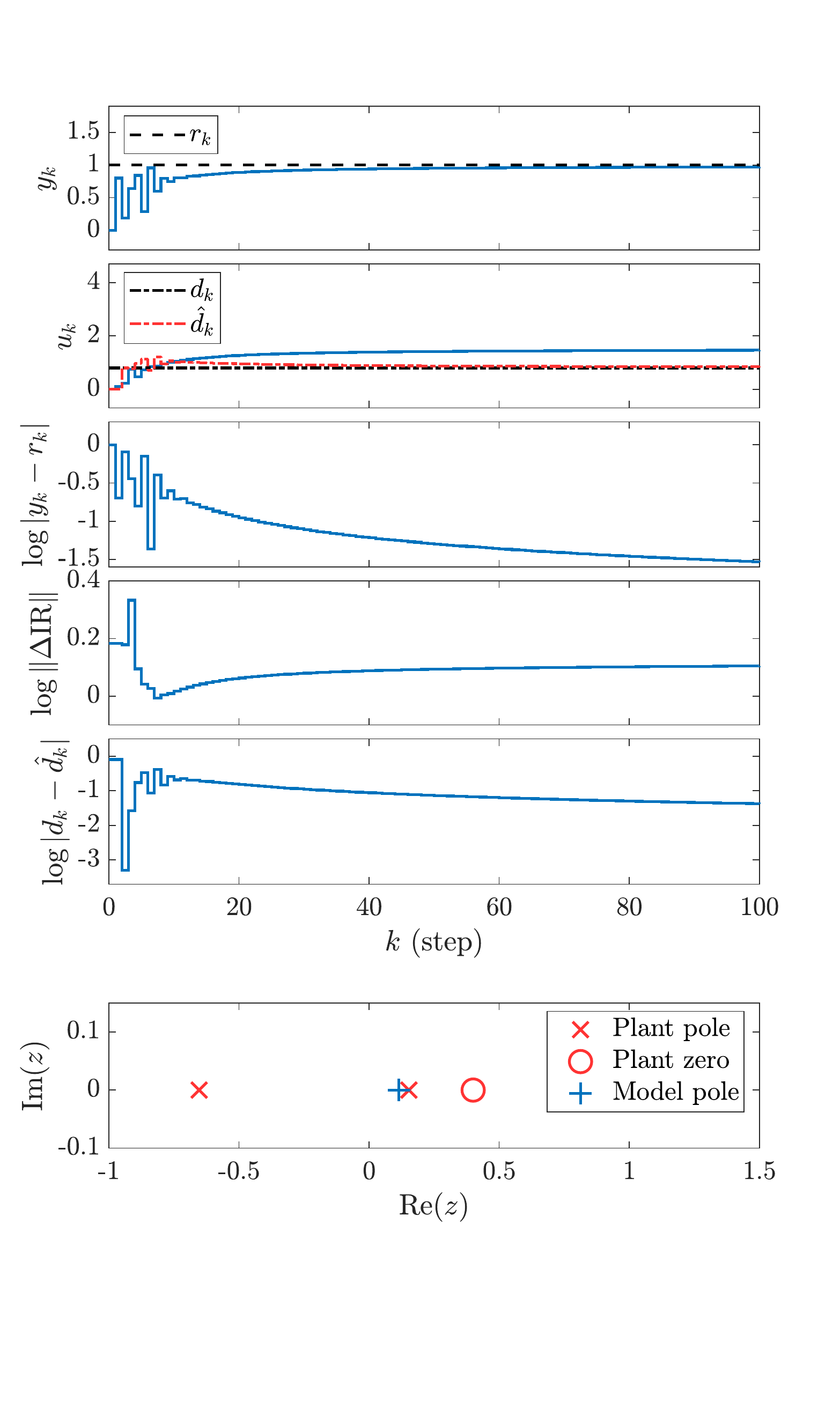}
        \caption{}
        \label{fig:ex1k}
    \end{subfigure}%
    ~ 
    \begin{subfigure}[t]{0.49\textwidth}
        \centering
        \includegraphics[trim = 10 100 10 30, width=\textwidth]{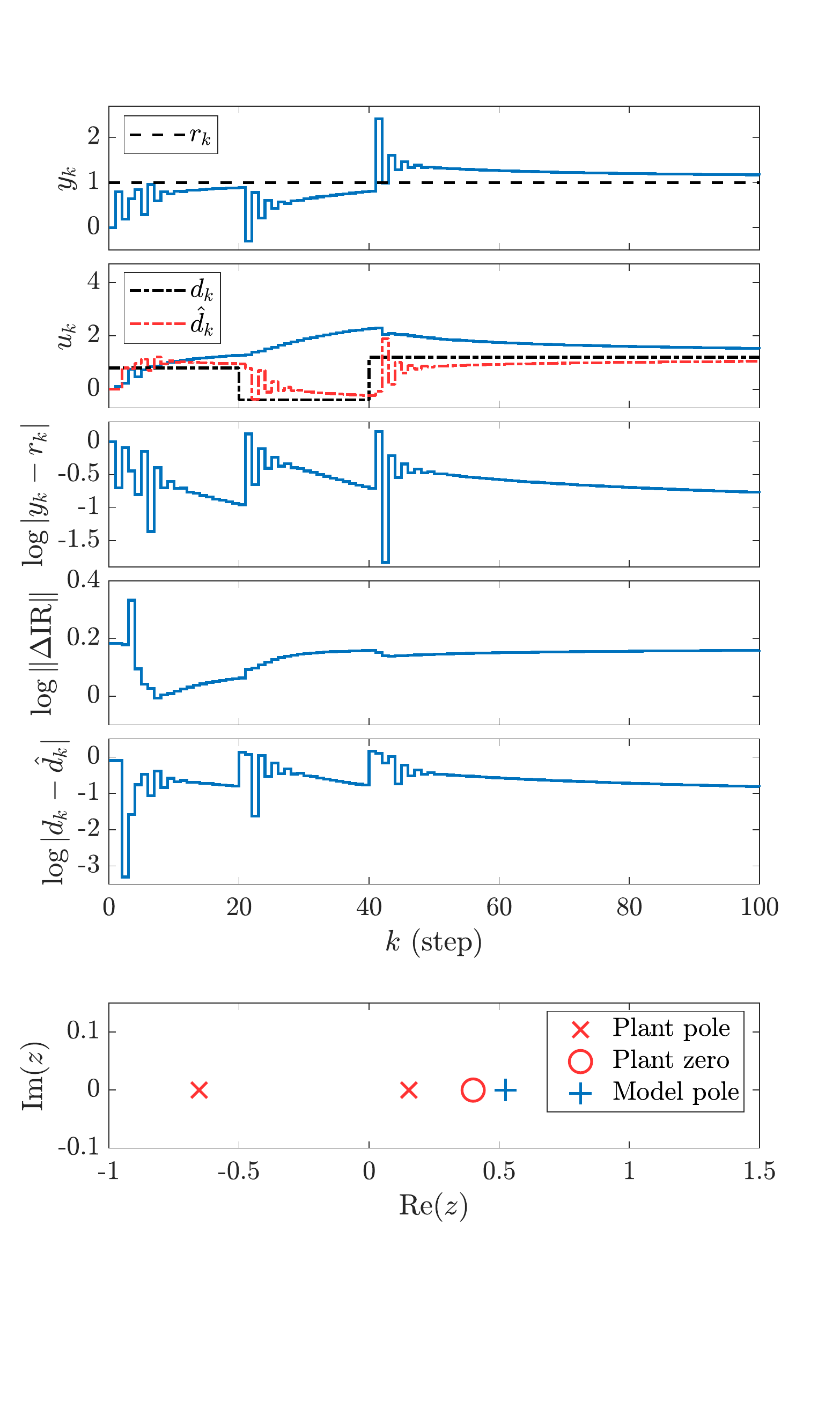}
        \caption{}
        \label{fig:ex1l}
    \end{subfigure}
    \caption{Example \ref{ex:ex1}. Command following and disturbance rejection for the SISO, asymptotically stable, discrete-time plant \eqref{eq:ex1} using $r_k\equiv1$ and $\hat{n}=1 < n = 2$ with a strictly proper model.
    (a) Single-step disturbance $d_k\equiv0.8$.
     The output $y_k$ approaches the step command with decreasing command-following error, which is $2.9 e\mbox{-2}$ at $k=100.$
    (b) Multi-step disturbance $d_k$ given by \eqref{eq:distex1}.
    For each step disturbance, $y_k$ approaches the command with decreasing command-following error, which is $1.2e\mbox{-1}$ at $k=19,$ $2.1e\mbox{-1}$ at $k=39,$ and $1.7e\mbox{-1}$ at $k=100.$
    Note that, in both (a) and (b), $y_k$ approaches the command more slowly than in the cases $\hat n = 2$ and $\hat n = 3$ considered in Figure \ref{fig:ex1gh} and Figure \ref{fig:ex1ij}, respectively.
    }
    \label{fig:ex1kl}
\end{figure*}

\begin{example}\label{ex:ex1}
\textit{Asymptotically stable, discrete-time plant.}
Consider the SISO, discrete-time input-output plant
\begin{align}
    y_k = -0.5 y_{k-1} + 0.1 y_{k-2} + u_{k-1} - 0.4 u_{k-2}. \label{eq:ex1}
\end{align}
Let $u_{\mathrm{min}} = -10,$ $u_{\mathrm{max}} = 10,$ $\Delta u_{\mathrm{min}} = -10,$ and $\Delta u_{\mathrm{max}} = 10.$
No output constraint is considered in this example.
The plant \eqref{eq:ex1} is initialized with $y_{-1} = y_{-2} = 0$ and $u_0 = u_{-1} = u_{-2} = 0.$ Note that the order of the plant is $n=2.$
At each time step, OFMPCOI uses  \eqref{eq:QPnoslack}--\eqref{eq:DeltaUcon} with $\ell=5,$ $\bar{Q}=2I_{\ell-1},$ $\bar{P}=5,$ and $R=I_\ell.$
Let $\lambda=1,$ $\theta_0 = 10^{-2} \cdot 1_{2\hat{n}\times 1},$ and $P_0=10^3I_{2\hat{n}}.$

Figures \ref{fig:ex1ab}, \ref{fig:ex1cd}, and \ref{fig:ex1ef} show the results for $\hat{n}=2,$ $\hat{n}=3,$ and $\hat{n}=1,$ respectively, using a strictly proper model, that is, using $\hat G_0 = 0.$
We compare the absolute value of the command-following error, the L2-norm difference between the 30-step impulse response of the plant and the model ($\Delta \mathrm{IR}$), the absolute value of the difference between the DC gain of the plant and the model ($\Delta \mathrm{DG}$), and the pole-zero location of the plant and the model identified at $k=60.$
Figures \ref{fig:ex1a}, \ref{fig:ex1c}, and \ref{fig:ex1e} show the results for the constant command $r_k\equiv1,$ and Figures \ref{fig:ex1b}, \ref{fig:ex1d}, and \ref{fig:ex1f} show for the three-step command
\begin{align}
    r_k &=
    \begin{cases}
        1 , & 0 \le k < 20, \\
        -1, & 20\le k < 40, \\
        3, &  k \ge 40.
    \end{cases} \label{eq:commandex1}
\end{align}

Next, a matched disturbance $d_k\in\mathbb{R}$ is applied to the plant \eqref{eq:ex1}, that is,
\begin{align}
    {y}_{k} = -\sum_{i=1}^{n}{F}_{i} y_{k-i} + \sum_{i=1}^{n}{G}_{i} (u_{k-i}+d_{k-i}), \label{eq:plantdistex1}
\end{align}
where $n=2,$ $F_1 = 0.5,$ $F_2=-0.1,$ $G_1=1,$ and $G_2 = -0.4.$
In the case of a constant matched disturbance $\hat{d}_k\in\mathbb{R},$ the disturbed plant is given by
\begin{align}
    {y}_{k} = -\sum_{i=1}^{n}{F}_{i} y_{k-i} + \sum_{i=1}^{n}{G}_{i} u_{k-i}+ \sum_{i=1}^{n}{G}_{i}\hat{d}_k, \label{eq:plantdistconstex1}
\end{align}
and for a given $\hat{n},$ the model identified by RLS at step $k$ is given by
\begin{align}
    {y}_{k} = -\sum_{i=1}^{\hat{n}}\hat{F}_{i,k} y_{k-i} + \sum_{i=1}^{\hat{n}}\hat{G}_{i,k} u_{k-i}. \label{eq:modelex1}
\end{align}
Using \eqref{eq:modelex1} in \eqref{eq:plantdistconstex1}, we extract the disturbance $\hat{d}_k$ estimated by RLS as
\begin{align}
    \hat{d}_k = \left[ -\sum_{i=1}^{\hat{n}}\hat{F}_{i,k} y_{k-i} + \sum_{i=1}^{\hat{n}}\hat{G}_{i,k} u_{k-i} + \sum_{i=1}^{n}{F}_{i} y_{k-i} - \sum_{i=1}^{n}{G}_{i} u_{k-i} \right]\sum_{i=1}^{n}{G}_{i}^{-1}. \label{eq:dhat}
\end{align}
Figures \ref{fig:ex1gh}, \ref{fig:ex1ij}, and \ref{fig:ex1kl} show the results for $\hat{n}=2,$ $\hat{n}=3,$ and $\hat{n}=1,$ respectively, with constant command $r_k\equiv1,$
where the error of the disturbance estimate $\hat{d}_k$ is computed.
In Figures \ref{fig:ex1g}, \ref{fig:ex1i}, and \ref{fig:ex1k}, we apply the constant disturbance $d_k \equiv 0.8,$ and in Figures \ref{fig:ex1h},  \ref{fig:ex1j}, and \ref{fig:ex1l}, we apply the three-step disturbance
\begin{align} \label{eq:distex1}
    d_k &=
    \begin{cases}
         0.8 , & 0 \le k < 20, \\
        -0.4, & 20\le k < 40, \\
        1.2, &  k \ge 40.
    \end{cases}
\end{align}

This example exhibits the manifestations of persistency, consistency, and exigency within closed-loop identification.
Persistency is manifested in this example through step-command changes and step disturbances.
In the case of a lack of persistency, we observe a clear bias in the coefficient estimates, which indicates a lack of consistency.
Additionally, it is seen that, although the impulse-response error $\Delta \mathrm{IR}$ stops decreasing after each transient, RLS keeps refining the DC gain of the model for the remainder of the simulation.
This indicates exigency within identification, as RLS prioritizes the identification of the DC gain over the impulse-response to achieve command following.

\end{example}

\clearpage

\begin{figure*}[t!]
    \centering
    \begin{subfigure}[t]{0.49\textwidth}
        \centering
        \includegraphics[trim = 10 120 10 30, width=\textwidth]{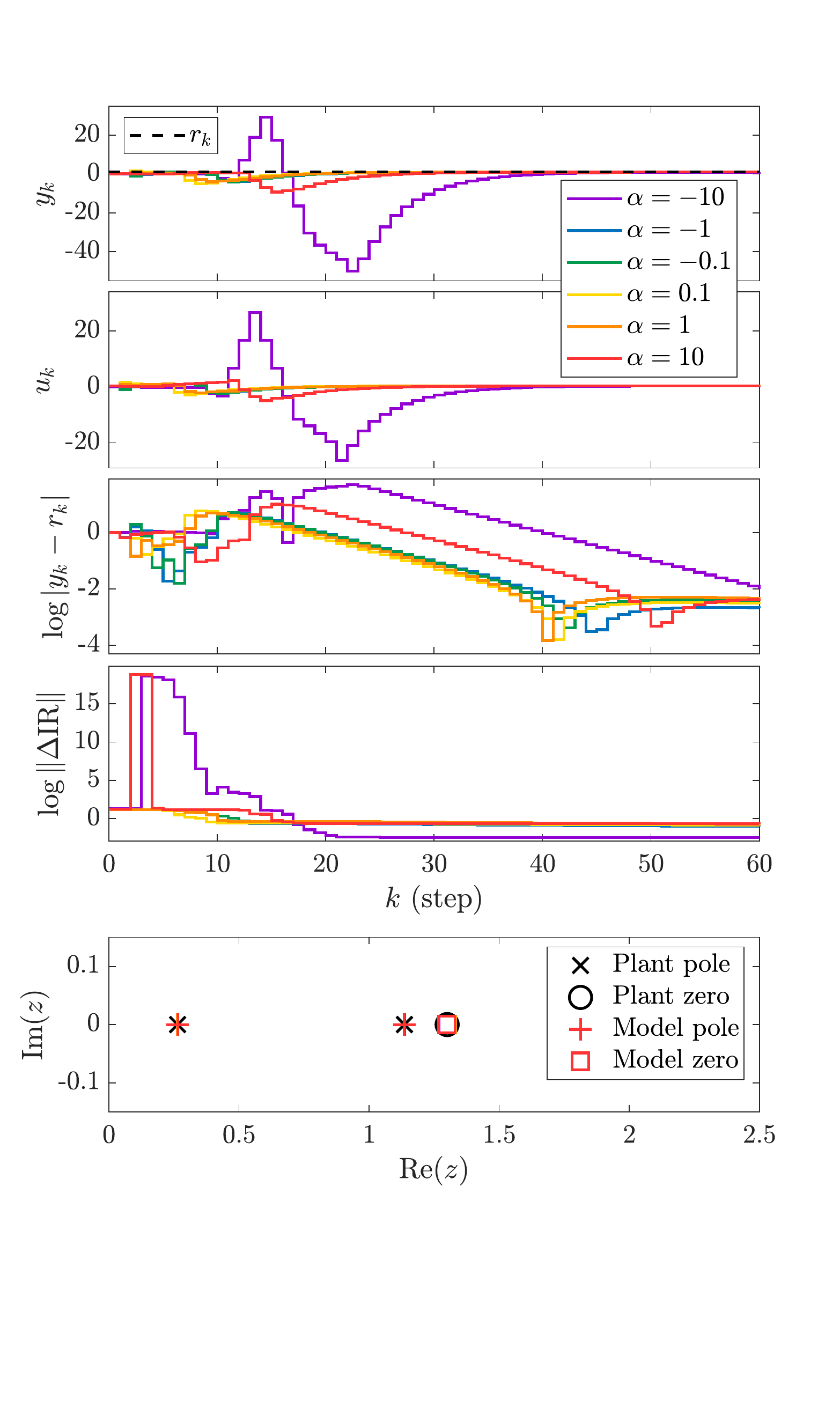}
        \caption{}
        \label{fig:ex2a}
    \end{subfigure}%
    ~ 
    \begin{subfigure}[t]{0.49\textwidth}
        \centering
        \includegraphics[trim = 10 120 10 30, width=\textwidth]{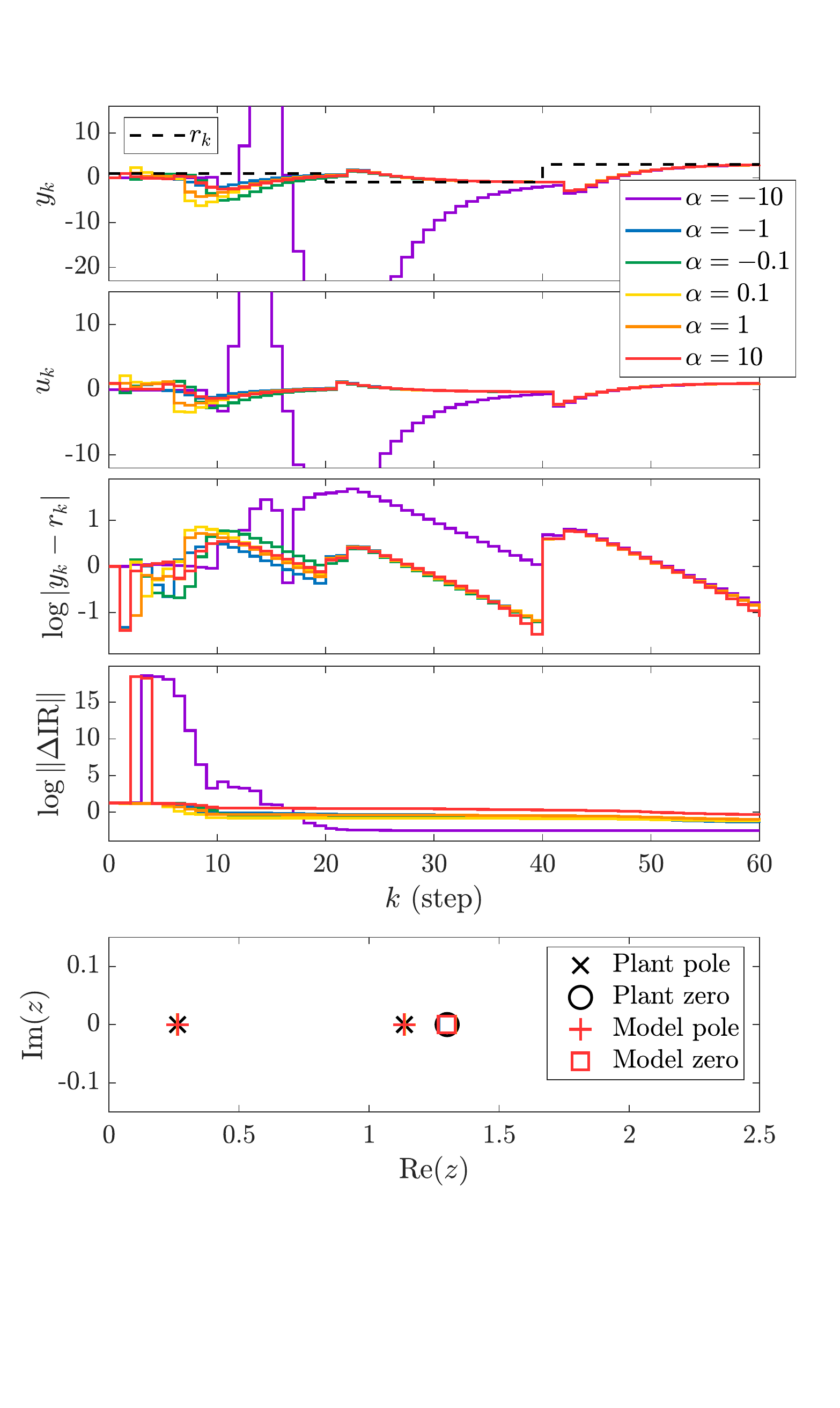}
        \caption{}
        \label{fig:ex2b}
    \end{subfigure}
    \caption{Example \ref{ex:ex2}. Command following for the SISO, unstable, NMP, discrete-time plant \eqref{eq:ex2} using $r_k\equiv1$ and $\hat{n}= n = 2$ with a strictly proper FIR model initialized with $\theta_0 = \alpha[ 0 \ 0 \ 0 \ 1]^\rmT,$ $\alpha\in\{-10,-1,-0.1,0.1,1,10\}.$
    (a) Single-step command $r_k\equiv1$.
    The output $y_k$ approaches the step command with decreasing command-following error, which is in the interval $[2.1e\mbox{-3},1.0e\mbox{-2}]$ at $k=60.$
    Note that, for all $\alpha,$ $\Delta \mathrm{IR}$ stops to substantially decrease after $k=25.$
    The bottom-most plot compares the poles and zero of the identified model at $k=60$ to the poles and zero of the plant.
    (b) Multi-step command $r_k$ given by \eqref{eq:commandex1}.
    For $\alpha\ne-10,$ $y_k$ approaches the three-step command with decreasing command-following error, which is in the interval $[4.3e\mbox{-1},1.1]$ at $k=19,$ $[3.3e\mbox{-2},6.7e\mbox{-2}]$ at $k=39,$ and $[8e\mbox{-2},1.2e\mbox{-1}]$ at $k=60.$
    For $\alpha=-10,$ $y_k$ approaches the last command with decreasing command-following error, which is $1.3e\mbox{-1}$ at $k=60.$
    Note that, for all $\alpha,$ $\Delta \mathrm{IR}$ stops to substantially decrease after $k=25.$
    The bottom-most plot compares the poles and zero of the identified model at $k=60$ to the poles and zero of the plant.
    }
    \label{fig:ex2ab}
\end{figure*}

\begin{figure*}[t!]
    \centering
    \begin{subfigure}[t]{0.49\textwidth}
        \centering
        \includegraphics[trim = 10 120 10 30, width=\textwidth]{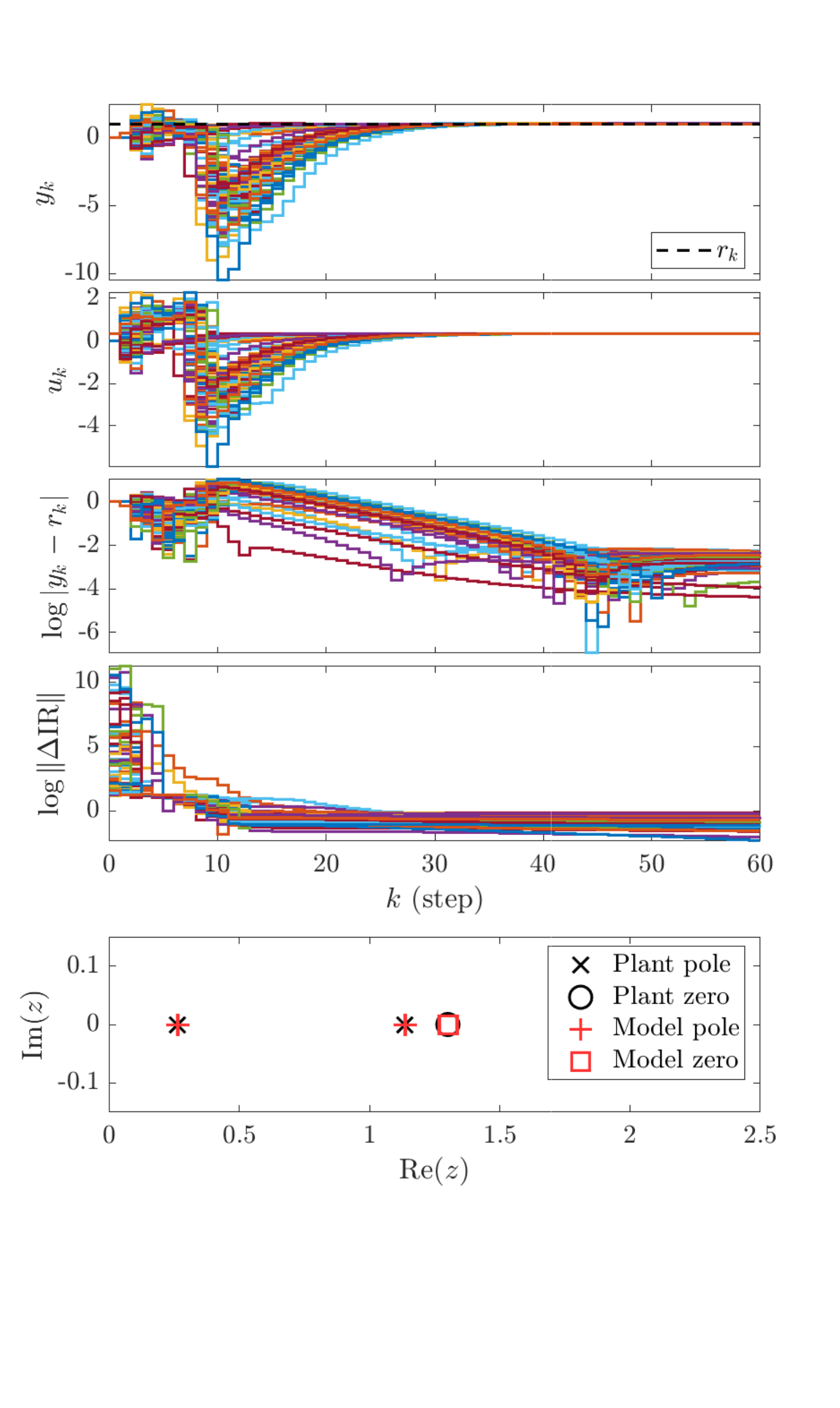}
        \caption{}
        \label{fig:ex2c}
    \end{subfigure}%
    ~ 
    \begin{subfigure}[t]{0.49\textwidth}
        \centering
        \includegraphics[trim = 10 120 10 30, width=\textwidth]{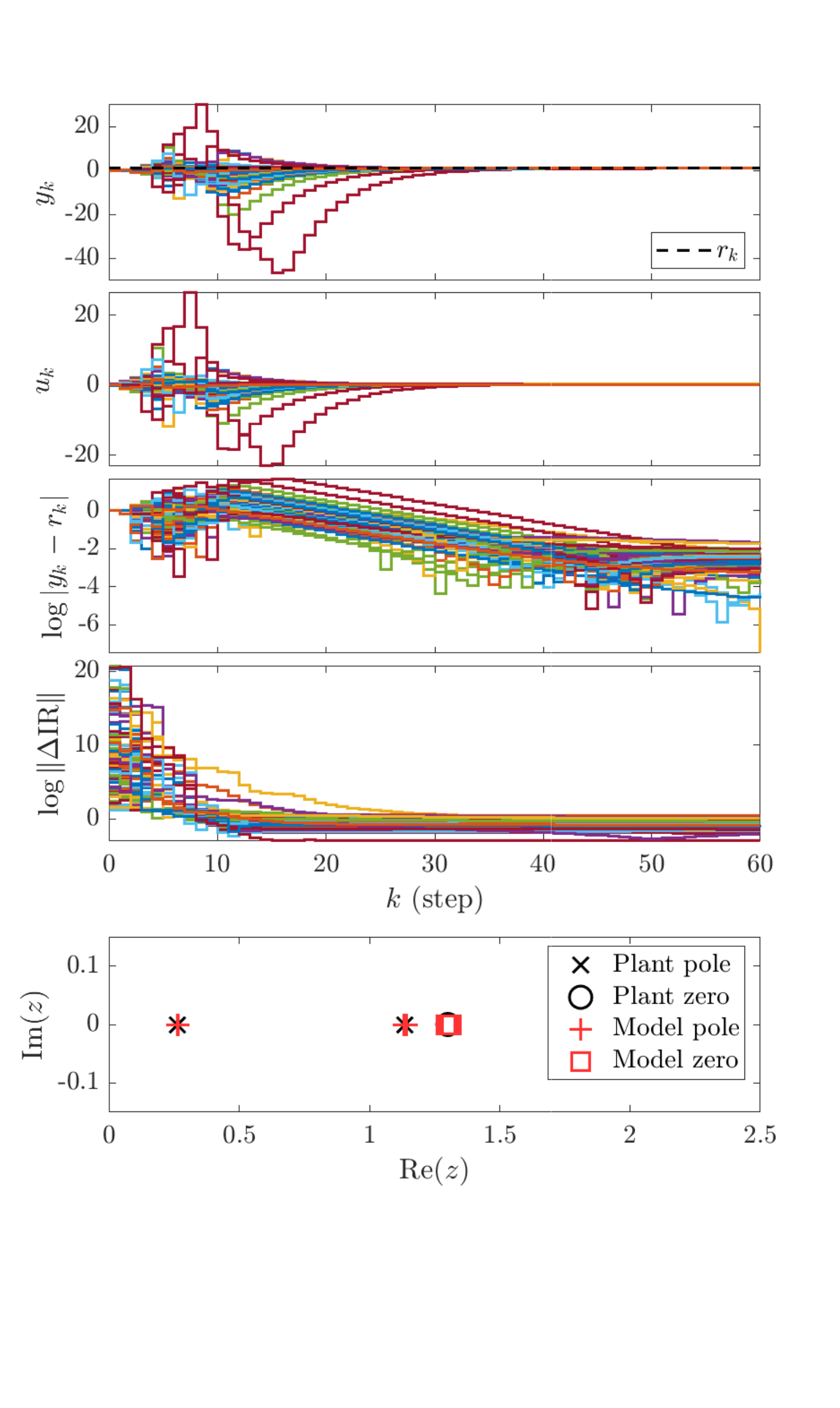}
        \caption{}
        \label{fig:ex2d}
    \end{subfigure}
    \caption{Example \ref{ex:ex2}. Command following for the SISO, unstable, NMP, discrete-time plant \eqref{eq:ex2} using $r_k\equiv1$ and $\hat{n}= n = 2$ with a strictly proper model, which is initialized with 100 Gaussian-distributed $\theta_0$ with standard deviation $\sigma.$
    (a) $\sigma=1.$
    The output $y_k$ approaches the step command with decreasing command-following error, which is in the interval $[9.5e\mbox{-4},5.5e\mbox{-3}]$ at $k=60.$
    (b) $\sigma=2.$
    The output $y_k$ approaches the step command with decreasing command-following error, which is in the interval $[6.7e\mbox{-3},2.1e\mbox{-2}]$ at $k=60.$
    Note that, in both (a) and (b), although $\Delta \mathrm{IR}$ stops decreasing after $k=25,$ indicating the presence of bias and thus the lack of consistency in the estimate of the model, the poles and zero of the model are close to the poles and zero of the plant at $k=60.$
    }
    \label{fig:ex2cd}
\end{figure*}

\begin{figure*}[t!]
    \centering
    \begin{subfigure}[t]{0.49\textwidth}
        \centering
        \includegraphics[trim = 10 70 10 30, width=\textwidth]{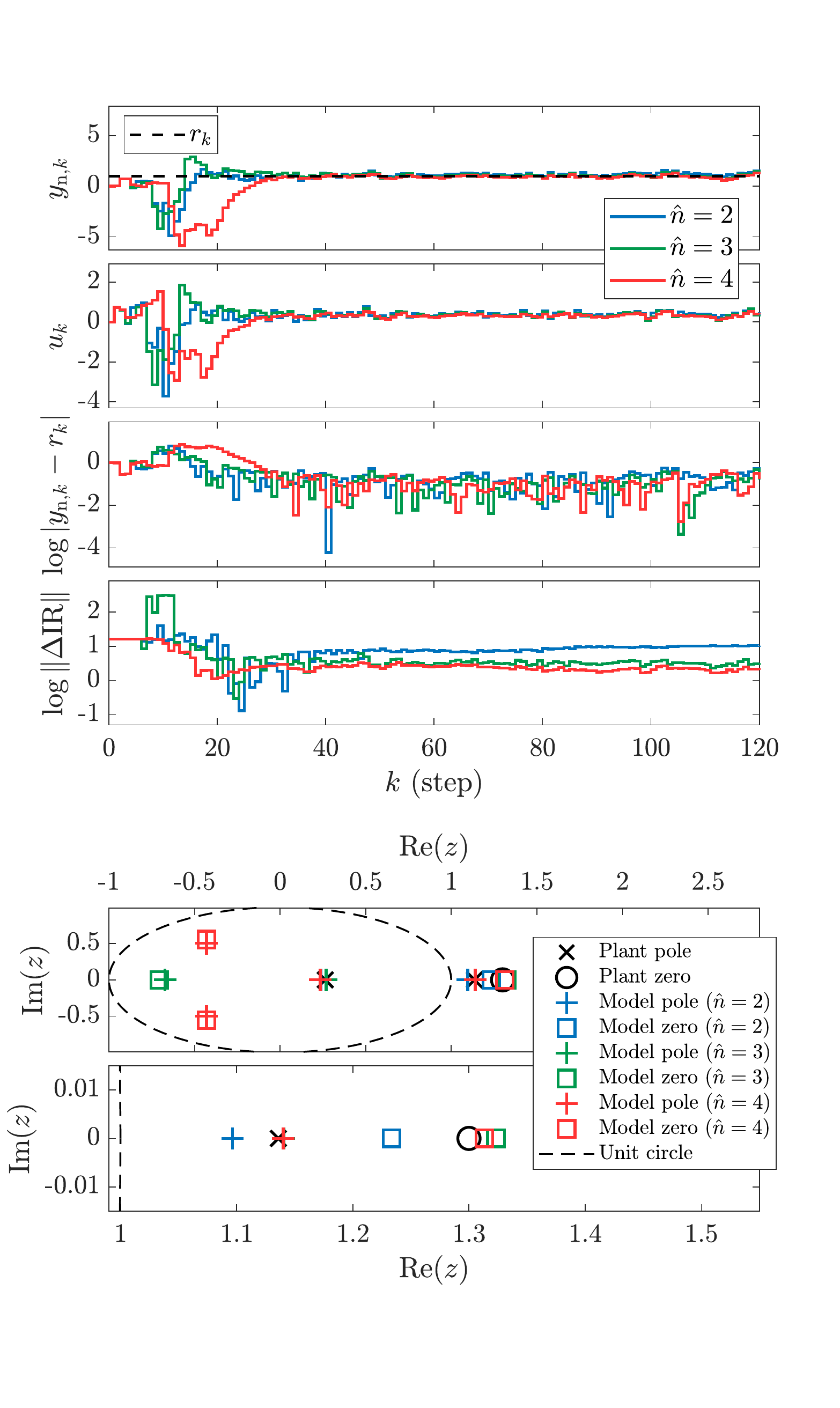}
        \caption{}
        \label{fig:ex2e}
    \end{subfigure}%
    ~ 
    \begin{subfigure}[t]{0.49\textwidth}
        \centering
        \includegraphics[trim = 10 70 10 30, width=\textwidth]{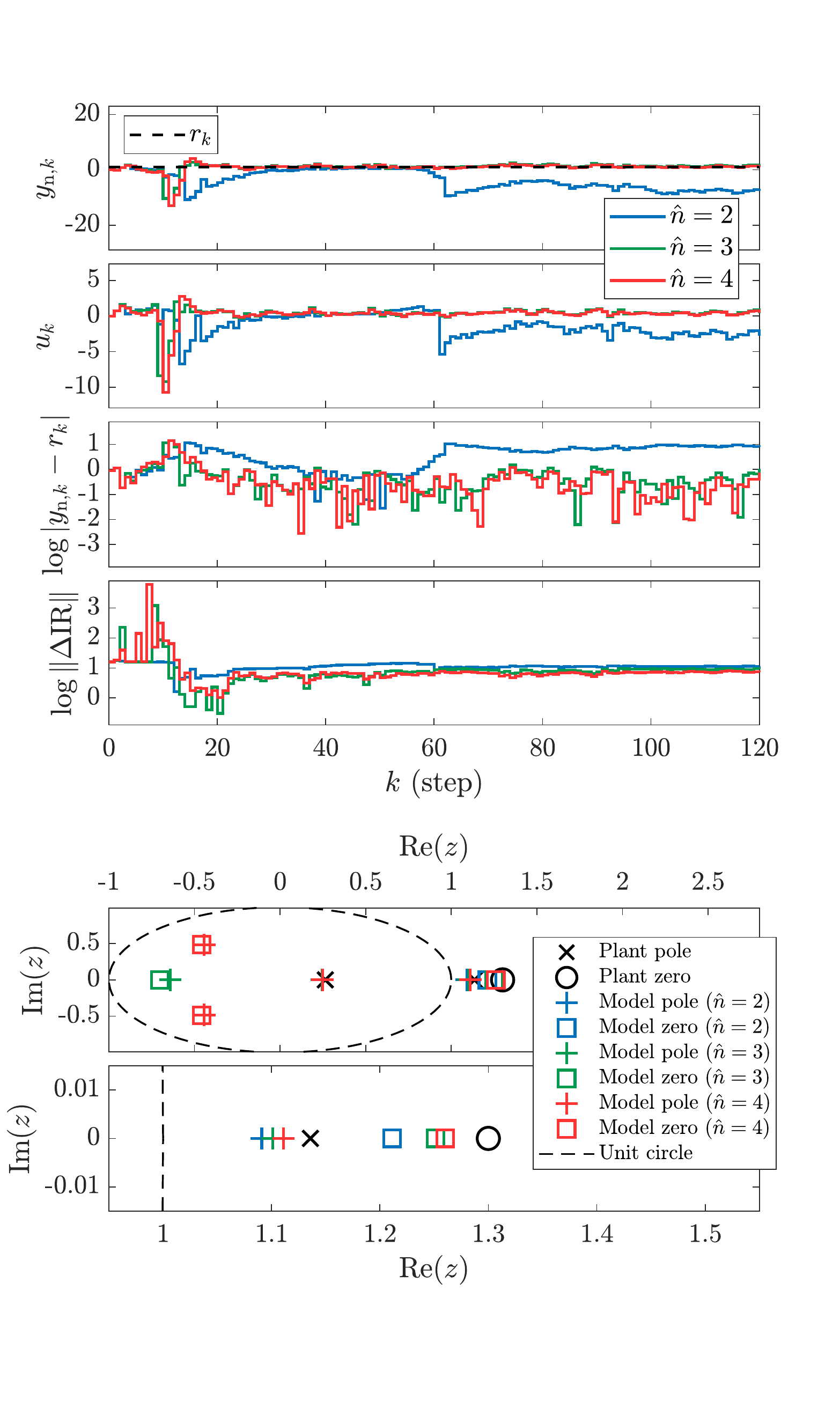}
        \caption{}
        \label{fig:ex2f}
    \end{subfigure}
    \caption{Example \ref{ex:ex2}. Command following for the SISO, unstable, NMP, discrete-time plant \eqref{eq:ex2} using $r_k\equiv1,$ the noisy measurement $y_{\rmn,k}$ given by \eqref{eq:ynoisy} with standard deviation $\sigma,$ and the FIR strictly proper initial model $\theta_0 = [0_{(2\hat{n}-1)\times1}^\rmT \ 1]^\rmT,$ where $\hat{n}=2,$ $\hat{n}=3,$ and $\hat{n}=4.$
    (a) For $\sigma = 0.05,$
    in all cases, $y_k$ approaches the step command with decreasing command-following error.
    (b) For $\sigma=0.15,$ 
    $y_k$ approaches the step command with decreasing command-following error for $\hat{n}=3$ and $\hat{n}=4.$
    For $\hat{n}=2,$ $y_k$ does not approach the step command, and the command-following error is $7.9$ at $k=120.$
    Note that, in both (a) and (b), $\Delta \mathrm{IR}$ decreases as $\hat{n}$ increases.
    The bottom-most plots show that the poles and zero of the model approach the poles and zero of the plant as $\hat{n}$ increases, while the extra poles and zeros of the model cancel out in the unit circle.
    }
    \label{fig:ex2ef}
\end{figure*}

\begin{figure*}[t!]
    \centering
    \begin{subfigure}[t]{0.49\textwidth}
        \centering
        \includegraphics[trim = 30 10 30 30, width=\textwidth]{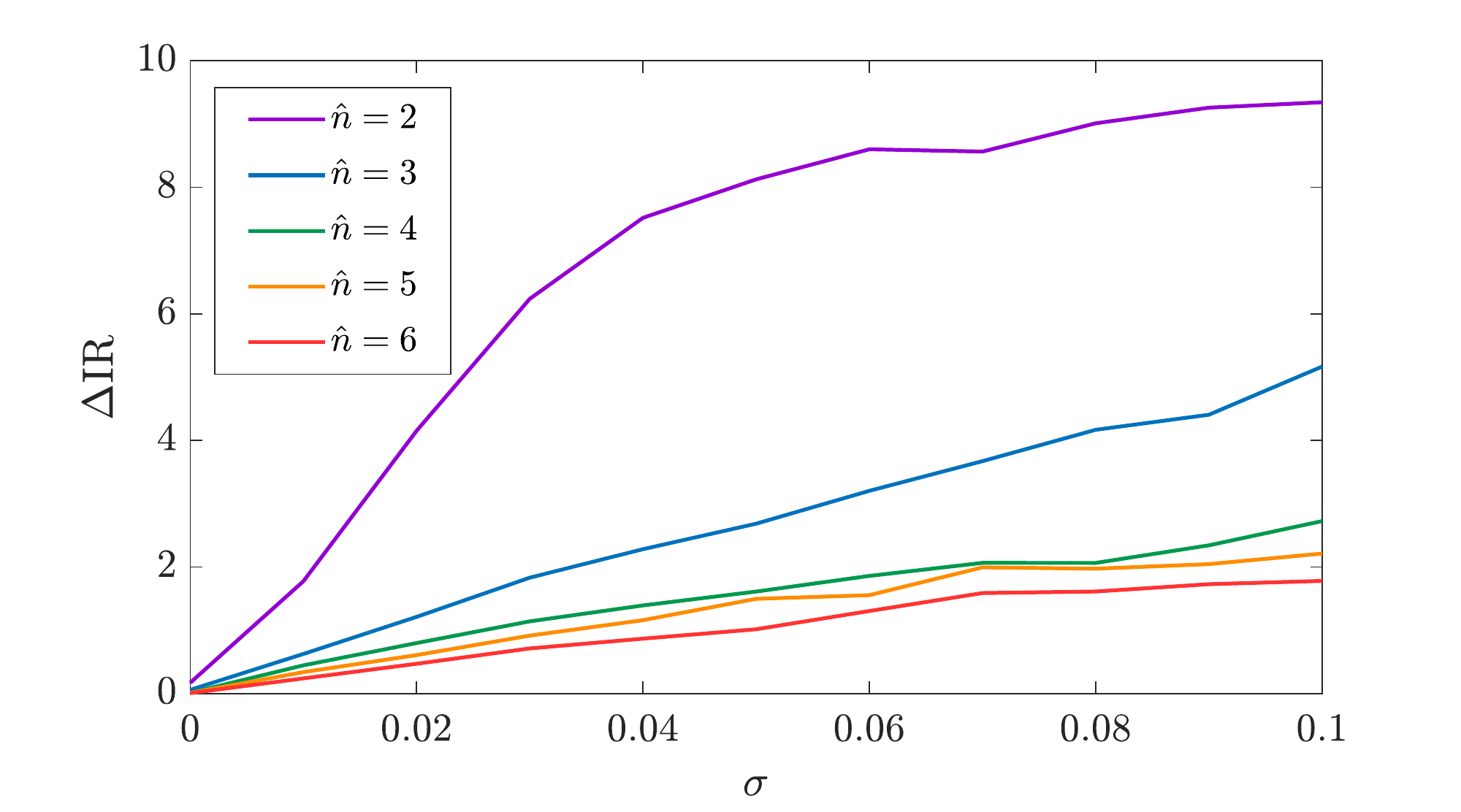}
        \caption{}
        \label{fig:ex2g}
    \end{subfigure}%
    ~ 
    \begin{subfigure}[t]{0.49\textwidth}
        \centering
        \includegraphics[trim = 30 10 30 30, width=\textwidth]{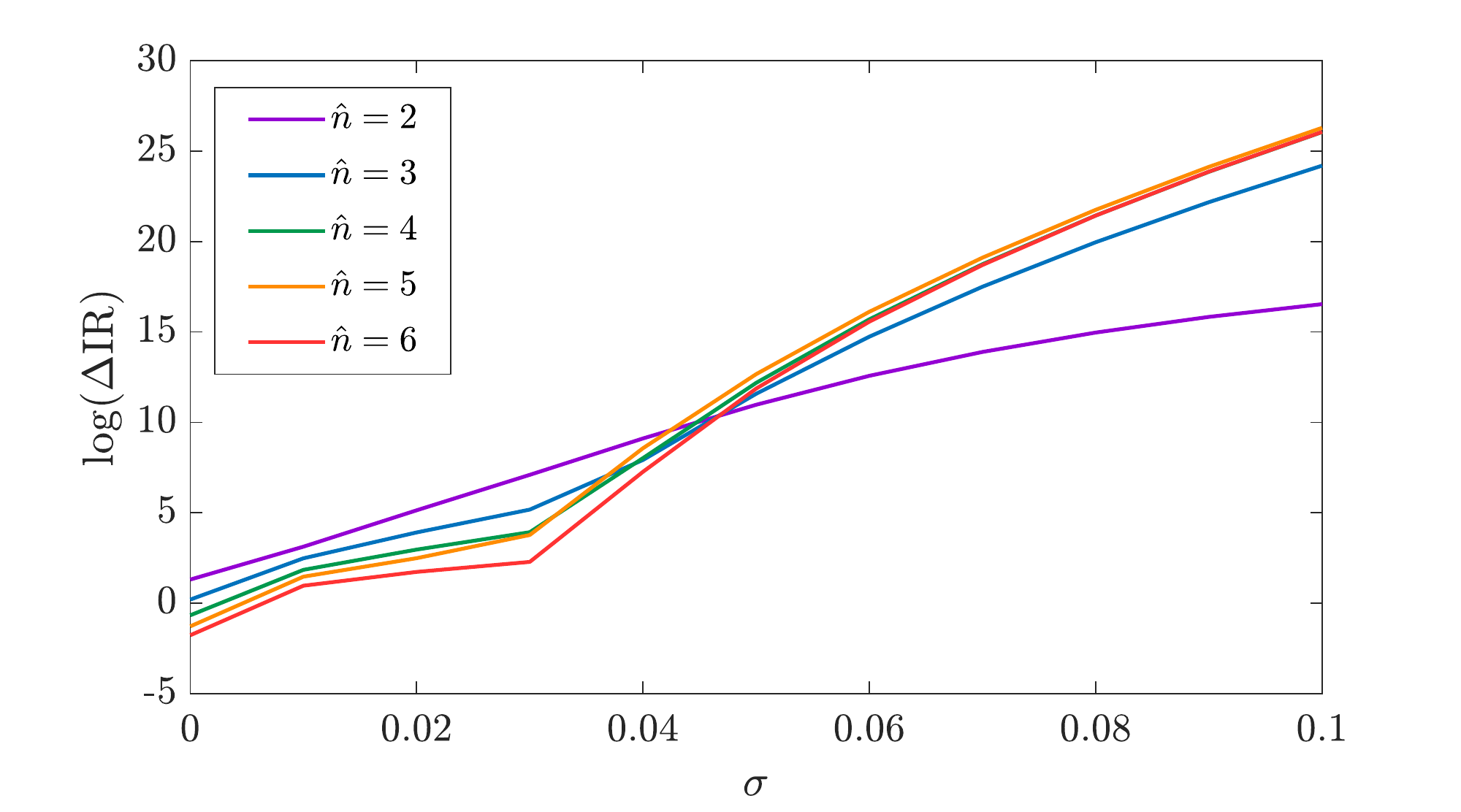}
        \caption{}
        \label{fig:ex2h}
    \end{subfigure}
    \caption{Example \ref{ex:ex2}. Accuracy of the  identified model versus the standard deviation $\sigma$ of the measurement noise for the SISO, unstable, NMP, discrete-time plant \eqref{eq:ex2} with $r_k\equiv1$ and with the controllers given by OFMPCOI and LQGI.
    The model accuracy is determined by $\Delta \mathrm{IR}$ at $k=140$ averaged over 100 simulations.
    OFMPCOI uses a strictly proper FIR initial model, and LQGI uses \eqref{eq:ex2} to compute the control gain with output and control weights $Q=2$ and $R=1$, respectively.
    For control and identification, OFMPCOI, LQGI, and RLS use the noisy measurement $y_{\rmn,k}.$
    (a) OFMPCOI.
    For each $\sigma,$ $\Delta \mathrm{IR}$ decreases as $\hat{n}$ increases.
    Therefore, at each measurement-noise level, the accuracy of the model improves as the model order $\hat{n}$ increases.
    (b) LQGI.
    For each $\hat{n},$ $\Delta \mathrm{IR}$ increases as $\sigma$ increases.
    Note that, unlike (a), for all $\sigma\ge0.05,$ an increase in the model order $\hat{n}$ results in a less accurate model.
    In addition, for each value of $\hat{n},$ the models in (b) are less accurate than those in (a).
    }
    \label{fig:ex2gh}
\end{figure*}

\begin{example}\label{ex:ex2}
\textit{Unstable, NMP, discrete-time plant.}
Consider the SISO, discrete-time input-output plant
\begin{align}
    y_k = 1.4 y_{k-1} - 0.3 y_{k-2} + u_{k-1} - 1.3 u_{k-2}. \label{eq:ex2}
\end{align}
Let $u_{\mathrm{min}} = -50,$ $u_{\mathrm{max}} = 50,$ $\Delta u_{\mathrm{min}} = -10,$ and $\Delta u_{\mathrm{max}} = 10.$
No output constraint is considered in this example.
The plant \eqref{eq:ex2} is initialized with $y_{-1} = y_{-2} = 0$ and $u_0 = u_{-1} = u_{-2} = 0.$
At each time step, OFMPCOI uses  \eqref{eq:QPnoslack}--\eqref{eq:DeltaUcon} with $\ell=20,$ $\bar{Q}=2I_{\ell-1},$ $\bar{P}=5,$ and $R=I_\ell.$
Let $\lambda=1$ and $P_0=10^3I_{2\hat{n}}.$

Figure \ref{fig:ex2ab} shows the response of OFMPCOI for $\hat{n}=n=2$ using a strictly proper model, where $\theta_0 = \alpha [0 \ 0 \ 0 \ 1]^\rmT,$ $\alpha\in\{-10,-1,-0.1,0.1,1,10\}.$
Note that the initial model is a finite-impulse-response (FIR) model.
Figure \ref{fig:ex2a} uses the constant command $r_k\equiv1,$ and Figure \ref{fig:ex2b} uses the three-step command $r_k$ given by \eqref{eq:commandex1}.

Figure \ref{fig:ex2cd} shows the response of OFMPCOI for $r_k\equiv1$ and $\hat{n}=n=2$ using a strictly proper model, which is initialized with 100 Gaussian-distributed $\theta_0$ with standard deviation $\sigma.$
Figure \ref{fig:ex2c} uses $\sigma=1$ and Figure \ref{fig:ex2d} uses $\sigma=2.$

Next, OFMPCOI uses the noisy measurement $y_{\rmn,k}\in\mathbb{R}^p$ instead of $y_k,$ that is,
\begin{align}
    y_{\rmn,k} = y_k + v_k, \label{eq:ynoisy}
\end{align}
where $v_k\in\mathbb{R}^p$ is a white Gaussian noise with standard deviation $\sigma.$
Figure \ref{fig:ex2ef} shows the response of OFMPCOI using $r_k\equiv1$ and the strictly proper FIR model $\theta_0 = [0_{(2\hat{n}-1)\times1}^\rmT \ 1]^\rmT$ with $\hat{n}=2,$ $\hat{n}=3,$ and $\hat{n}=4.$
Figure \ref{fig:ex2gh} shows the accuracy of the model identified by RLS as $\sigma$ increases for $\hat{n}\in\{1,2,\ldots,6\}$ with the controllers given by OFMPCOI and LQGI, that is, LQG with a built-in integrator.
The accuracy of the model is determined by $\Delta \mathrm{IR}$ at $k=140$ averaged over 100 simulation runs.

In this example, it is seen that, for unstable plants, although there is a bias and thus a lack of consistency in the model coefficients, the output approaches the step commands.
Additionally, we show that, even in the presence of sensor noise, with a single step command and, thus, in the absence of substantial persistency, the interplay between identification and control within OFMPCOI produces a more accurate model than LQGI as $\hat{n}$ increases, indicating more consistency within OFMPCOI than LQGI.

\end{example}

\clearpage

\begin{figure*}[t!]
    \centering
    \begin{subfigure}[t]{0.49\textwidth}
        \centering
        \includegraphics[trim = 10 110 10 30, width=\textwidth]{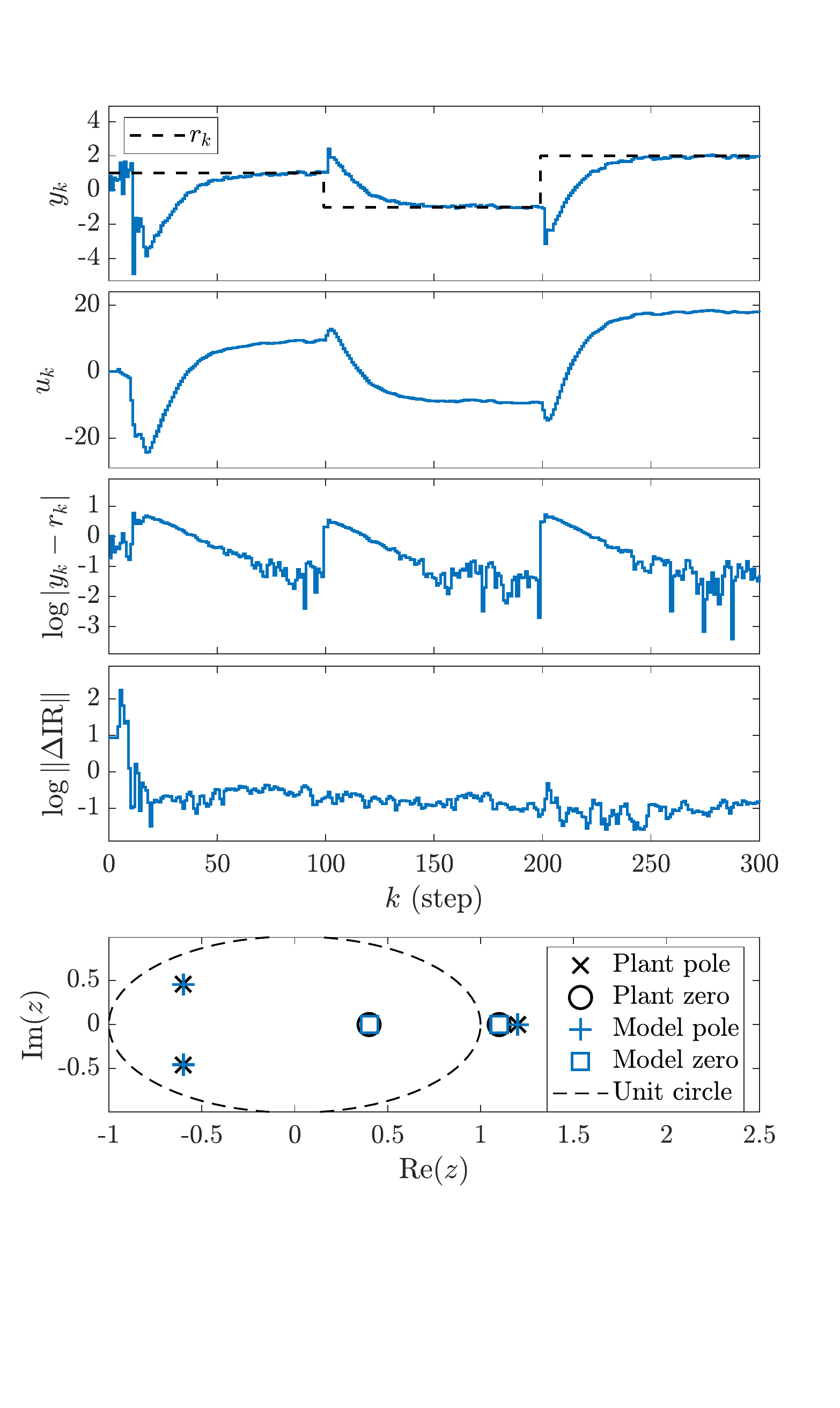}
        \caption{}
        \label{fig:ex2bisa}
    \end{subfigure}%
    ~ 
    \begin{subfigure}[t]{0.49\textwidth}
        \centering
        \includegraphics[trim = 10 110 10 30, width=\textwidth]{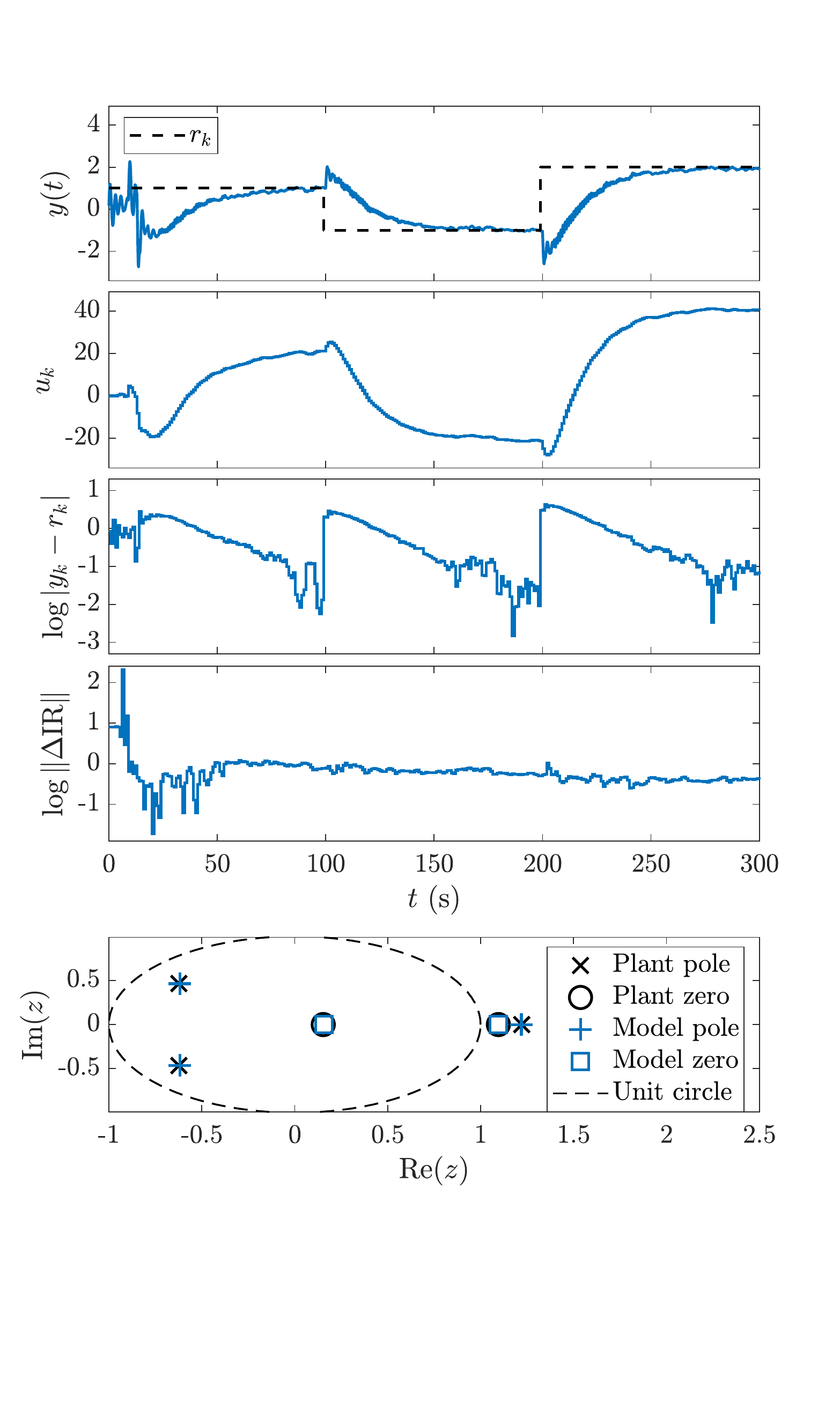}
        \caption{}
        \label{fig:ex2bisb}
    \end{subfigure}
    \caption{Example \ref{ex:ex2bis}. Command following for the SISO, unstably stabilizable plants \eqref{eq:ex2bisDT} and \eqref{eq:ex2bisCT} using $\hat{n}=n=3$ and the
    three-step command \eqref{eq:commandex2bis}.
    (a) Discrete-time plant \eqref{eq:ex2bisDT}.
    The output $y_k$ approaches each step command with decreasing command-following error.
    The bottom-most plot compares the poles and zeros of the identified model at $k=300$ to the poles and zeros of the plant.
    It was observed (not shown) that the response time was larger than $100$ steps for $32 \le \ell \le 40,$ and, OFMPCOI was unable to follow the step commands for $\ell \le 31$. 
    For $\ell > 85,$ it was observed (not shown) that the output diverged.
    (b) Continuous-time plant \eqref{eq:ex2bisCT}.
    The output  $y(t)$ approaches each step command with decreasing command-following error.
    The bottom-most plot compares the poles and zeros of the identified model at $t=300$ s to the poles and zeros of the exact discretization of the continuous-time plant \eqref{eq:ex2bisCT}.
    Note that the discretized plant is unstably stabilizable.
    For $\ell=50,$ on a Macbook Pro 2015 13' Retina, the average computation time of each iteration of OFMPCOI is approximately $t_\rmc = 0.019$ s, which is implementable in practice because $t_\rmc < T_s = 1$ s.
    For $\ell = 84,$ the average computation time of each iteration is approximately $t_\rmc = 0.024$ s.
    }
    \label{fig:ex2bisab}
\end{figure*}

\begin{example}\label{ex:ex2bis}
\textit{Unstably stabilizable plant.}
Consider the SISO, discrete-time plant
\begin{align}
    G(z) = \dfrac{(z-1.1)(z-0.4)}{(z-1.2)(z^2+1.2z+0.57)}. \label{eq:ex2bisDT}
\end{align}
Note that the order of the plant is $n=3$ and its relative degree is 1.
The damping ratio and natural frequency of the complex poles of \eqref{eq:ex2bisDT} are approximately $0.1$ and $0.8\pi$ rad/step, respectively.
Since there is an unstable pole on the right side of an NMP zero, pole-zero interlacing fails in the open right-half plane, and thus the plant is unstably stabilizable, that is, it can  be stabilized only by an unstable controller \cite{vidyasagarMITpressbook}.

Let $u_{\mathrm{min}} = -50,$ $u_{\mathrm{max}} = 50,$ $\Delta u_{\mathrm{min}} = -10,$ $\Delta u_{\mathrm{max}} = 10,$ $\ell=50,$ $\bar{Q}=4I_{\ell-1},$ $\bar{P}=4,$ $R=I_\ell,$ $\lambda=1,$ $P_0=10^3I_{2\hat{n}},$ and $\theta_0 = [0_{(2\hat{n}-1)\times 1}^\rmT \ \ 1]^\rmT.$
No output constraint is considered in this example.
The plant is initialized with $y_{-1} = -0.4,$ $y_{-2}=0.3,$ $y_{-3}=0.8,$ and $u_{-1}=u_{-2}=u_{-3}=0.$
OFMPCOI uses \eqref{eq:QPnoslack}--\eqref{eq:DeltaUcon} with the noisy measurement $y_{\rmn,k}$ given by \eqref{eq:ynoisy}, where the standard deviation of the noise $v_k$ is $\sigma=0.02.$
In this and the following examples, identification and control do not commence until the regressor matrix $\phi_k$ given by \eqref{eq:regressor} is populated with $\hat{n}$ measurements.
Figure \ref{fig:ex2bisa} shows the response of OFMPCOI for the discrete-time plant \eqref{eq:ex2bisDT} for $\hat{n}=n=3$ using the three-step command
\begin{align}
    r_k & =
    \begin{cases}
        1, & 0 \le k < 100, \\
        -1, & 100 \le k < 200, \\
        2, & k \ge 200.
    \end{cases} \label{eq:commandex2bis}
\end{align}

Next, consider the SISO, continuous-time plant
\begin{align}
    G(s) = \dfrac{(s-0.1)(s+0.6)}{(s-0.2)(s^2+2\zeta\omega_\rmn s+\omega_\rmn^2)}, \label{eq:ex2bisCT}
\end{align}
where $\zeta=0.1$ and $\omega_\rmn=0.8\pi.$
The data are sampled with sample period $T_\rms = 1$ s.
Note that the order and relative degree of \eqref{eq:ex2bisCT} are $n=3$ and 1, respectively.
Furthermore, note that \eqref{eq:ex2bisCT} and the sampled-data plant are unstably stabilizable.
The plant is initialized with $x(0)=[0.4 \ -\mspace{-5mu}0.8 \ 1.1]^\rmT,$ where $x(0)$ is the initial state of the realization
\begin{align}
    \dot{x} & = \left[
    \begin{matrix}
        -0.3027  &  -3.1   &    0.6317 \\
            2    &     0   &      0 \\
            0    &     1   &      0
    \end{matrix}
    \right] x +
    \left[
    \begin{matrix}
        1 \\ 0 \\ 0
    \end{matrix}
    \right] u, \\
    y & = \left[
    \begin{matrix}
        1  &  0.25 &  -0.03
    \end{matrix}
    \right]x.
\end{align}
Figure \ref{fig:ex2bisb} shows the response of OFMPCOI for the continuous-time plant \eqref{eq:ex2bisCT} using the same setup as for the discrete-time plant \eqref{eq:ex2bisDT}.

\end{example}

\clearpage

\begin{figure*}[t!]
    \centering
    \begin{subfigure}[t]{0.49\textwidth}
        \centering
        \includegraphics[trim = 10 110 10 30, width=\textwidth]{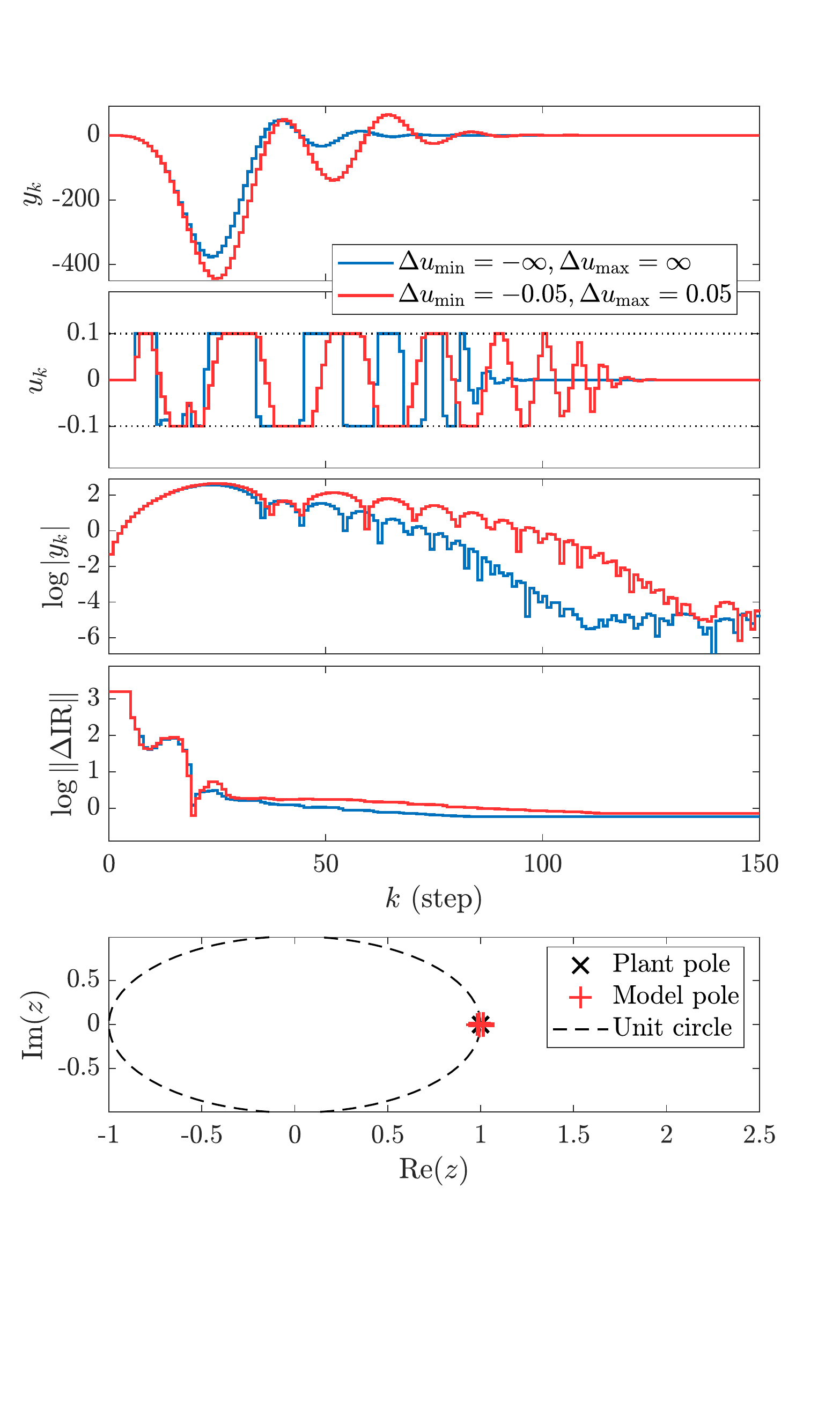}
        \caption{}
        \label{fig:ex2tera}
    \end{subfigure}%
    ~ 
    \begin{subfigure}[t]{0.49\textwidth}
        \centering
        \includegraphics[trim = 10 110 10 30, width=\textwidth]{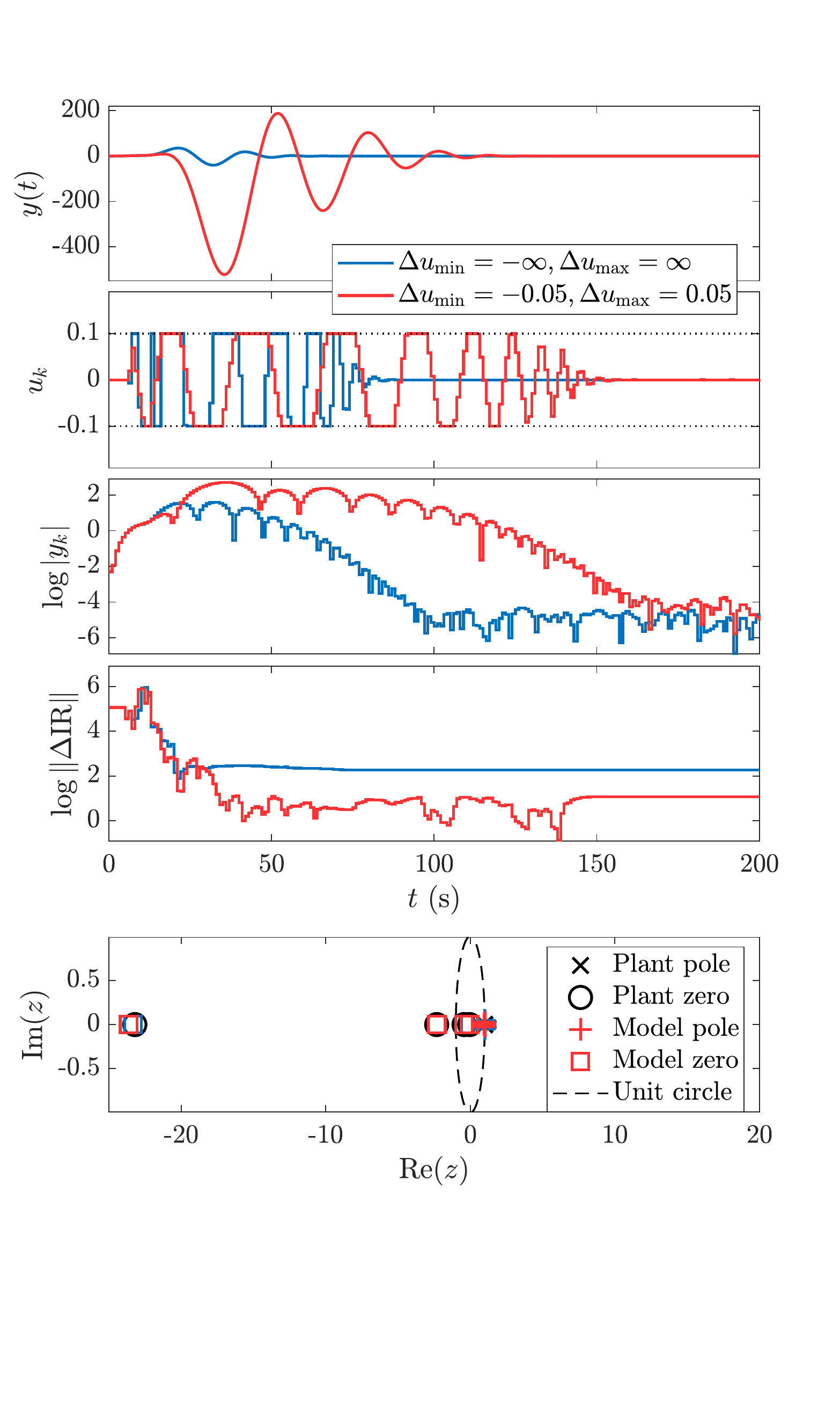}
        \caption{}
        \label{fig:ex2terb}
    \end{subfigure}
    \caption{Example \ref{ex:ex2ter}. Stabilization of the SISO, chain-of-integrators plants \eqref{eq:ex2terDT} and \eqref{eq:ex2terCT} using $u_{\mathrm{min}} = -0.1$ and $u_{\mathrm{max}} = 0.1$ with and without the move-size control constraint \eqref{eq:movesat}, where $\Delta u_{\mathrm{min}} = -0.05$ and $\Delta u_{\mathrm{max}} = 0.05.$
    (a) Discrete-time plant \eqref{eq:ex2terDT}.
    Note that, both with and without the move-size control constraint \eqref{eq:movesat}, the plant is stabilized by OFMPCOI.
    The bottom-most plot compares the poles of the identified model at $k=150$ to the poles of the plant.
    (b) Continuous-time plant \eqref{eq:ex2terCT}.
    Note that, both with and without the move-size control constraint \eqref{eq:movesat}, the plant is stabilized by OFMPCOI.
    The bottom-most plot compares the poles and zeros of the identified model at $t=200$ s to the poles and zeros of the exact discretization of the continuous-time plant.
    Note that, compared to (a), the discretized plant has four additional sampling zeros due to the discretization of \eqref{eq:ex2terCT}.
    }
    \label{fig:ex2terab}
\end{figure*}

\begin{example}\label{ex:ex2ter}
\textit{Chain-of-integrators plant.}
Consider the SISO, discrete-time plant
\begin{align}
    G(z) = \dfrac{1}{(z-1)^5}, \label{eq:ex2terDT}
\end{align}
which consists of a chain of five integrators with relative degree 5.
Let $u_{\mathrm{min}} = -0.1,$ $u_{\mathrm{max}} = 0.1,$ $\ell=40,$ $\bar{Q}=10I_{\ell-1},$ $\bar{P}=10,$ $R=2I_\ell,$ $\hat{n}=n=5,$ $\lambda=1,$ $P_0=10^5I_{2\hat{n}},$ $r_k\equiv0,$ and $\theta_0 = [-\mspace{-5mu}4.995 \ 9.99 \ -\mspace{-5mu}9.99 \ 4.995 \ -\mspace{-5mu}0.995 \ 0 \ 0 \ 0 \ 0 \ 0.999]^\rmT.$
No output constraint is considered in this example.
The plant is initialized with $y_{-1} = -1.37e\mbox{-5},$ $y_{-2}=-1.69e\mbox{-5},$ $y_{-3}=0.63e\mbox{-5},$ $y_{-4}=1.02e\mbox{-5},$ $y_{-5}=2.12e\mbox{-5},$ and $u_{-1}=u_{-2}=u_{-3}=u_{-4}=u_{-5}=0.$
OFMPCOI uses \eqref{eq:QPnoslack}--\eqref{eq:DeltaUcon} with the measurement $y_{k}.$
Figure \ref{fig:ex2tera} shows the response of OFMPCOI
with and without the move-size control constraint \eqref{eq:movesat}, where $\Delta u_{\mathrm{min}} = -0.05$ and $\Delta u_{\mathrm{max}} = 0.05.$

Next, consider the SISO, continuous-time plant
\begin{align}
    G(s) = \dfrac{1}{s^5}, \label{eq:ex2terCT}
\end{align}
which consists of a chain of five integrators with relative degree 5.
The data are sampled with sample period $T_\rms = 1$ s.
The plant is initialized with $x(0) = e\mbox{-2}[0.5 \ -\mspace{-5mu}1.0 \ 2.0 \ 4.5 \ -\mspace{-5mu}0.15 ]^\rmT,$ where $x(0)$ is the initial state of the realization
\begin{align}
    \dot{x} & = \left[
    \begin{matrix}
        0  &   1  &   0  &   0   &  0 \\
        0  &   0  &   1  &   0   &  0 \\
        0  &   0  &   0  &   1   &  0 \\
        0  &   0  &   0  &   0   &  1 \\
        0  &   0  &   0  &   0   &  0
    \end{matrix}
    \right] x + 
    \left[
    \begin{matrix}
        0 \\ 0 \\ 0 \\ 0 \\ 1
    \end{matrix}
    \right]u, \\
    y & = \left[
    \begin{matrix}
        1 & 0 & 0 & 0 & 0
    \end{matrix}
    \right]x.
\end{align}
Using the same setup as for the discrete-time plant \eqref{eq:ex2terDT}, Figure \ref{fig:ex2terb} shows the response of OFMPCOI with and without the move-size control constraint \eqref{eq:movesat}.
\end{example}

\clearpage

\begin{figure*}[t!]
    \centering
    \begin{subfigure}[t]{0.49\textwidth}
        \centering
        \includegraphics[trim = 10 110 10 30, width=\textwidth]{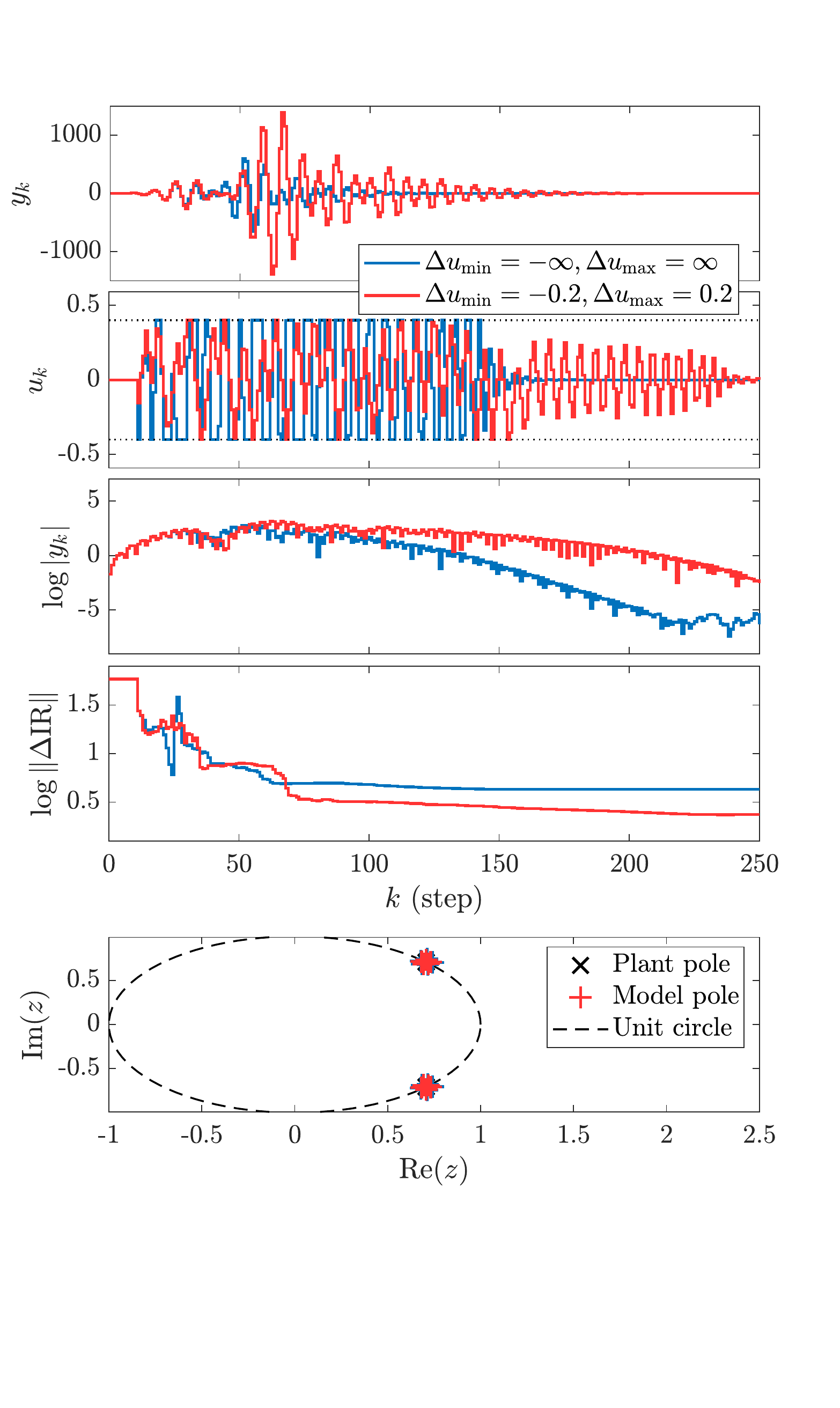}
        \caption{}
        \label{fig:ex2quata}
    \end{subfigure}%
    ~ 
    \begin{subfigure}[t]{0.49\textwidth}
        \centering
        \includegraphics[trim = 10 110 10 30, width=\textwidth]{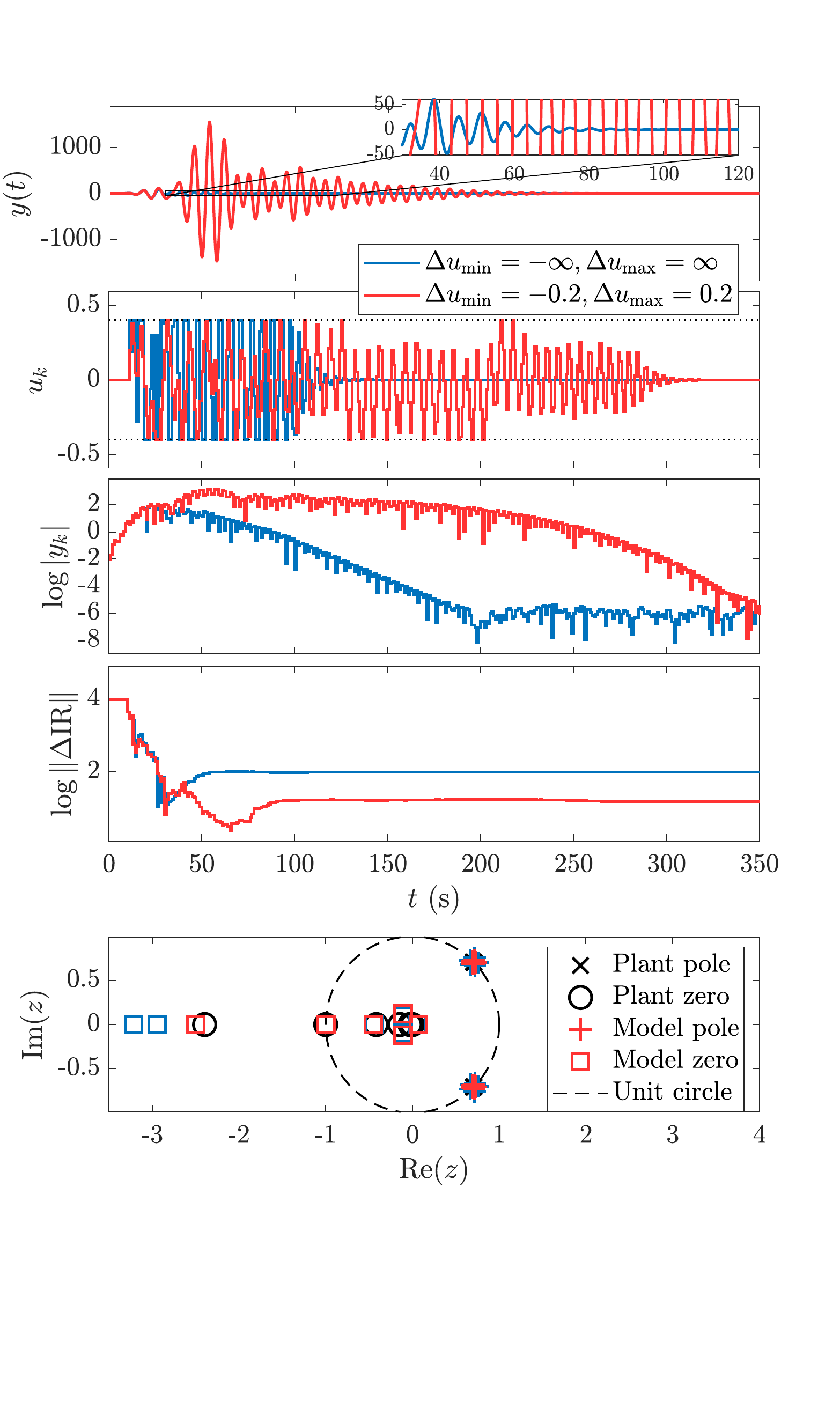}
        \caption{}
        \label{fig:ex2quatb}
    \end{subfigure}
    \caption{Example \ref{ex:ex2quat}. Stabilization of the SISO, chain-of-undamped-oscillators plants \eqref{eq:ex2quatDT} and \eqref{eq:ex2quatCT} using $u_{\mathrm{min}} = -0.4$ and $u_{\mathrm{max}} = 0.4$ with and without the move-size control constraint \eqref{eq:movesat}, where $\Delta u_{\mathrm{min}} = -0.2$ and $\Delta u_{\mathrm{max}} = 0.2.$
    (a) Discrete-time plant \eqref{eq:ex2quatDT}.
    Note that, both with and without the move-size control constraint \eqref{eq:movesat}, the plant is stabilized by OFMPCOI.
    The bottom-most plot compares the poles of the identified model at $k=250$ to the poles of the plant.
    (b) Continuous-time plant \eqref{eq:ex2quatCT}.
    Note that, both with and without the move-size control constraint \eqref{eq:movesat}, the plant is stabilized by OFMPCOI.
    The bottom-most plot compares the poles and zeros of the identified model at $t=350$ s to the poles and zeros of the exact discretization of the continuous-time plant.
    }
    \label{fig:ex2quatab}
\end{figure*}

\begin{example}\label{ex:ex2quat}
\textit{Chain-of-undamped-oscillators plant.}
Consider the SISO, discrete-time plant
\begin{align}
    G(z) = \dfrac{1}{(z^2 - 1.414z + 1)^5}, \label{eq:ex2quatDT}
\end{align}
which consists of a chain of five undamped oscillators with natural frequency $\omega_\rmn = 0.25\pi$ rad/step.
Let $u_{\mathrm{min}} = -0.4,$ $u_{\mathrm{max}} = 0.4,$ $\ell=60,$ $\bar{Q}=10I_{\ell-1},$ $\bar{P}=10,$ $R=2I_\ell,$ $\hat{n}=n=10,$ $\lambda=1,$ $P_0=10^5I_{2\hat{n}},$ $r_k\equiv0,$ and $\theta_0 = [-\mspace{-5mu}7.064 \ 24.98 \ -\mspace{-5mu}56.51 \ 89.91 \ -\mspace{-5mu}104.55 \ 89.91 \ -\mspace{-5mu}56.51 \ 24.98 \ -\mspace{-5mu}7.064 \ 0.999 \ 0_{1\times 9} \ 0.999]^\rmT.$
No output constraint is considered in this example.
The plant is initialized with $y_{-1} = -0.29e\mbox{-2},$ $y_{-2}=0.49e\mbox{-2},$ $y_{-3}=0.45e\mbox{-2},$ $y_{-4}=-0.68e\mbox{-2},$ $y_{-5}=-0.81e\mbox{-2},$ $y_{-6}=0.88e\mbox{-2},$ $y_{-7}=1.56e\mbox{-2},$ $y_{-8}=-0.80e\mbox{-2},$ $y_{-9}=-2.75e\mbox{-2},$ $y_{-5}=-0.17e\mbox{-2},$ and $u_{-1}=u_{-2}=u_{-3}=u_{-4}=u_{-5}=u_{-6}=u_{-7}=u_{-8}=u_{-9}=u_{-10}=0.$
OFMPCOI uses \eqref{eq:QPnoslack}--\eqref{eq:DeltaUcon} with the measurement $y_{k}.$
Figure \ref{fig:ex2quata} shows the response of OFMPCOI with and without the move-size control constraint
\eqref{eq:movesat}, where $\Delta u_{\mathrm{min}} = -0.2$ and $\Delta u_{\mathrm{max}} = 0.2.$

Next, consider the SISO, continuous-time plant
\begin{align}\label{eq:ex2quatCT}
    G(s) = \dfrac{1}{(s^2 + (0.25\pi)^2)^5},
\end{align}
which consists of a chain of five undamped oscillators with natural frequency $\omega_\rmn = 0.25\pi$ rad/s.
The data are sampled with sample period $T_\rms = 1$ s.
The plant is initialized with $x(0) =  [0.01 \ -\mspace{-5mu}0.02 \ 0.04 \ 0.09 \ -\mspace{-5mu}0.03 \ 0.08 \ -\mspace{-5mu}0.07 \ -\mspace{-5mu}0.10 \ 0.02 \ 0.06]^\rmT,$ where $x(0)$ is the initial state of the realization
\begin{align}
    \dot{x} & = \left[
    \begin{matrix}
        0  &  1    &     0    &     0     &    0     &    0    &     0     &    0     &    0    &     0 \\
   -0.6169    &     0  &  1     &    0    &     0     &    0    &     0    &     0     &    0    &     0 \\
         0    &     0    &     0  &  1     &    0     &    0     &    0     &    0     &    0     &    0 \\
         0    &     0  & -0.6169    &     0  &  1    &     0   &      0     &    0      &   0     &    0 \\
         0    &     0    &     0     &    0    &     0  &  1     &    0     &    0     &    0    &     0 \\
         0    &     0     &    0     &    0 &  -0.6169     &    0  &  1     &    0     &    0     &    0 \\
         0    &     0     &    0     &    0    &     0     &    0    &     0 &   1     &    0      &   0 \\
         0    &     0     &    0     &    0     &    0     &    0  & -0.6169    &     0  &  1    &     0 \\
         0    &     0      &   0     &    0     &    0    &     0    &     0     &    0    &     0 &   1 \\
         0    &     0     &    0     &    0    &     0     &    0     &    0    &     0 &  -0.6169    &     0
    \end{matrix}
    \right] x + 
    \left[
    \begin{matrix}
        0 \\ 0 \\ 0 \\ 0 \\ 0 \\ 0 \\ 0 \\ 0 \\ 1
    \end{matrix}
    \right]u, \\
    y & = \left[
    \begin{matrix}
        1 & 0 & 0 & 0 & 0 & 0 & 0 & 0 & 0 & 0
    \end{matrix}
    \right] x.
\end{align}
Using $\theta_0 = [-\mspace{-5mu}7.06 \allowbreak \ 24.98 \allowbreak \ -\mspace{-5mu}56.51 \allowbreak \ 89.91 \allowbreak \ -\mspace{-5mu}104.55 \allowbreak \ 89.91 \allowbreak \ -\mspace{-5mu}56.51 \allowbreak \ 24.98 \allowbreak \ -\mspace{-5mu}7.06 \allowbreak \ 0.999 \allowbreak \ 0.00 \allowbreak \ 0.00 \allowbreak \ 0.01 \allowbreak \ 0.10 \allowbreak \ 0.28 \allowbreak \ 0.28 \allowbreak \ 0.10 \allowbreak \ 0.01 \allowbreak \ 0.00 \allowbreak \ 0.00]^\rmT$ and the same setup as for \eqref{eq:ex2quatDT}, Figure \ref{fig:ex2quatb} shows the response of OFMPCOI with and without the move-size control constraint \eqref{eq:movesat}.

\end{example}

\clearpage

\begin{figure*}[t!]
    \centering
    \begin{subfigure}[t]{0.49\textwidth}
        \centering
        \includegraphics[trim = 10 80 10 340, width=\textwidth]{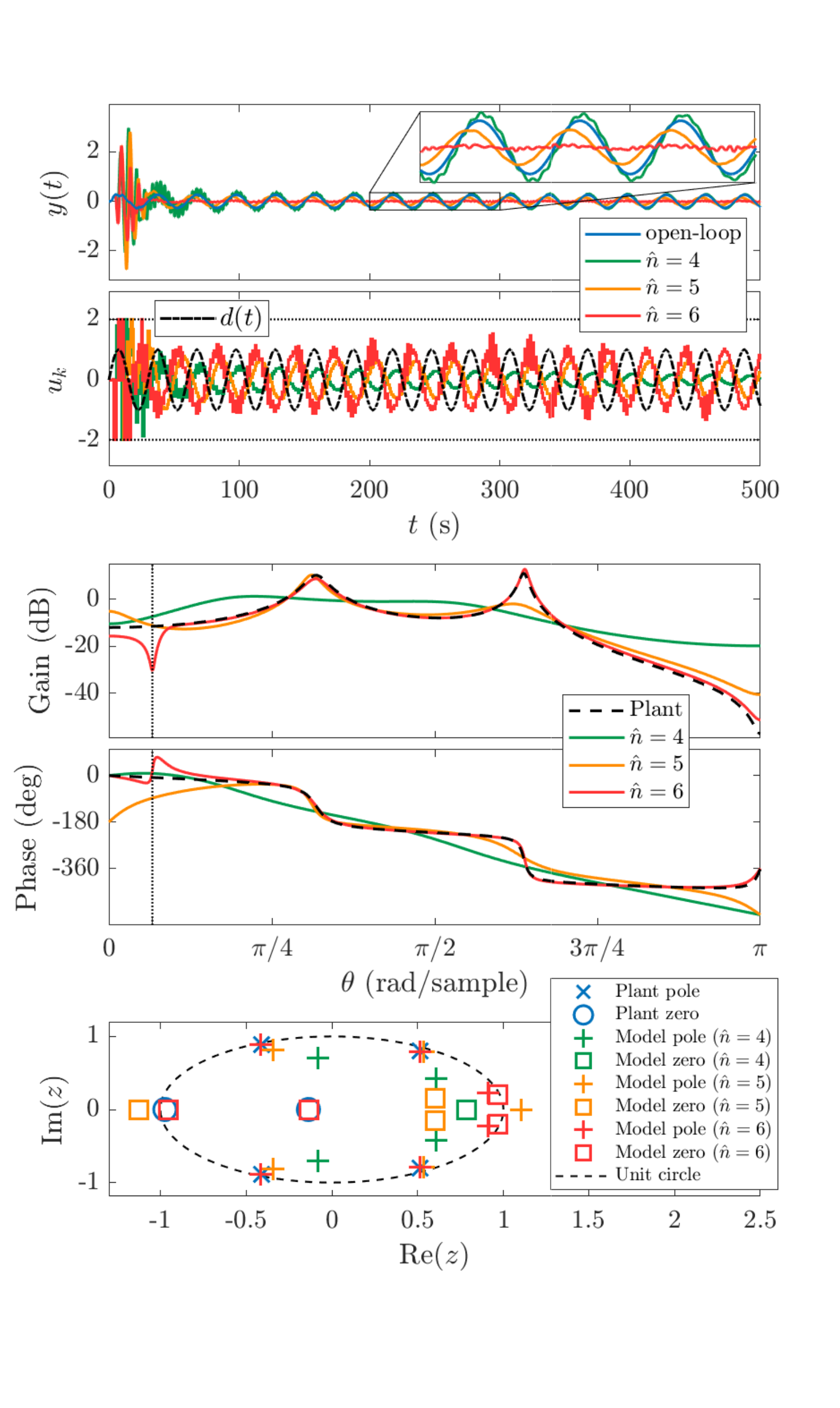}
        \caption{}
        \label{fig:ex3a}
    \end{subfigure}%
    ~ 
    \begin{subfigure}[t]{0.49\textwidth}
        \centering
        \includegraphics[trim = 10 80 10 30, width=\textwidth]{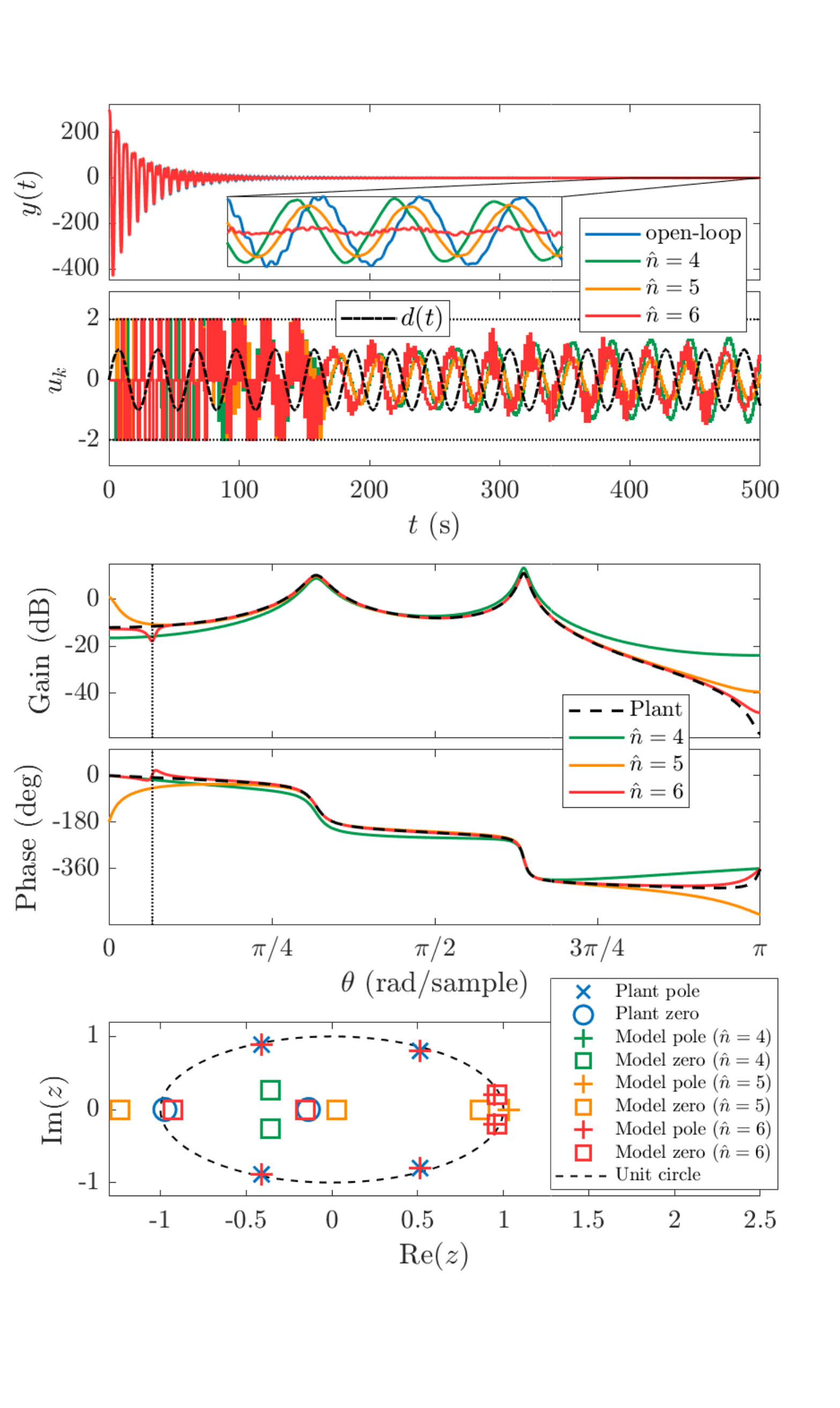}
        \caption{}
        \label{fig:ex3b}
    \end{subfigure}
    \caption{Example \ref{ex:ex3}.
    Matched disturbance rejection for the SISO, asymptotically stable, continuous-time plant \eqref{eq:ex3} and the harmonic disturbance \eqref{eq:HDR} with $\hat{n}=n=4,$ $\hat{n}=5,$ and $\hat{n}=6.$
    (a)
    Zero initial conditions.
    Note that the performance improves as $\hat{n}$ increases, and that $u_k$ approximately cancels $d(t)$ for $\hat{n}  = 6.$
    Furthermore, as $\hat{n}$ increases, the frequency response of the identified models at $t=500$ s become more accurate.
    For $\hat{n}=6,$ the identified model has an approximate pole-zero cancellation at the disturbance frequency $\theta_\rmd = \frac{\pi}{15}$ rad/sample, which is denoted by the vertical dash-dotted line; the spurious pole in the identified model allows MPC to predict the future harmonic response of the system, which facilitates optimization of the future control inputs.
    This pole-zero cancellation serves as an implicit internal model of the unknown disturbance.
    The bottom-most plot compares the poles and zeros of the identified model at $t=500$ s to the poles and zeros of the exact discretization of the plant.
    For $\hat{n}=6,$ the approximate pole-zero cancellations on the unit circle occur at the frequency $\theta_\rmd$ of the harmonic disturbance.
    (b)
    Nonzero initial conditions $x(0) = [10 \ -\mspace{-5mu}5 \ -\mspace{-5mu}8 \ 300]^\rmT$ for the state-space realization \eqref{eq:ex3ss1}--\eqref{eq:ex3ss2}. Compared to the case of zero initial conditions, the transient response is more pronounced, and the control saturation occurs over a longer interval of time.
    Moreover, since nonzero initial conditions result in more persistency during the transient,  the identified models in (b) are more accurate than those in (a).
    }
    \label{fig:ex3ab}
\end{figure*}

\begin{figure*}[t!]
    \centering
    \begin{subfigure}[t]{0.49\textwidth}
        \centering
        \includegraphics[trim = 10 120 10 30, width=\textwidth]{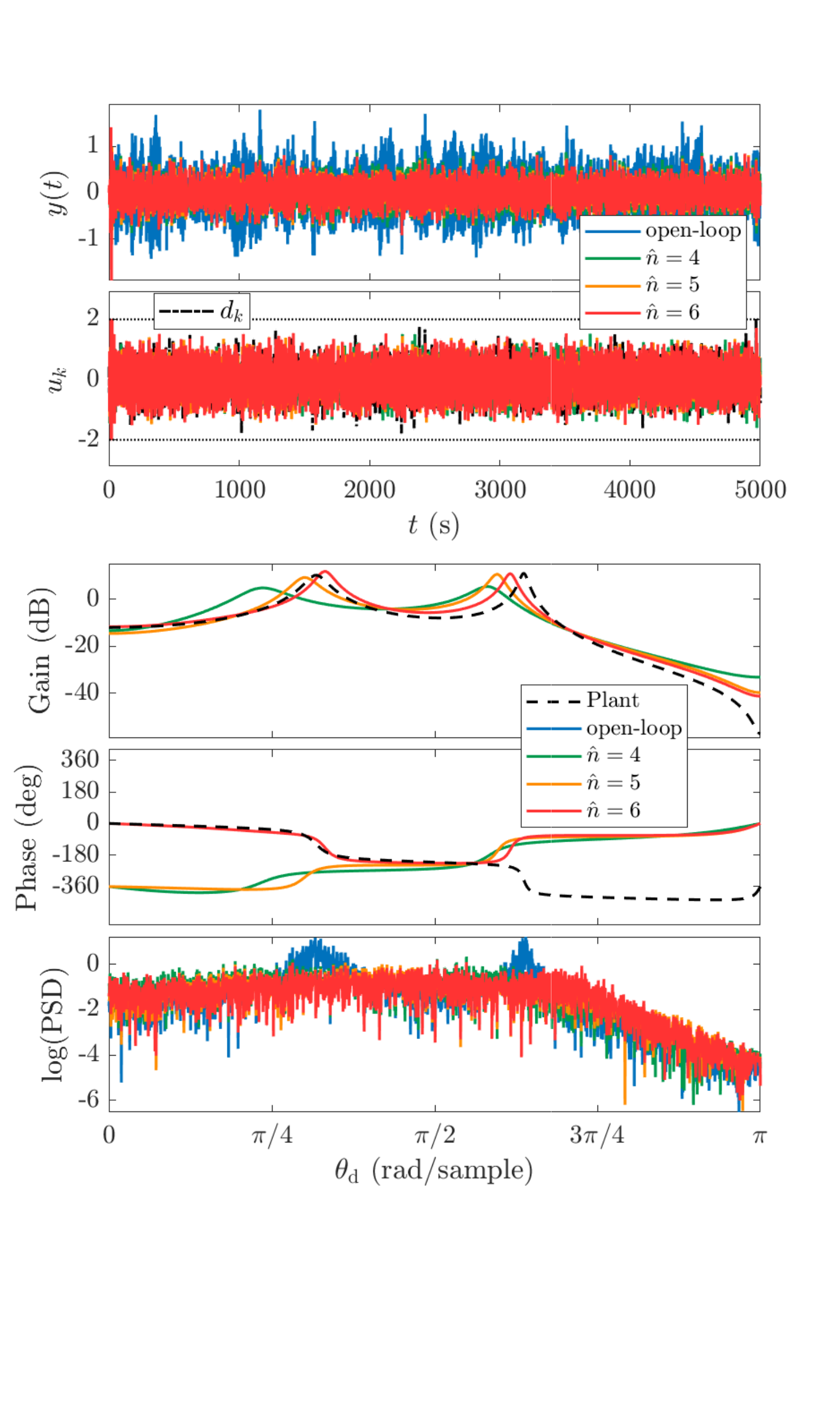}
        \caption{}
        \label{fig:ex3c}
    \end{subfigure}%
    ~ 
    \begin{subfigure}[t]{0.49\textwidth}
        \centering
        \includegraphics[trim = 10 120 10 30, width=\textwidth]{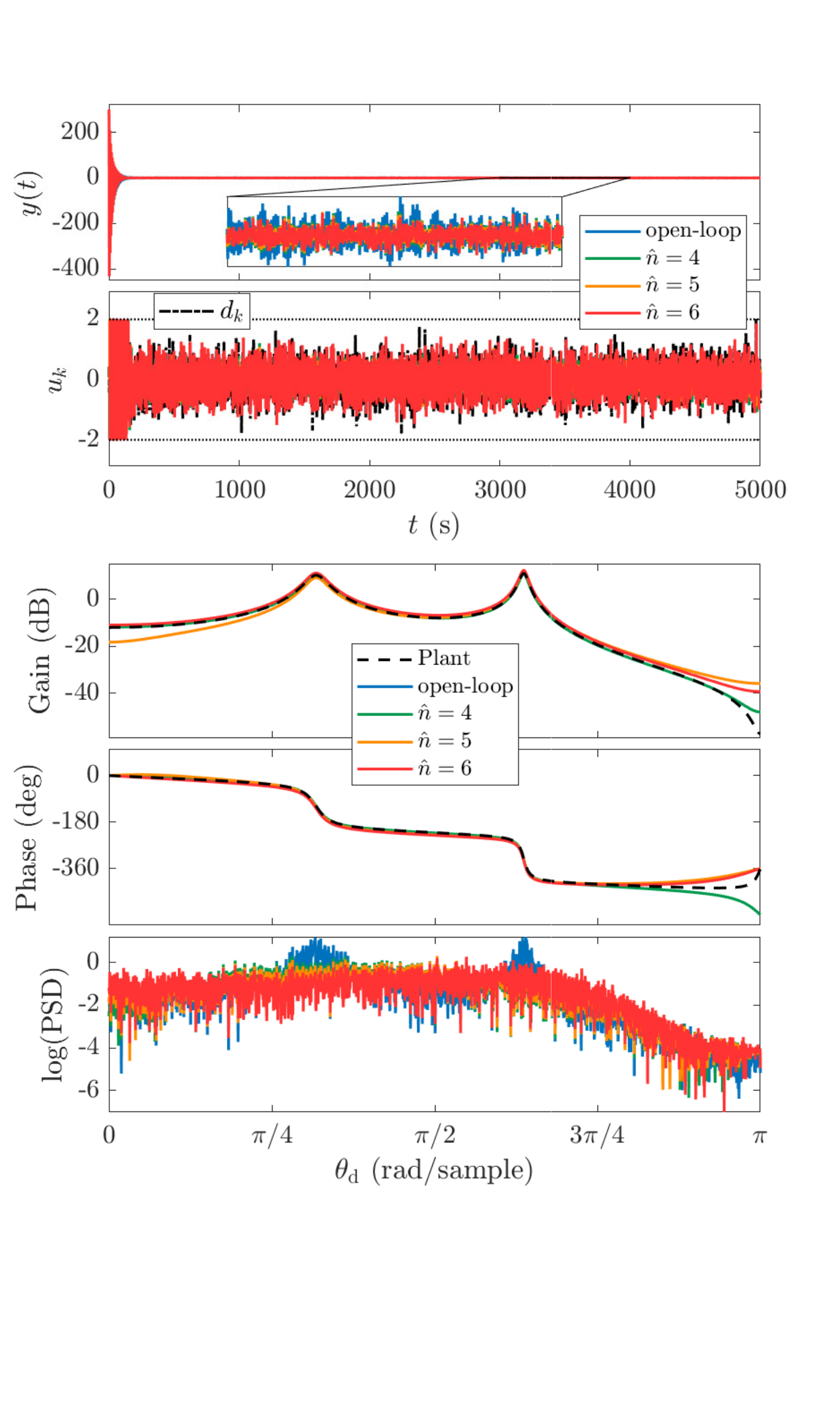}
        \caption{}
        \label{fig:ex3d}
    \end{subfigure}
    \caption{Example \ref{ex:ex3}.
    Matched disturbance rejection for the SISO, asymptotically stable, continuous-time plant \eqref{eq:ex3} and the discrete-time broadband disturbance $d_k$ with standard deviation $\sigma=0.5$ for $\hat{n}=n=4,$ $\hat{n}=5,$ and $\hat{n}=6.$
    (a)
    Zero initial conditions.
    Note that, as $\hat{n}$ increases, the magnitudes of the frequency response of the identified models at $t=5000$ s become more accurate.
    The bottom-most plot shows that, for all $\hat{n},$ OFMPCOI reduces the two peaks of the open-loop power spectral density (PSD) of $y_k,$ which indicates that OFMPCOI rejects  the broadband disturbance $d_k.$
    (b)
    Nonzero initial conditions $x(0) = [10 \ -\mspace{-5mu}5 \ -\mspace{-5mu}8 \ 300]^\rmT$ for the state-space realization \eqref{eq:ex3ss1}--\eqref{eq:ex3ss2}.
   As in (a), the bottom-most plot shows that, for all $\hat{n},$ OFMPCOI reduces the two peaks of the open-loop PSD of $y_k.$
    Note that, for all $\hat{n},$ the frequency responses of the identified models at $t=5000$ s are more accurate in (b) than in (a) due to the nonzero initial conditions.
    }
    \label{fig:ex3cd}
\end{figure*}

\begin{example} \label{ex:ex3}
\textit{Asymptotically stable, continuous-time plant.}
Consider the SISO, continuous-time plant
\begin{align}
    G(s) = \dfrac{1}{(s^2+2\zeta_1\omega_{\rmn,1} s + \omega_{\rmn,1}^2)(s^2+2\zeta_2\omega_{\rmn,2} s + \omega_{\rmn,2}^2)}, \label{eq:ex3}
\end{align}
where $\zeta_1 = 0.05,$ $\omega_{\rmn,1} = 1,$ $\zeta_2 = 0.01,$ and $\omega_{\rmn,2} = 2.$
The data are sampled with sample period $T_s = 1$ s.
Note that the order of the plant is $n=4.$
Let $u_{\mathrm{min}} = -2,$ $u_{\mathrm{max}} = 2,$ $\Delta u_{\mathrm{min}} = -2,$ $\Delta u_{\mathrm{max}} = 2,$  $\theta_0 = [0_{(2\hat{n}-1)\times1}^\rmT \ 1]^\rmT,$ $r_k \equiv0,$ $\lambda=1$ and $P_0=10^3I_{2\hat{n}}.$
No output constraint is considered in this example.
At each time step, OFMPCOI solves  \eqref{eq:QPnoslack}--\eqref{eq:DeltaUcon} with $\ell=10,$ $\bar{Q}=50I_{\ell-1},$ $\bar{P}=50,$ and $R=0.1I_\ell.$
OFMPCOI uses the noisy measurement $y_{\rmn,k}$ given by \eqref{eq:ynoisy}, where the standard deviation of the noise $v_k$ is $\sigma=0.005.$

Figure \ref{fig:ex3ab} shows harmonic disturbance rejection and Figure \ref{fig:ex3cd} shows broadband disturbance rejection using OFMPCOI with $\hat{n}=n=4,$ $\hat{n}=5,$ and $\hat{n}=6.$
The matched, continuous-time, harmonic disturbance is given by
\begin{align}
    d(t) = A_\rmd \sin(\omega_\rmd t), \label{eq:HDR}
\end{align}
where $A_\rmd = 1$ and $\omega_\rmd = \pi/15.$
The matched, discrete-time, broadband disturbance is given by $d_k,$ which is Gaussian-distributed with standard deviation $\sigma=0.5.$
Figures \ref{fig:ex3a} and \ref{fig:ex3c} consider zero initial conditions for \eqref{eq:ex3}, and Figures \ref{fig:ex3b} and \ref{fig:ex3d} consider nonzero initial conditions, where $x(0) = [10 \ -\mspace{-5mu}5 \ -\mspace{-5mu}8 \ 300]^\rmT$ is used as the initial state of the realization
\begin{align}
    \dot{x} & = \left[
    \begin{matrix}
       -0.14 &   -2.5  &   -0.2 &  -2 \\
        2    &     0     &    0     &    0 \\
         0    &      1    &     0    &     0 \\
         0    &     0     &   1      &   0
    \end{matrix} \right] x + \left[\begin{matrix}0.5 \\ 0 \\ 0 \\ 0\end{matrix}\right] (u+d), \label{eq:ex3ss1} \\
    y & = \left[\begin{matrix}0 & 0 & 0 & 1\end{matrix}\right]x. \label{eq:ex3ss2}
\end{align}

This example exhibits the manifestations of exigency within closed-loop identification.
In fact, in the presence of a harmonic disturbance, a pole-zero cancellation occurs at the frequency of the harmonic disturbance, which serves as an implicit internal model of the unknown disturbance.
This pole-zero cancellation indicates exigency, whereby RLS captures the frequency and amplitude of the harmonic disturbance in order to predict and cancel the harmonic disturbance for MPC.
For broadband disturbances, as the order of the model increases, the frequency response of the identified model becomes more accurate, thus manifesting exigency.

\end{example}

 \clearpage
 
\begin{figure*}[t!]
    \centering
    \begin{subfigure}[t]{0.49\textwidth}
        \centering
        \includegraphics[trim = 10 100 10 60, width=\textwidth]{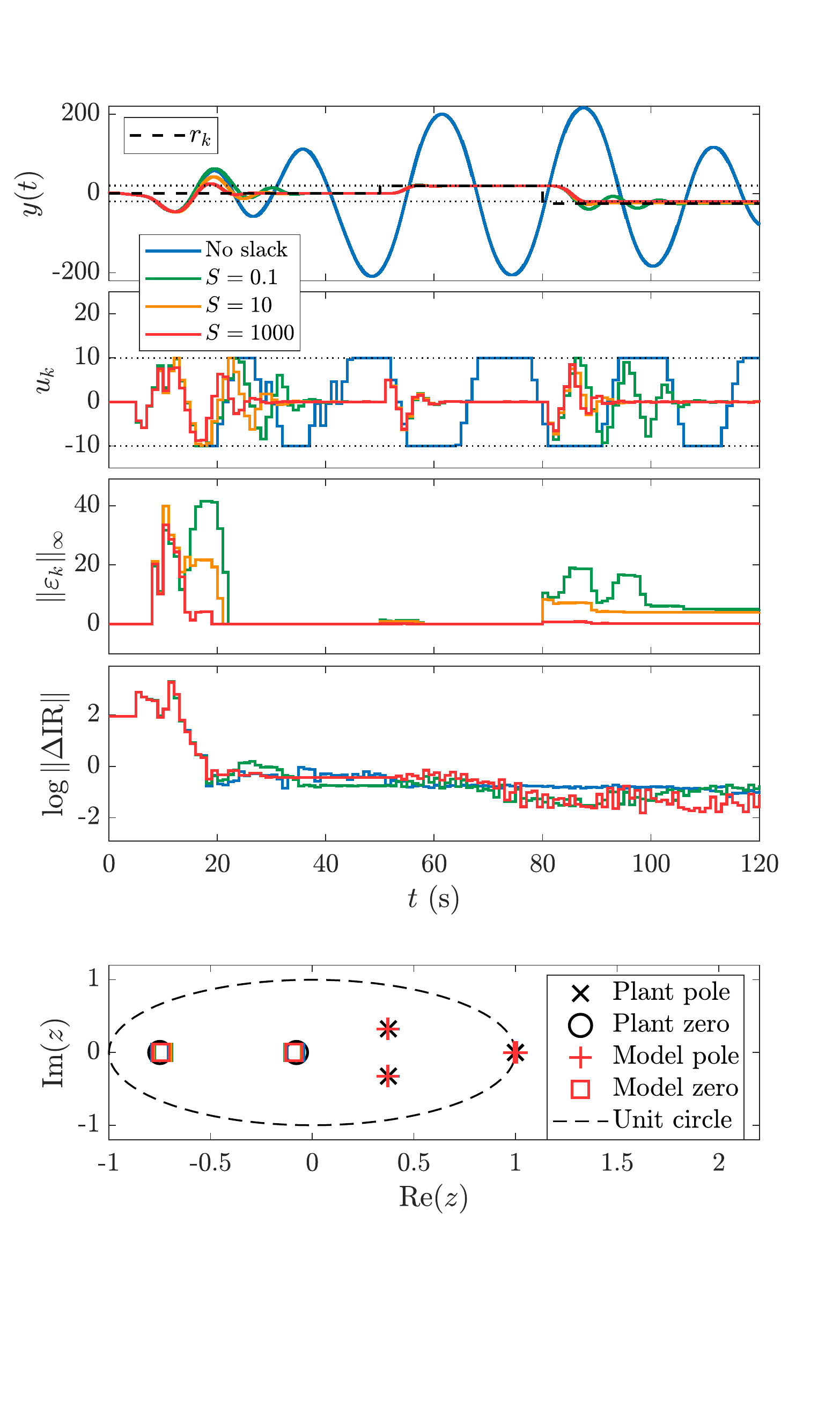}
        \caption{}
        \label{fig:ex4a}
    \end{subfigure}%
    ~ 
    \begin{subfigure}[t]{0.49\textwidth}
        \centering
        \includegraphics[trim = 10 100 10 60, width=\textwidth]{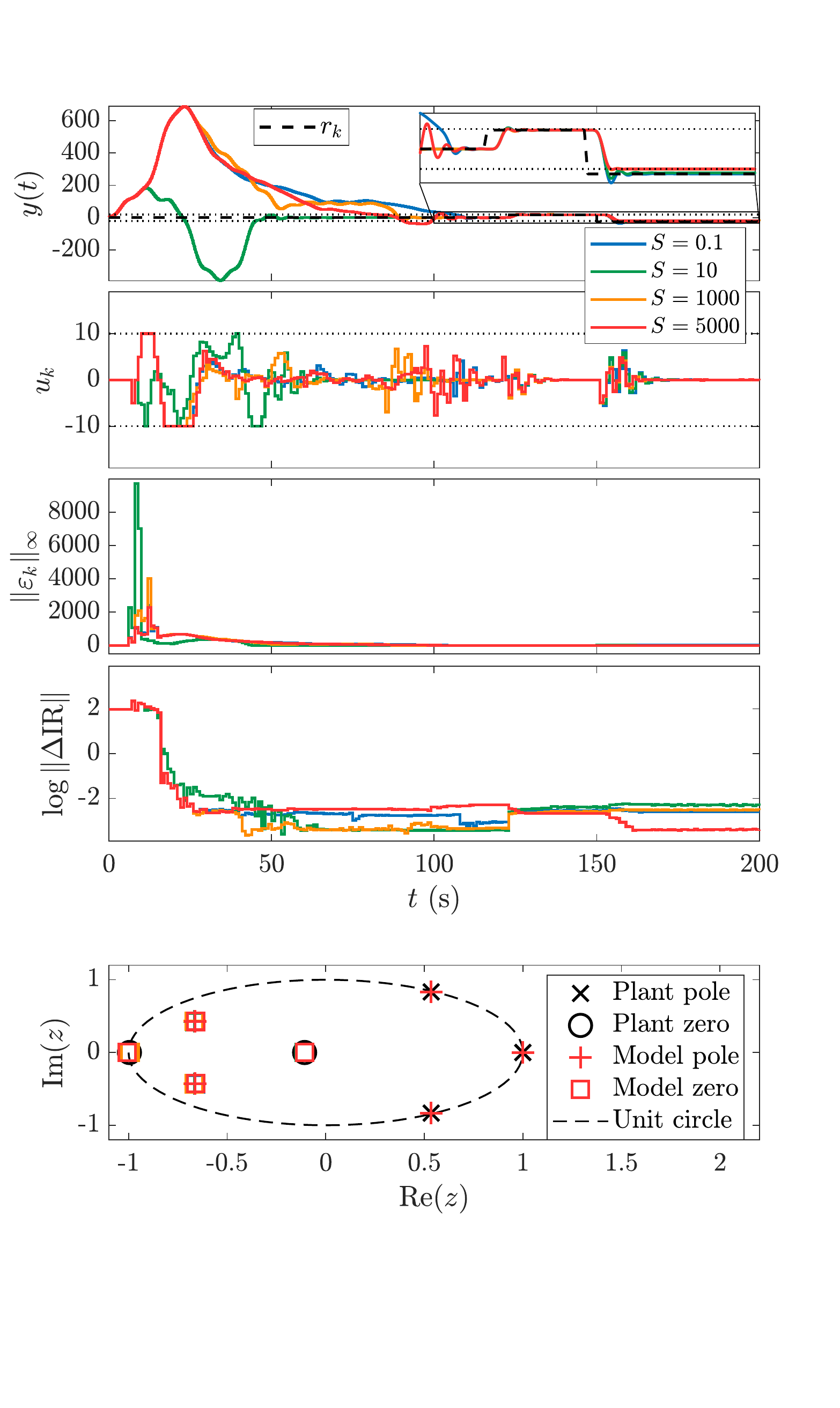}
        \caption{}
        \label{fig:ex4b}
    \end{subfigure}
    \caption{
    Example \ref{ex:ex4}.
    Command following with the output constraint $-20 \le y_k \le 20$ and the SISO, unstable, continuous-time plant \eqref{eq:ex4} for various values of $S.$
    (a)
    OFMPCOI obtained from
    \eqref{eq:QPnoslack}--\eqref{eq:DeltaUcon} and \eqref{eq:QPwithslack}--\eqref{eq:slackconstraint} with $\hat{n}=n=4,$ $\zeta=0.7,$ $x(0)=[1 \ -\mspace{-5mu}3 \ \ 2 \ \ 0.5]^\rmT,$ and the three-step command \eqref{eq:commandex4}. 
    At $t=9$ s,  \eqref{eq:QPnoslack}--\eqref{eq:DeltaUcon} becomes infeasible, and the three-step command cannot be followed.
    In contrast, with  \eqref{eq:QPwithslack}--\eqref{eq:slackconstraint}, $y(t)$ approaches the first two step commands, and the slack is activated at $t=8$ s, $t=50$ s, and $t=80$ s, as can be seen in the third plot.  
    Since the last step command is not achievable, $y(t)$ moves further away from the command as $S$ increases, where the command-following error at $t=120$ s is approximately 0.02 for $S=0.1,$ 1.00 for $S=10,$ and 4.81 for $S=1000.$
    Furthermore, note that the overshoot at $t=19$ s violates the output constraint due to the poor initial model and the nonzero initial condition.
    The distance between $y(t)$ and the constraint at $t=19$ s for $S=0.1$ is 41.47, which is reduced by 47\% for $S=10$ and 89\% for $S=1000.$
    (b)
    OFMPCOI obtained from
    \eqref{eq:QPwithslack}--\eqref{eq:slackconstraint} with $\hat{n}=6>n=4,$ $\zeta=0.01,$ $x(0)=[10 \ -\mspace{-5mu}5 \ \ 8 \ \ 3]^\rmT,$ and the three-step command \eqref{eq:commandex4b}.
    Compared to (a), note that, for all $S,$ the output constraint violation is more severe during the first transient.
    However, as the model becomes more accurate, $y(t)$ approaches the first and second step commands, where the constraint is enforced after $t=106$ s.
    For the third step command, as in (a), $y(t)$ moves away from the step command as $S$ increases.
    }
    \label{fig:ex4ab}
\end{figure*}

 \begin{example}\label{ex:ex4}
 \textit{Unstable, continuous-time plant.}
 Consider the SISO, continuous-time plant
 \begin{align}
     G(s) = \dfrac{1}{s^2(s^2+2\zeta s + 1)}. \label{eq:ex4}
 \end{align}
The data are sampled with sample period $T_s = 1$ s.
Let $u_{\mathrm{min}} = -10,$ $u_{\mathrm{max}} = 10,$ $\Delta u_{\mathrm{min}} = -5,$ $\Delta u_{\mathrm{max}} = 5,$ $\lambda=1,$ $P_0=10^3I_{2\hat{n}},$ $\ell=20,$ $\bar{Q}=40I_{\ell-1},$ $\bar{P}=40,$ and $R=10I_\ell.$
OFMPCOI is initialized with the strictly proper FIR model $\theta_0 = [0_{(2\hat{n}-1)\times 1}^\rmT \ 1]^\rmT,$ and uses the noisy measurement $y_{\rmn,k}$ given by \eqref{eq:ynoisy}, where the standard deviation of the noise $v_k$ is $\sigma=0.001.$
We consider the output constraint  \eqref{eq:constraint}, where $\SC = [1 \ -\mspace{-5mu}1]^\rmT$ and $\SD = [-\mspace{-5mu}20 \ -\mspace{-5mu}20]^\rmT.$
A realization of \eqref{eq:ex4} is given by
\begin{align}
    \dot{x} & = \left[
    \begin{matrix}
        -2\zeta  & -1    &     0     &    0 \\
        1    &    0     &    0    &     0 \\
         0  &  1    &     0    &     0 \\
         0   &      0  &  1     &    0
    \end{matrix}
    \right]x + \left[\begin{matrix}1 \\ 0 \\ 0 \\ 0\end{matrix}\right]u, \label{eq:ex4ss1} \\
    y & = \left[\begin{matrix}0 & 0 & 0 & 1\end{matrix}\right] x. \label{eq:ex4ss2}
\end{align}
Denote the $i$th component of $u_k,$ $\Delta u_{\rm{max}},$ $\Delta u_{\rm{min}},$ and $u_{1|k},$ by $u_{k,i},$ $\Delta u_{\rm{max},i}$ $\Delta u_{\rm{min},i},$ and $u_{1|k,i},$ respectively.
In this example, for each $i=1,\ldots,m,$ we apply the saturation
\begin{align}
    u_{k+1,i} = &
    \begin{cases}
        \min(\max(u_{k,i} + \Delta u_{\mathrm{max},i},u_{\mathrm{min},i}),u_{\mathrm{max},i}), & u_{1|k,i}-u_{k,i} > \Delta u_{\mathrm{max},i} \\
         \min(\max(u_{k,i} + \Delta u_{\mathrm{min},i},u_{\mathrm{min},i}),u_{\mathrm{max},i}), & u_{1|k,i}-u_{k,i} < \Delta u_{\mathrm{min},i}, \\
        \min(\max(u_{1|k,i},u_{\mathrm{min},i}),u_{\mathrm{max},i}), & \mbox{otherwise.}
    \end{cases}
\end{align}

Figure \ref{fig:ex4ab} shows the response of OFMPCOI for various values of $S.$
Figure \ref{fig:ex4a} shows the response of OFMPCOI obtained from \eqref{eq:QPnoslack}--\eqref{eq:DeltaUcon} as well as from \eqref{eq:QPwithslack}--\eqref{eq:slackconstraint} with $\hat{n}=n=4,$ $\zeta=0.7,$ $x(0) = [1 \ -\mspace{-5mu}3 \ \ 2 \ \ 0.5]^\rmT,$ and the three-step command
\begin{align}
    r_k & =
    \begin{cases}
        0, & 0 \le k < 50, \\
        19, & 50 \le k < 80, \\
        -25, & k \ge 80.
    \end{cases} \label{eq:commandex4}
\end{align}
Figure \ref{fig:ex4b} shows the response of OFMPCOI obtained from \eqref{eq:QPwithslack}--\eqref{eq:slackconstraint} with $\hat{n}=6>n=4,$ $\zeta=0.01,$ $x(0) = [10 \ -\mspace{-5mu}5 \ \ 8 \ \ 3]^\rmT,$ and the three-step command
\begin{align}
    r_k & =
    \begin{cases}
        0, & 0 \le k < 120, \\
        19, & 120 \le k < 150, \\
        -25, & k \ge 150.
    \end{cases} \label{eq:commandex4b}
\end{align}
\end{example}

 \clearpage
 
 \section{Linear MIMO Examples}
This section considers linear, time-invariant, multi-input multi-output (MIMO) plants with the state-space realization
\begin{align}
    \dot{x} & = Ax + Bu + D_1 w, \label{eq:mimo1} \\
    \tilde{y} & = Cx + Du, \label{eq:mimo2} \\
    \tilde{z} & = E x, \label{eq:mimo3}
\end{align}
where $x(t)\in\mathbb{R}^{l_x}$ is the state, $w(t)\in\mathbb{R}^{l_w}$ is an unmatched disturbance, $\tilde{y}(t)\in\mathbb{R}^{l_{\tilde{y}}},$ $\tilde{z}(t)\in\mathbb{R}^{l_{\tilde{z}}},$ $A\in\mathbb{R}^{l_x\times l_x},$ $B\in\mathbb{R}^{l_x\times m},$ $C\in\mathbb{R}^{l_{\tilde{y}}\times l_x},$ $D\in\mathbb{R}^{l_{\tilde{y}}\times m},$ $D_1\in\mathbb{R}^{l_x\times l_w},$ and $E\in\mathbb{R}^{l_{\tilde{z}}\times l_x}.$
At each time step, OFMPCOI uses the sampled measurement \eqref{eq:ynoisy}, where $y_k=[\tilde{y}_k^\rmT \ \ \tilde{z}_k^\rmT]^\rmT \in \mathbb{R}^{l_{\tilde{y}}+l_{\tilde{z}}},$ and the tracking output \eqref{eq:yt} and the constrained output \eqref{eq:yc} are based on $\tilde{z}_k.$
In other words, both $\tilde y_k$ and $\tilde z_k$ are noisy measurements, and $\tilde z_k$ is used to define the performance metric.

\begin{figure*}[t!]
    \centering
    \begin{subfigure}[t]{0.49\textwidth}
        \centering
    \includegraphics[trim = 20 220 20 40, width=\textwidth]{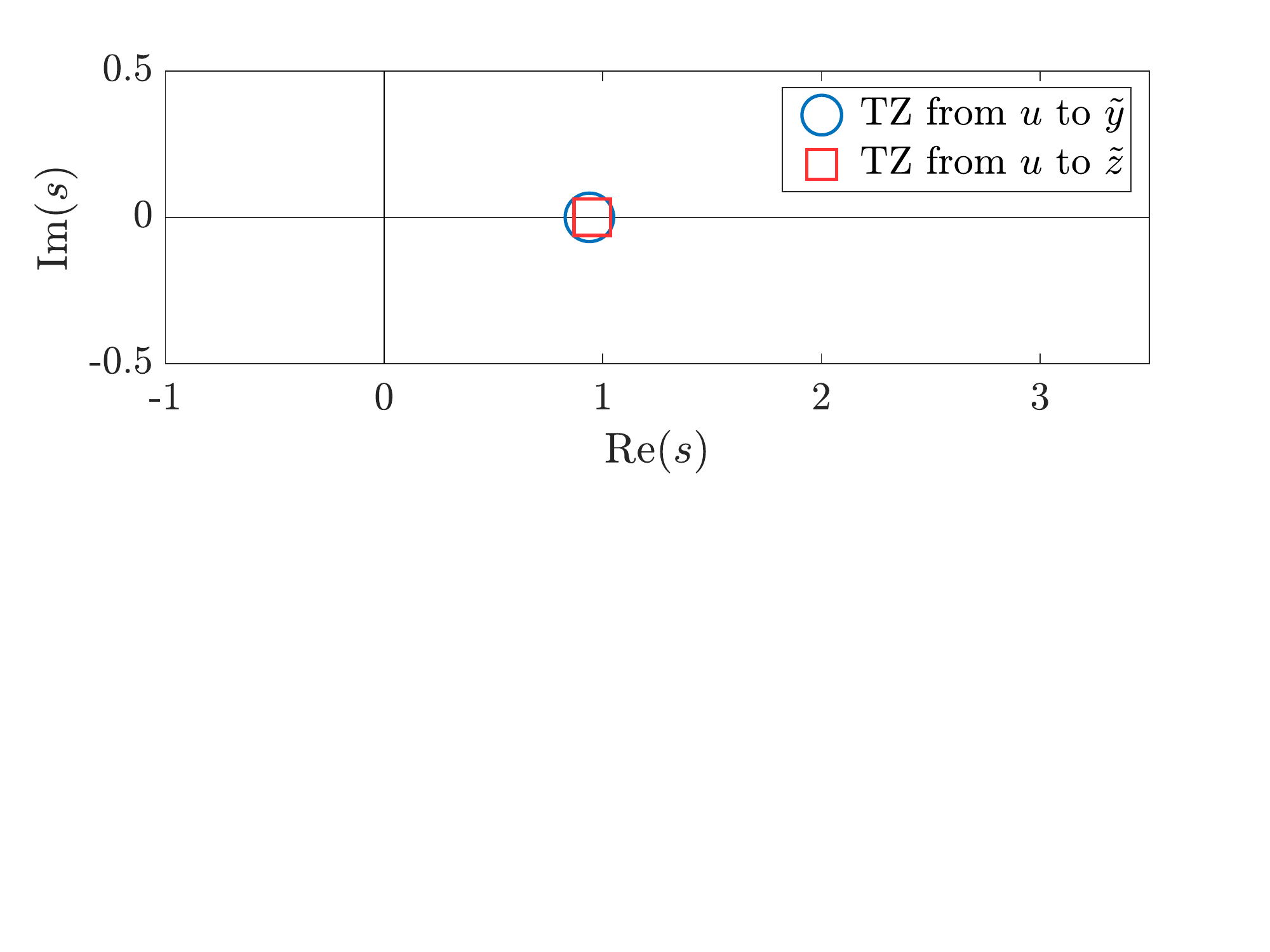}
        \caption{}
        \label{fig:ex5_tzcz_a}
    \end{subfigure}%
    ~ 
    \begin{subfigure}[t]{0.49\textwidth}
        \centering
        \includegraphics[trim = 20 300 20 40, width=\textwidth]{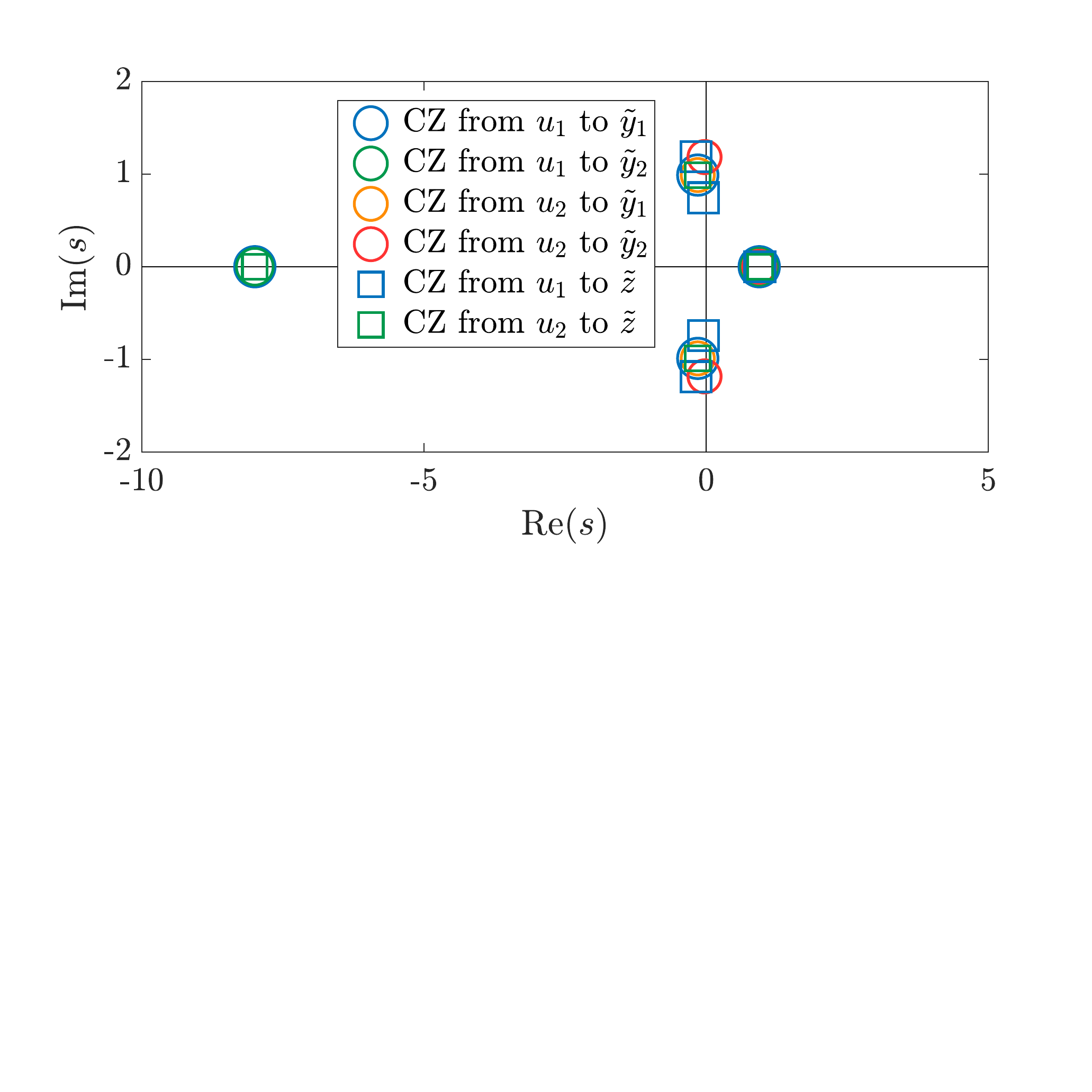}
        \caption{}
        \label{fig:ex5_tzcz_b}
    \end{subfigure}
    \caption{
    Example \ref{ex:ex5}. Transmission zeros (TZ) and channel zeros (CZ) of the MIMO, asymptotically stable, continuous-time plant \eqref{eq:ex5AB}--\eqref{eq:ex5D1E}.
    (a) The transfer function from $u$ to $[\tilde{y}^\rmT \ \tilde{z}^\rmT]^\rmT$ has two NMPTZs approximately at $z_\rmt = 0.94.$
    (b) Each transfer function has one NMPCZ approximately at $z_\rmc = 0.94.$
    }
    \label{fig:ex5_TZCZ}
\end{figure*}

\begin{figure*}[t!]
    \centering
    \includegraphics[trim = 10 80 10 40, width=.8\textwidth]{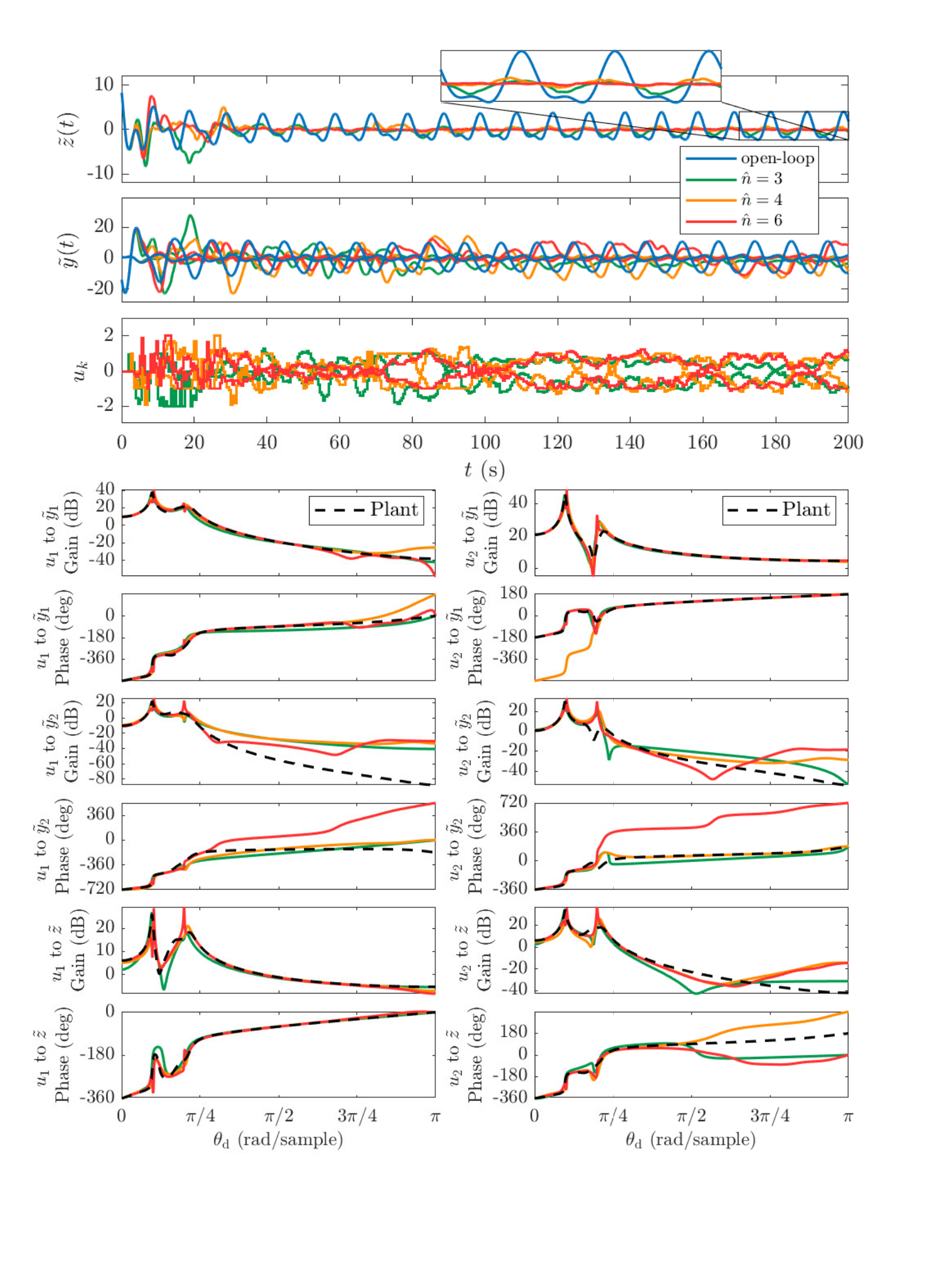}
    \caption{
    Example \ref{ex:ex5}.
    Unmatched harmonic disturbance rejection for the MIMO, asymptotically stable, continuous-time plant \eqref{eq:ex5AB}--\eqref{eq:ex5D1E} with NMPTZ and NMPCZ for $\hat{n}=3,$ $\hat{n}=4,$ and $\hat{n}=6$ using the continuous-time harmonic disturbance \eqref{eq:HDRMIMO}.
    Since RLS identifies the harmonic disturbance more accurately as $\hat{n}$ increases, MPC is able to predict and reject the disturbance more accurately for larger values of $\hat{n}$. 
    Accordingly, the resulting closed-loop performance improves as $\hat n$ increases.
    For all six channels, namely, from $u_1$ and $u_2$ to $\tilde y_1,$ $\tilde y_2,$ and $\tilde z,$ the twelve bottom-most plots compare the frequency response of the identified model at $t=200$ s to the frequency response of the plant.
    This case indicates exigency within closed-loop identification.
    In fact, similarly to Example \ref{ex:ex3}, RLS is able to identify the frequencies and amplitudes of the unmatched harmonic disturbance in order to predict and cancel the disturbance for MPC.
    }
    \label{fig:ex5a}
\end{figure*}

\begin{figure*}[t!]
    \centering
    \includegraphics[trim = 10 75 10 10, width=.8\textwidth]{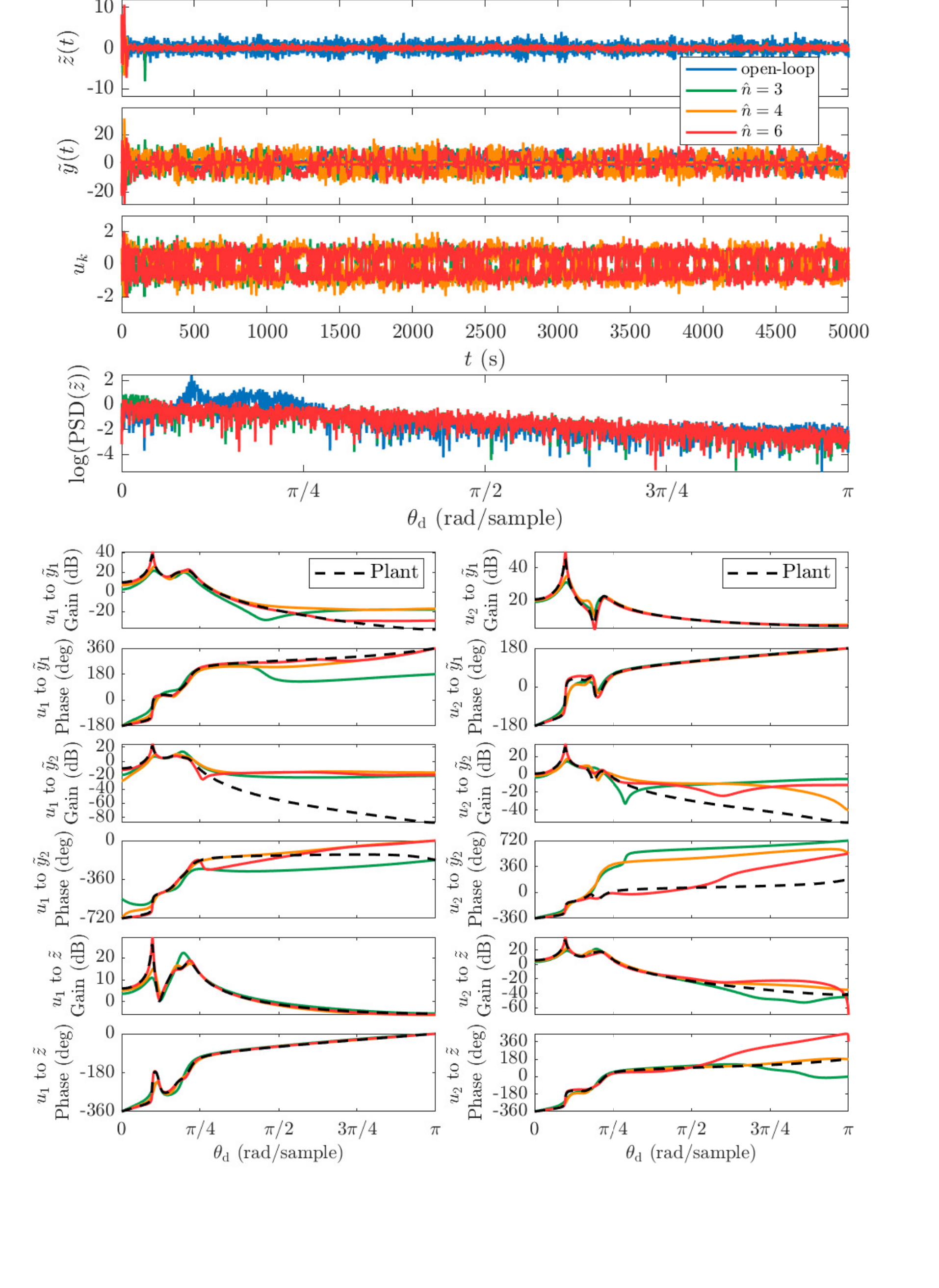}
    \caption{
    Example \ref{ex:ex5}.
    Unmatched broadband disturbance rejection for the MIMO, asymptotically stable, continuous-time plant \eqref{eq:ex5AB}--\eqref{eq:ex5D1E} with NMPTZ and NMPCZ 
    for $\hat{n}=3,$ $\hat{n}=4,$ and $\hat{n}=6$ using the discrete-time broadband disturbance $w_k=[w_{1,k} \ w_{2,k}]^\rmT\in\mathbb{R}^2,$ where $\sigma(w_{1,k})=1.7$ and $\sigma(w_{2,k})=2.5.$
    Note that, for all $\hat{n},$ the magnitude of the open-loop $\mathrm{PSD}(\tilde{z})$ in the range $\pi/16 \le \theta_d \le \pi/4$ is reduced, which indicates suppression of the broadband disturbance $w_k.$
    For all six channels, namely, from $u_1$ and $u_2$ to $\tilde y_1,$ $\tilde y_2,$ and $\tilde z,$ the twelve bottom-most plots compare the frequency response of the identified model at $t=5000$ s to the frequency response of the plant.
    This case indicates exigency within closed-loop identification.
    In fact, similarly to Example \ref{ex:ex3}, as the order of the model increases, RLS is able to more accurately extract the plant dynamics from the broadband disturbance in order to suppress the disturbance for MPC.
    }
    \label{fig:ex5b}
\end{figure*}

\begin{example}\label{ex:ex5}
\textit{Asymptotically stable plant with NMP transmission zeros and NMP channel zeros.} Consider the MIMO, continuous-time plant \eqref{eq:mimo1}--\eqref{eq:mimo3}, where
\begin{align}
    A & = \left[
    \begin{matrix}
         0  &  1   &   0    &     0     &    0      &   0 \\
       -1.4 &  -0.06  &   0.4   &  0.05    &     0    &     0 \\
         0     &    0    &     0  &  1    &    0     &    0 \\
      0.8   &  0.1  &  -1   &  -0.16  &  0.2  &  0.06 \\
         0    &     0     &    0     &    0     &    0  &  1 \\
         0    &     0    &    1   &   0.3  &  -1  &   -0.3
    \end{matrix}
    \right], \quad
    B = \left[
    \begin{matrix}
        0 & 0 \\
        1 & 0 \\
        0 & 0 \\
        0 & 2 \\
        0 & 0 \\
        0 & 0
    \end{matrix}
    \right], \label{eq:ex5AB} \\
    C & = \left[
    \begin{matrix}
         0    &     0   & -3.1  &  3.3    &     0    &     0 \\
         0    &     0   &      0    &     0  &  0.31  &  -0.33
    \end{matrix}
    \right], \quad
    D = 0_{2\times2}, \label{eq:ex5CD} \\
    D_1 & = \left[
    \begin{matrix}
    -0.1 & 0 \\
    0 & 0 \\
    0 & 0 \\
    0 & 0 \\
    0 & 0 \\
    0 & -0.5
    \end{matrix}
    \right], \quad
    E = \left[
    \begin{matrix}
        2  &   -2.1  &    0     &    0     &    0    &     0
    \end{matrix}
    \right]. \label{eq:ex5D1E}
\end{align}
The transmission zeros (TZ) and the channel zeros (CZ) are shown in Figure \ref{fig:ex5_TZCZ}.
The data are sampled with sample period $T_\rms = 0.5$ s.
Let $u_{\mathrm{min}} = [-\mspace{-5mu}2 \ -\mspace{-5mu}1]^\rmT,$ $u_{\mathrm{max}} = [2 \ 1]^\rmT,$ $\Delta u_{\mathrm{min}} = [-\mspace{-5mu}1 \ -\mspace{-5mu}0.5]^\rmT,$ $\Delta u_{\mathrm{max}} = [1 \ 0.5]^\rmT,$ $\lambda=1,$ $P_0=10^3I_{2\hat{n}},$ $\ell=10,$ $\bar{Q}=40I_{\ell-1},$ $\bar{P}=40,$ $R=\left[\begin{smallmatrix} 10 & 0 \\ 0 & 10 \end{smallmatrix}\right] \otimes I_{\ell},$ and $r_k\equiv 0.$
The standard deviation of the measurement noise $v_k$ is $\sigma=0.02.$
The plant is initialized with $x(0) = [1 \ \ -\mspace{-5mu}3 \ \ 5 \ \ 0.5 \ \ 2 \ \ -\mspace{-5mu}1]^\rmT,$ and OFMPCOI uses \eqref{eq:QPnoslack}--\eqref{eq:DeltaUcon} with the strictly proper FIR initial model $\theta_0 = [0_{(\hat{n}(m+p)p-mp)\times 1}^\rmT \ 1_{mp\times 1}]^\rmT.$
No output constraint is considered in this example, and the tracking output $y_{\rmt,k}$ is given by \eqref{eq:yt} with $C_\rmt = [0 \ 0 \ 1],$ that is, $y_{\rmt,k} = \tilde{z}_k.$

Figure \ref{fig:ex5a} shows harmonic disturbance rejection and Figure \ref{fig:ex5b} shows broadband disturbance rejection for $\hat{n}=3,$ $\hat{n}=4,$ and $\hat{n}=6.$
The unmatched, continuous-time, harmonic disturbance is given by
\begin{align}
    w(t) = \left[
    \begin{matrix}
        A_{w_1} \sin(\omega_{w_1}t + \phi_{w_1}) \\
        A_{w_2} \sin(\omega_{w_2}t + \phi_{w_2})
    \end{matrix}
    \right], \label{eq:HDRMIMO}
\end{align}
where $A_{w_1} = 3,$ $A_{w_2} = 2,$ $\omega_{w_1} = \pi/5,$ $\omega_{w_2} = 2\pi/5,$ $\phi_{w_1} = 0,$ and $\phi_{w_2} = \pi/4.$
The unmatched, discrete-time, broadband disturbance is given by the random vector $w_k = [w_{1,k} \ w_{2,k}]^\rmT\in\mathbb{R}^2,$ where the components $w_{1,k}$ and $w_{2,k}$ are Gaussian with the standard deviation $\sigma(w_{1,k})=1.7$ and $\sigma(w_{2,k})=2.5.$

\end{example}

\begin{figure*}[t!]
    \centering
    \begin{subfigure}[t]{0.49\textwidth}
        \centering
    \includegraphics[trim = 20 220 20 40, width=\textwidth]{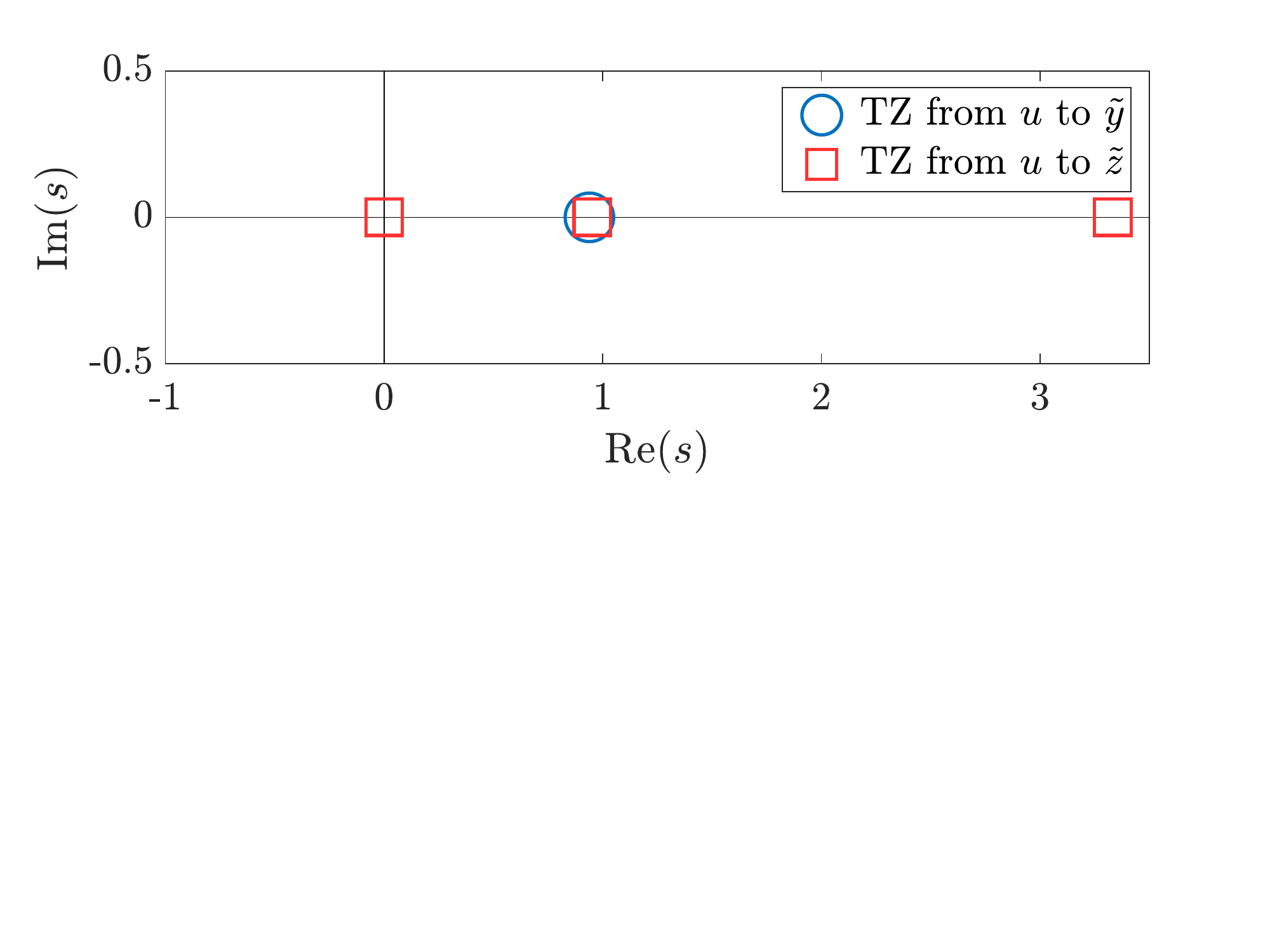}
        \caption{}
        \label{fig:ex6_tzcz_a}
    \end{subfigure}%
    ~ 
    \begin{subfigure}[t]{0.49\textwidth}
        \centering
        \includegraphics[trim = 20 300 20 40, width=\textwidth]{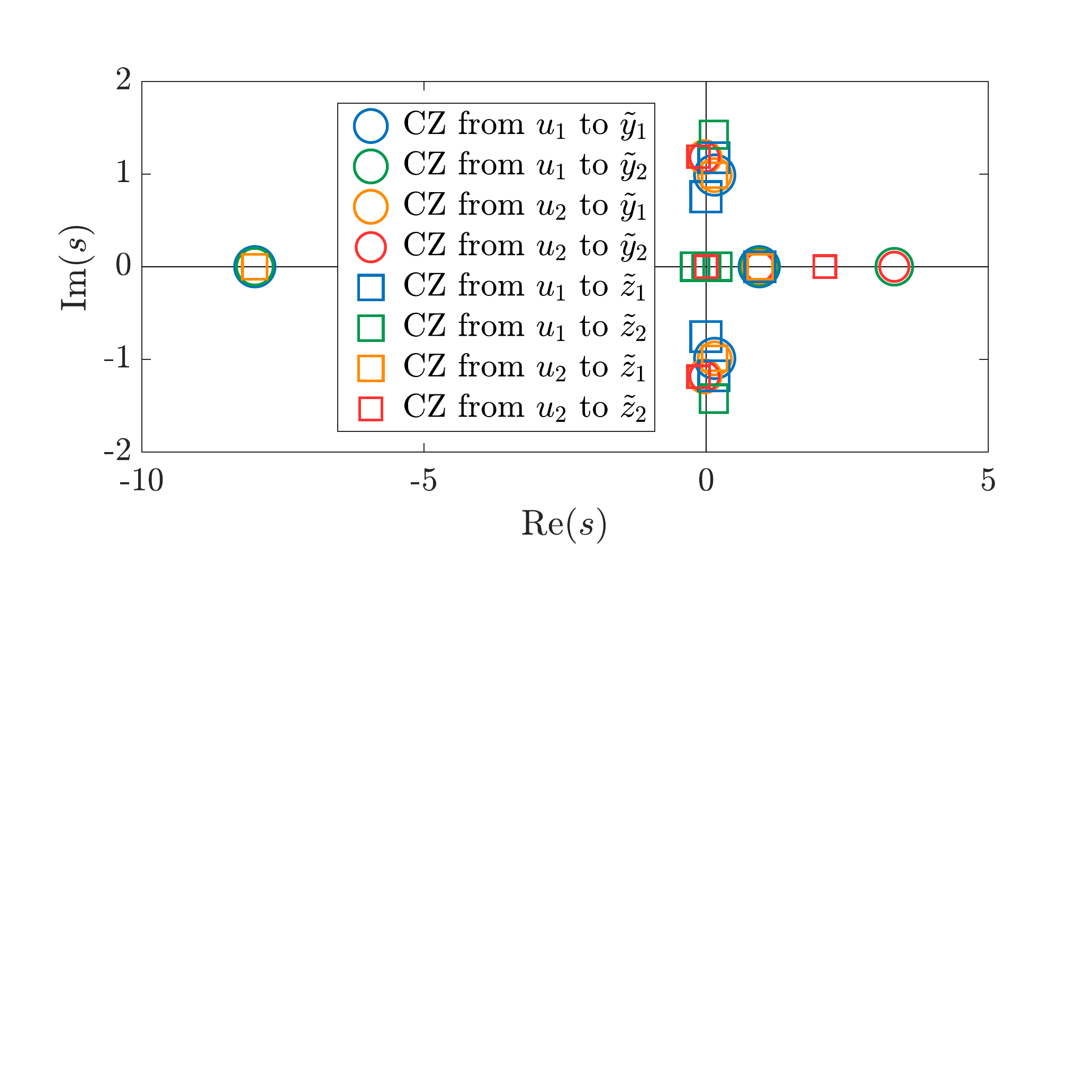}
        \caption{}
        \label{fig:ex6_tzcz_b}
    \end{subfigure}
    \caption{
    Example \ref{ex:ex6}. TZ and CZ of the MIMO, unstable, continuous-time plant \eqref{eq:ex6AE1}.
    (a) The transfer function from $u$ to $[\tilde{y} \ \tilde{z}^\rmT]^\rmT$ has four NMPTZs approximately at $z_{\rmt,1} = 0.00,$ $z_{\rmt,2} = 0.94,$ and $z_{\rmt,3} = 3.33.$
    (b) The eight transfer functions have 21 NMPCZs approximately at $z_{\rmc,1} = 0.00,$ $z_{\rmc,2} = 0.13 \pm 1.42 \jmath,$ $z_{\rmc,3} = 0.14\pm 1.18\jmath,$ $z_{\rmc,4}=0.15\pm0.99 \jmath,$ $z_{\rmc,5}=0.19,$ $z_{\rmc,6} = 0.95,$ $z_{\rmc,7}=2.10,$ and $z_{\rmc,8}=3.33.$
    }
    \label{fig:ex6_TZCZ}
\end{figure*}

\begin{figure*}[t!]
    \centering
    \includegraphics[trim = 10 20 10 10, width=.75\textwidth]{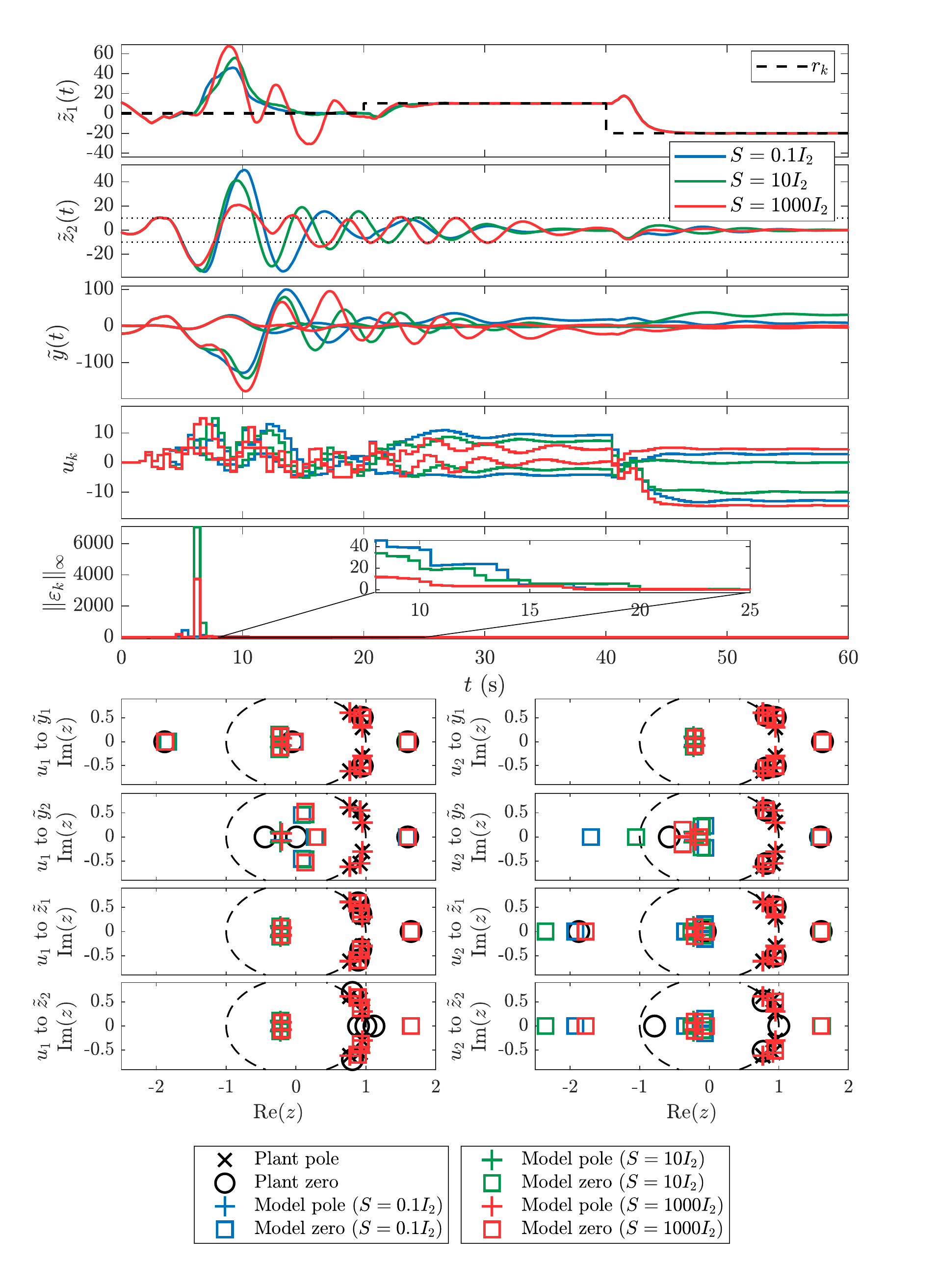}
    \caption{
    Example \ref{ex:ex6}.
    Command following for the MIMO, unstable, continuous-time plant \eqref{eq:ex6AE1} with NMPTZ and NMPCZ for $\hat{n}=2$ and various values of $S$ using the three-step command \eqref{eq:commandex6}.
    For each $S,$ the tracking output approaches the three-step command.
    Note that, as $S$ increases, the output-constraint violation becomes less severe.
    Due to the poor initial model, the output constraint is relaxed by the slack $\varepsilon_k$ until $t=20$ s.
    For all $t\ge20$ s, note that, for each $S,$ as the model becomes more accurate and $\tilde{z}_1(t)$ approaches zero, 
    the output constraint is enforced. 
    The eight bottom-most plots compare the poles and zeros of the identified model at $t=60$ s to the poles and zeros of the plant.
    %
    %
    %
    }
    \label{fig:ex6}
\end{figure*}

\clearpage
\begin{example}\label{ex:ex6}
\textit{Unstable plant with NMP transmission zeros and NMP channel zeros.}
Consider the MIMO, continuous-time plant \eqref{eq:mimo1}--\eqref{eq:mimo3}, where $B$ is given by \eqref{eq:ex5AB}, $C$ and $D$ are given by \eqref{eq:ex5CD}, $D_1=0_{6\times 2},$ and
\begin{align}
    A & =
    \left[
    \begin{matrix}
        0  &  1    &     0    &     0    &     0     &    0 \\
   -1.4 &  -0.06  &  0.4  &  0.05   &      0     &    0 \\
         0    &     0    &     0  &  1   &      0     &    0 \\
    0.8  &  0.1  & -1 &  -0.04  &  0.2 &  -0.06 \\
         0    &     0     &    0     &    0     &    0  &  1 \\
         0    &     0  &  1  & -0.3 &  -1  &  0.3
    \end{matrix}
    \right], \quad
    E = \left[
    \begin{matrix}
         2  & -2.1    &     0    &     0    &     0     &    0 \\
         0  &  1    &     0     &    0     &    0  &  -1.1
    \end{matrix}
    \right]. \label{eq:ex6AE1}
\end{align}
The TZ and the CZ of \eqref{eq:ex6AE1} are shown in Figure \ref{fig:ex6_TZCZ}.
The data are sampled with sample period $T_\rms = 0.5$ s.
Let $u_{\mathrm{min}} = [-\mspace{-5mu}15 \ \ -\mspace{-5mu}5]^\rmT,$ $u_{\mathrm{max}} = [15 \ \ 5]^\rmT,$ $\Delta u_{\mathrm{min}} = [-\mspace{-5mu}5 \ -\mspace{-5mu}2]^\rmT,$ $\Delta u_{\mathrm{max}} = [5 \ \ 2]^\rmT,$ $\lambda=1,$ $P_0=10^3I_{2\hat{n}},$ $\ell=40,$ $\bar{Q}=40I_{\ell-1},$ $\bar{P}=40,$ and $R=\left[\begin{smallmatrix} 10 & 0 \\ 0 & 10 \end{smallmatrix}\right] \otimes I_{\ell}.$
The standard deviation of the measurement noise $v_k$ is $\sigma=0.02.$
The plant is initialized with $x(0) = [2 \ \ -\mspace{-5mu}3.3 \ \ 4 \ \ -\mspace{-5mu}2.6 \ \ 2 \ \ -\mspace{-5mu}1.3]^\rmT,$ and OFMPCOI uses \eqref{eq:QPwithslack}--\eqref{eq:slackconstraint} with the strictly proper FIR initial model $\theta_0 = [0_{(\hat{n}(m+p)p-mp)\times 1}^\rmT \ 1_{mp\times 1}]^\rmT.$
The tracking output $y_{\rmt,k}$ is given by \eqref{eq:yt} with $C_\rmt = [1 \ 0],$ and the constrained output is given by \eqref{eq:yc} with $C_c = [0 \ 1],$ that is,  $y_{\rmt,k} = \tilde{z}_{1,k}$ and $y_{\rmc} = \tilde{z}_{2,k}.$
The output constraint is given by \eqref{eq:constraint}, where $\SC = [1 \ -\mspace{-5mu}1]^\rmT$ and $\SD = [-\mspace{-5mu}10 \ -\mspace{-5mu}10]^\rmT.$

Figure \ref{fig:ex6} shows the response of OFMPCOI for $\hat{n}=2$ and various values of $S$ using the three-step command
\begin{align}\label{eq:commandex6}
    r_k & =
    \begin{cases}
        0, & 0 \le k < 40, \\
        10, & 40 \le k < 80, \\
        -20, & k \ge 80.
    \end{cases}
\end{align}
\end{example}

\clearpage

\section{Linear Time-Varying Examples}

This section considers linear, time-varying, SISO plants, where $y_{\rmt,k} = y_{\rmc,k} = y_k.$

\begin{figure*}[t!]
    \centering
    \begin{subfigure}[t]{0.49\textwidth}
        \centering
        \includegraphics[trim = 10 25 10 40, width=\textwidth]{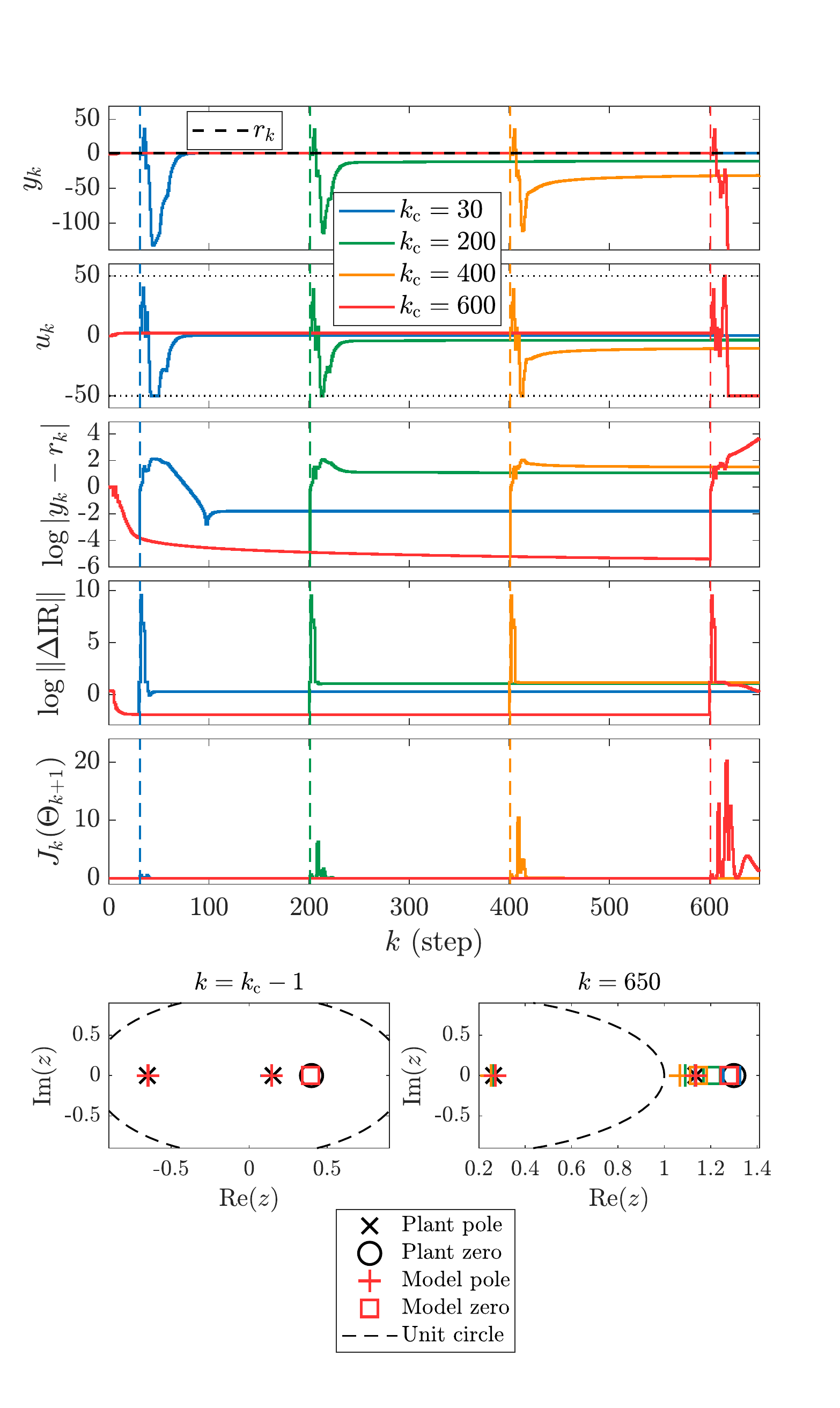}
        \caption{}
        \label{fig:ex1LTVa}
    \end{subfigure}%
    ~ 
    \begin{subfigure}[t]{0.49\textwidth}
        \centering
        \includegraphics[trim = 10 10 10 40, width=\textwidth]{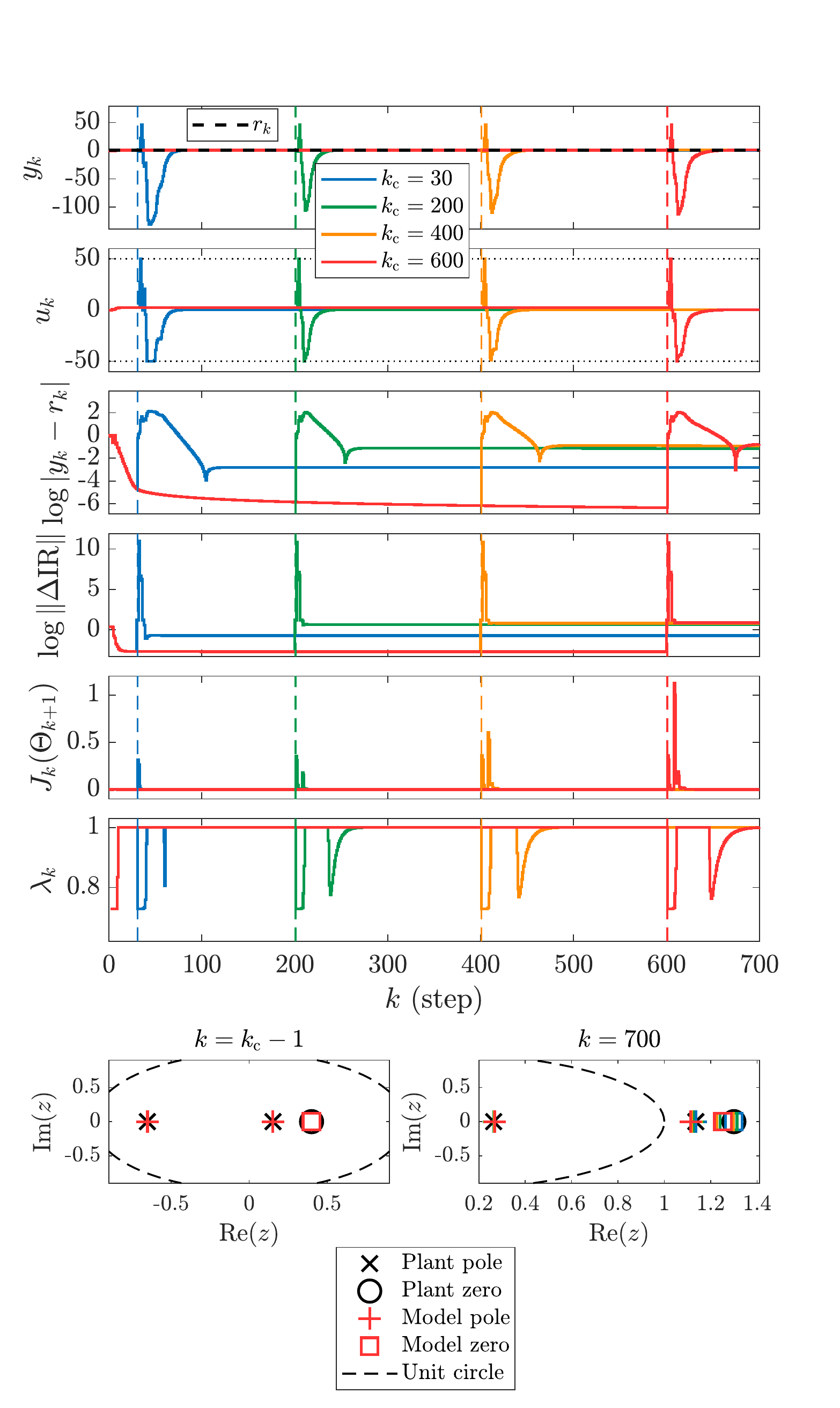}
        \caption{}
        \label{fig:ex1LTVb}
    \end{subfigure}
    \caption{Example \ref{ex:ex1LTV}.
    Command following with abruptly changing dynamics occurring at step $k_\rmc$ represented by the dashed vertical line for $r_k\equiv 1$ and various values of $k_\rmc.$
    At step $k_\rmc,$ the SISO, asymptotically stable, discrete-time plant \eqref{eq:ex1} changes to the unstable, NMP, discrete-time plant \eqref{eq:ex2}.
    (a) CRF with $\lambda=1.$
    Note that, as $k_\rmc$ increases, $y_k$ moves further away from the command after the abrupt change.
    In particular, for $k_\rmc = 600,$ the slow reidentification after the abrupt change causes the control input to saturate and the output error to diverge.
    Note that the maximum value of the RLS cost $J_k(\theta_{k+1})$ \eqref{eq:J} increases as $k_\rmc$ increases.
    This is due to the fact that $\lambda=1,$ and thus the RLS cost accumulates all instantaneous costs from the beginning of the simulation, making it more difficult for RLS to reidentify the model coefficients.
    The two bottom-most plots compare the poles and zeros of the identified model with the poles and zeros of the plant at $k=k_\rmc-1$ and $k=650.$
    (b) VRF \eqref{eq:vrf} with $\eta=0.9,$ $\tau_\rmn=5,$ and $\tau_\rmd=10.$
    In contrast with (a), for each $k_\rmc,$ $y_k$ approaches the command after the abrupt change.
    The sixth plot shows that the variable-rate forgetting factor $\lambda_k$ is activated at each abrupt change.
    }
    \label{fig:ex1LTV}
\end{figure*}

\begin{example}\label{ex:ex1LTV}
\textit{Abrupt change from an asymptotically stable plant to an unstable, NMP plant.}
Consider the SISO, asymptotically stable, discrete-time plant \eqref{eq:ex1}, which abruptly changes at step $k_\rmc$ to the SISO, unstable, NMP, discrete-time plant \eqref{eq:ex2}.
Note that the order of both plants are $n=2$ with relative degree 1.

Let $u_{\mathrm{min}} = -50,$ $u_{\mathrm{max}} = 50,$ $\Delta u_{\mathrm{min}} = -25,$ $\Delta u_{\mathrm{max}} = 25,$ $\ell=20,$ $\bar{Q}=2I_{\ell-1},$ $\bar{P}=5,$ $R=I_\ell,$ $\hat{n}=n=2,$ and $P_0=10^3I_{2\hat{n}}.$
No output constraint is considered in this example.
The plant is initialized with $y_{-1} = -0.2,$ $y_{-2} = 0.4,$ and $u_{-1}=u_{-2}=0.$
OFMPCOI uses the strictly proper FIR initial model $\theta_0 = [0_{1\times 3} \ 1]^\rmT,$ and uses \eqref{eq:QPnoslack}--\eqref{eq:DeltaUcon} with the measurement $y_{k}.$

Figure \ref{fig:ex1LTV} shows the response of OFMPCOI for $r_k\equiv1$ using CRF with $\lambda=1$ and VRF \eqref{eq:vrf} with $\eta=0.9,$ $\tau_\rmn=5,$ and $\tau_\rmd=10$ for $k_\rmc = 30,$ $k_\rmc = 200,$ $k_\rmc = 400,$ and $k_\rmc = 600.$
It can be seen in Figure \ref{fig:ex1LTVa} that, due to the fact that $\lambda=1,$ the RLS cost accumulates all instantaneous costs from the beginning of the simulation, making it more difficult for RLS to reidentify the model coefficients after the abrupt change.
In contrast, in Figure \ref{fig:ex1LTVb}, the effect of the abrupt change is independent of the time at which the change occurs, as can be seen by the reconvergence of $y_k$ to the command after each abrupt plant change.
\end{example}

\clearpage

\begin{figure*}[t!]
    \centering
    \begin{subfigure}[t]{0.49\textwidth}
        \centering
        \includegraphics[trim = 10 10 10 40, width=\textwidth]{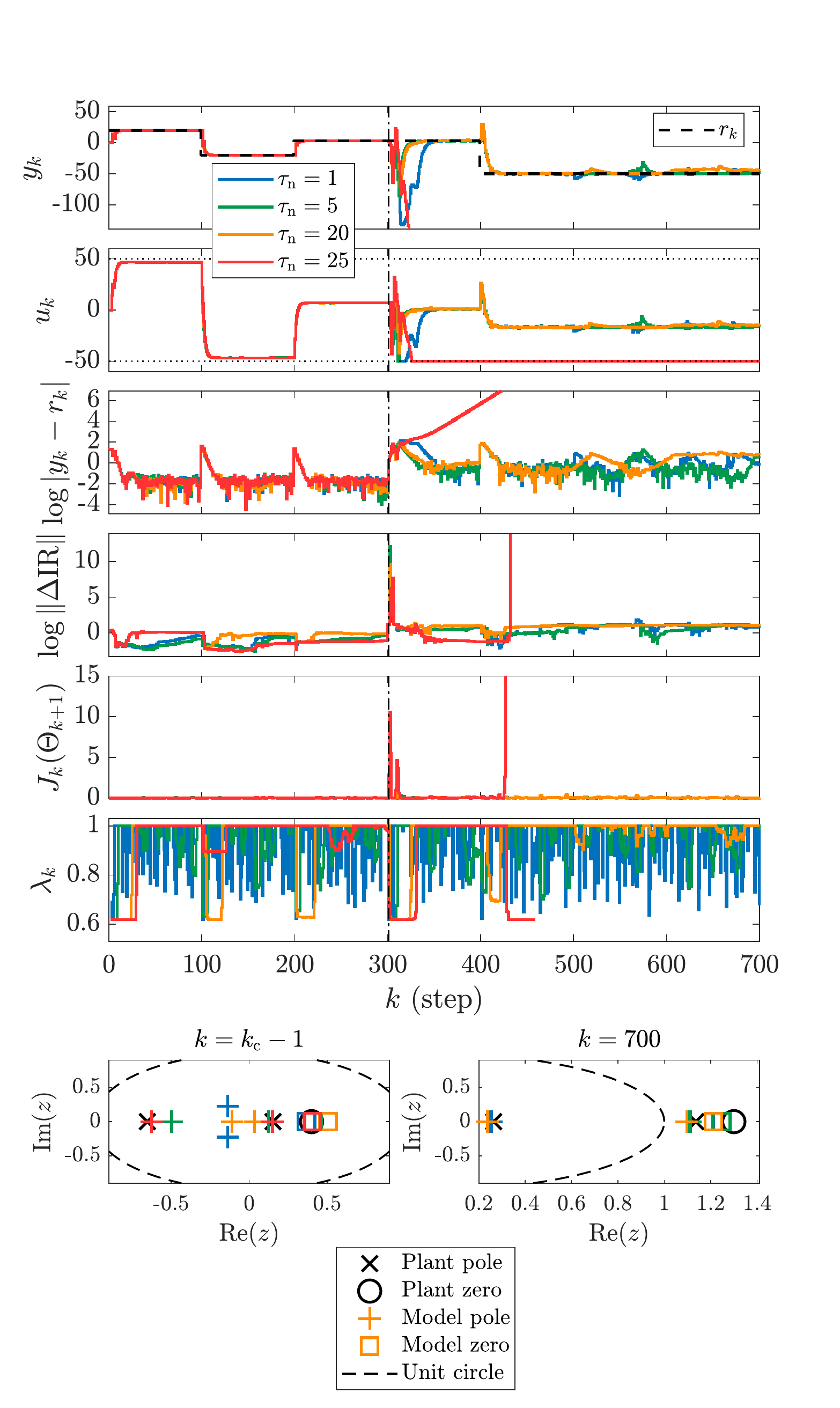}
        \caption{}
        \label{fig:ex2LTVa}
    \end{subfigure}%
    ~ 
    \begin{subfigure}[t]{0.49\textwidth}
        \centering
        \includegraphics[trim = 10 10 10 40, width=\textwidth]{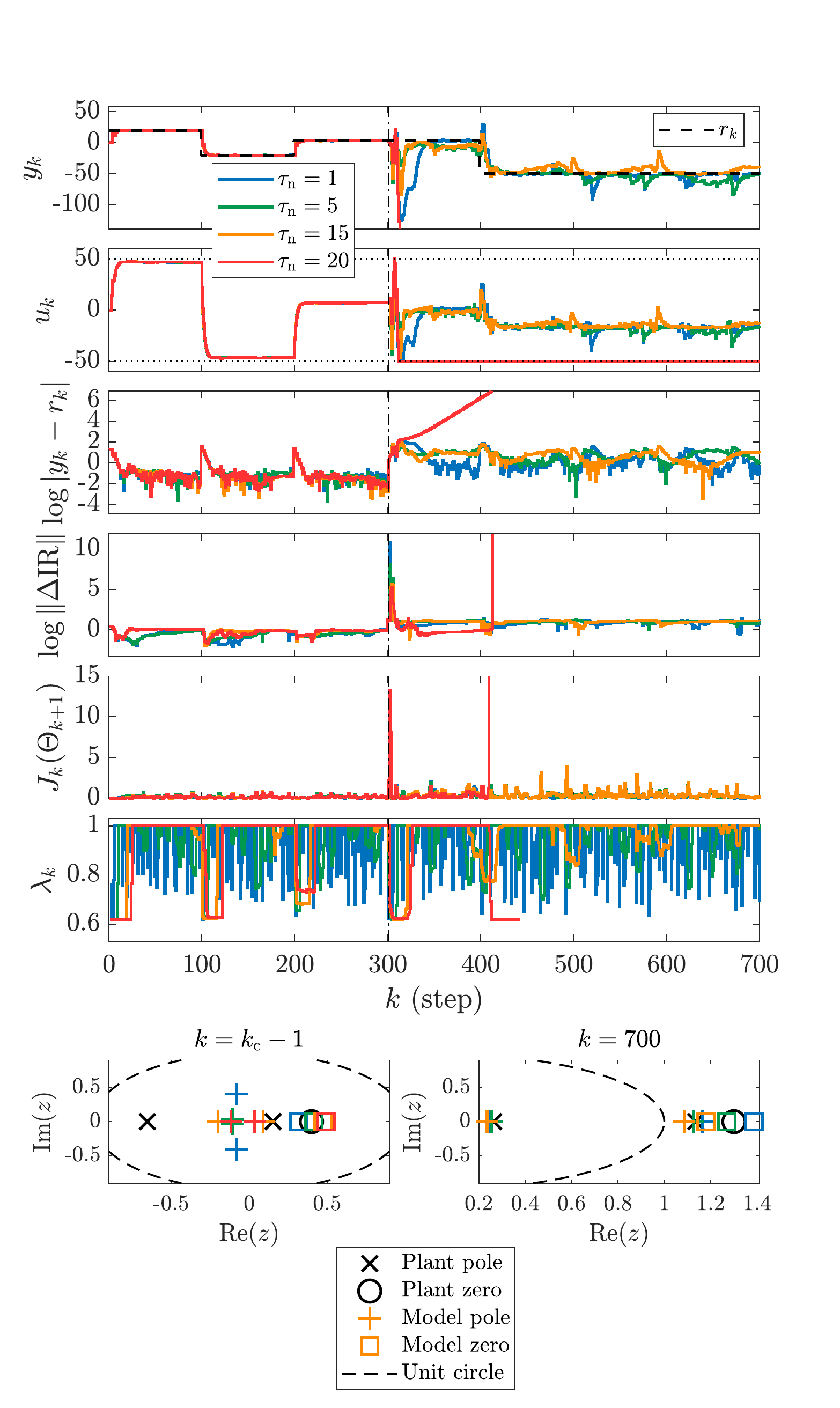}
        \caption{}
        \label{fig:ex2LTVb}
    \end{subfigure}
    \caption{Example \ref{ex:ex2LTV}.
    Command following with dynamics that abruptly change from \eqref{eq:ex1} to \eqref{eq:ex2} at step $k_\rmc=300$ with measurement noise.
    OFMPCOI uses the command profile \eqref{eq:commandex2LTV} and VRF \eqref{eq:vrf} with $\eta=0.5,$ $\tau_\rmd = 5\tau_\rmn,$ and various values of $\tau_\rmn.$
    (a) The standard deviation of the measurement noise is $\sigma = 0.1.$
    Note that, before the abrupt change and for all values of $\tau_\rmn,$ $y_k$ approaches each step command.
    After the abrupt change, however, $y_k$ approaches the command for $\tau_\rmn=1,$ $\tau_\rmn=5,$ and $\tau_\rmn=20,$ and diverges for $\tau_\rmn=25.$
    Furthermore, for $\tau_\rmn=25,$ the impulse-response error and RLS cost diverge approximately at $k=430.$
    Note that, as $\tau_\rmn$ increases, the variable-rate forgetting factor $\lambda_k$ becomes less sensitive to sensor noise and is activated only at the abrupt change and at each step-command change.
    (b) The standard deviation of the measurement noise is  $\sigma = 0.3.$
    Note that, compared to (a), since the measurement-noise level is larger, the value of $\tau_\rmn$ for which the output,  impulse-response error, and  RLS cost diverge is lower, namely, $\tau_\rmn=20.$
    }
    \label{fig:ex2LTV}
\end{figure*}

\begin{example}\label{ex:ex2LTV}
\textit{Abrupt change from an asymptotically stable plant to an unstable, NMP plant with measurement noise.}
This example uses the same setup as Example \ref{ex:ex1LTV} except that OFMPCOI now uses the noisy measurement $y_{\rmn,k}$ \eqref{eq:ynoisy} instead of $y_k$ with the command profile
\begin{align}\label{eq:commandex2LTV}
    r_k & = 
    \begin{cases}
        20, & 0 \le k < 100, \\
        -20, & 100 \le k < 200, \\
        3, & 200 \le k < 400, \\
        -50, & k \ge 400.
    \end{cases}
\end{align}
Figure \ref{fig:ex2LTV} shows the response of OFMPCOI for two different levels of sensor noise using $k_\rmc=300$ and VRF \eqref{eq:vrf} with $\eta=0.5,$ $\tau_\rmd = 5\tau_\rmn,$ and various values of $\tau_\rmn.$
As shown in Figure \ref{fig:ex2LTVa} and Figure \ref{fig:ex2LTVb}, as $\tau_\rmn$ increases, the time given to forgetting increases.
Hence, as $\tau_\rmn$ increases, reidentification becomes slower and, for large values of $\tau_\rmn,$ OFMPCOI is not able to return $y_k$ to the command after the abrupt change.
However, note that, as $\tau_\rmn$ increases, the variable-rate forgetting $\lambda_k$ becomes less sensitive to sensor noise.
This example emphasizes the importance of consistency and speed of convergence of the coefficients after the abrupt change.
A too slow reconvergence of the coefficients after the abrupt change causes a longer bias in time in the coefficient estimates, and, hence, a longer lack of consistency in time, which in turn causes divergence of the output.

\end{example}

\clearpage

\begin{figure*}[t!]
    \centering
    \begin{subfigure}[t]{0.49\textwidth}
        \centering
        \includegraphics[trim = 10 100 10 40, width=\textwidth]{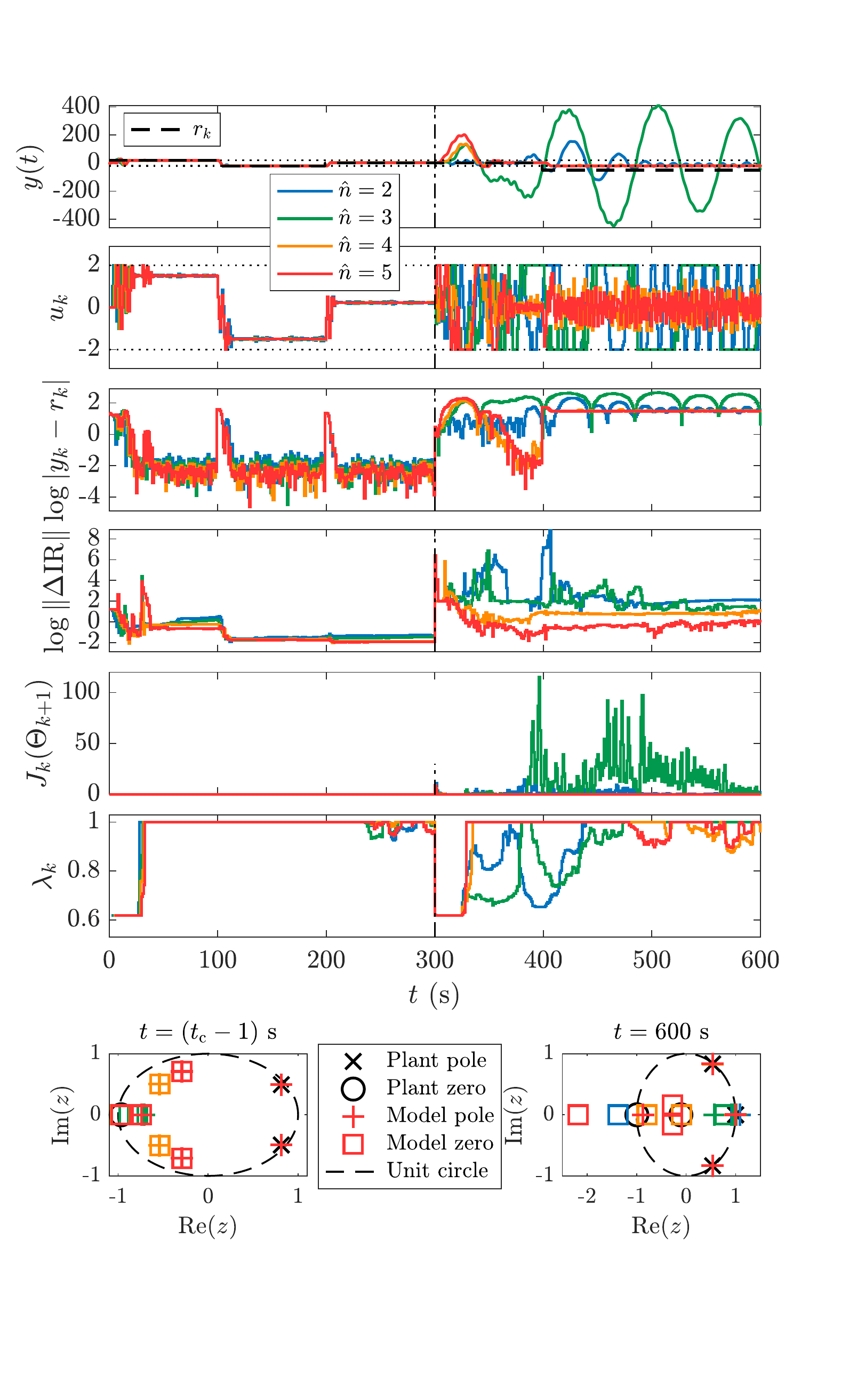}
        \caption{}
        \label{fig:ex3LTVa}
    \end{subfigure}%
    ~ 
    \begin{subfigure}[t]{0.49\textwidth}
        \centering
        \includegraphics[trim = 10 100 10 40, width=\textwidth]{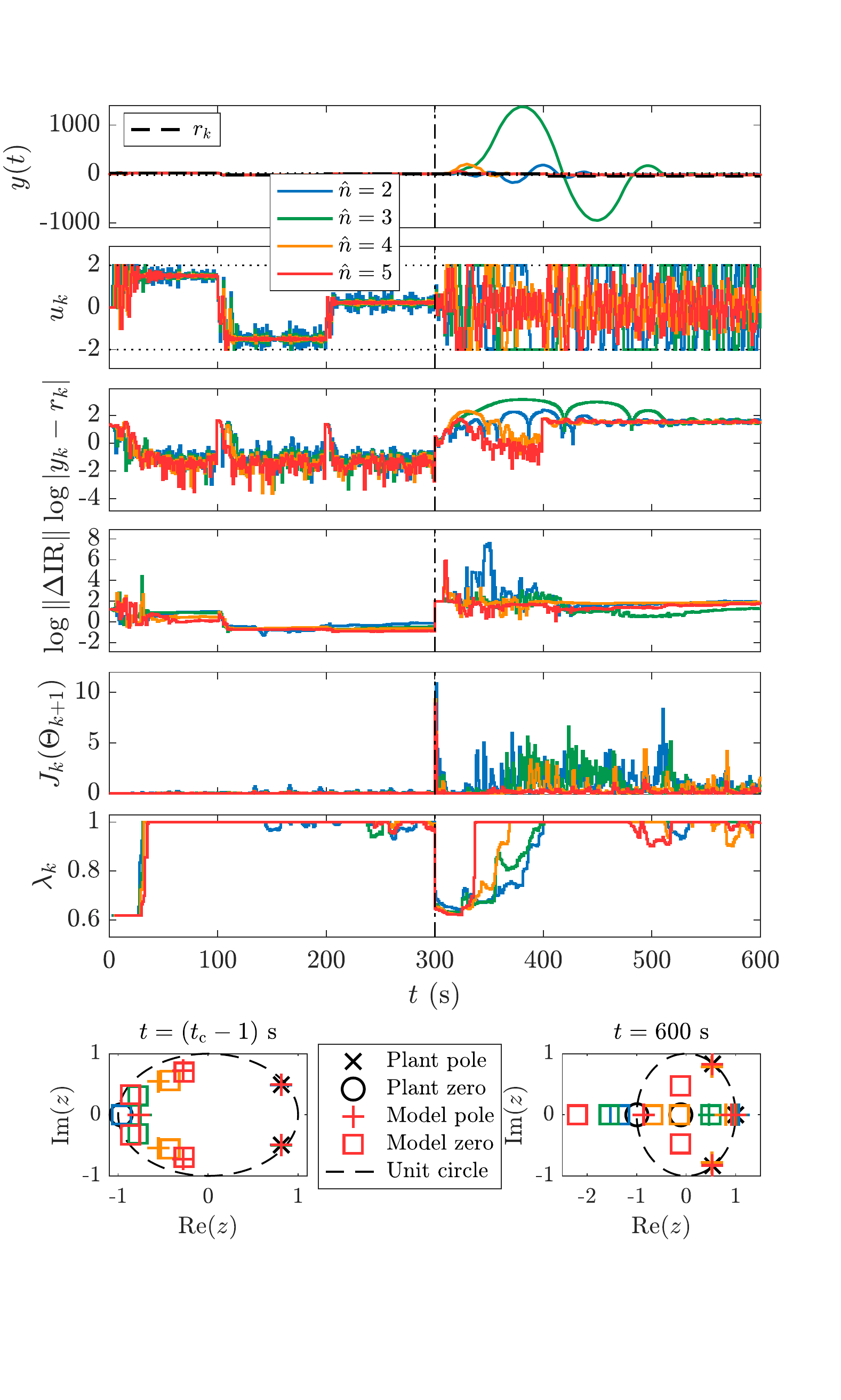}
        \caption{}
        \label{fig:ex3LTVb}
    \end{subfigure}
    \caption{Example \ref{ex:ex3LTV}.
    Command following with dynamics that abruptly change from \eqref{eq:ex3LTVAS} to \eqref{eq:ex4} at time $t_\rmc=300$ s with measurement noise and an output constraint for various values of $\hat{n}.$
    OFMPCOI uses the command profile \eqref{eq:commandex2LTV} and VRF \eqref{eq:vrf} with $\eta=0.5,$ $\tau_\rmn=25,$ and $\tau_\rmd = 125.$
    (a) The standard deviation of the measurement noise is $\sigma = 0.01.$
    Note that, before the abrupt change occurs and for each value of $\hat{n},$ $y(t)$ approaches the step commands and the output constraint is satisfied at all times.
    After the abrupt change occurs, however, the double integrator and lightly damped poles are poorly identified by RLS in the underparameterized cases $\hat{n}<n=4,$ causing $y(t)$ to oscillate and violate the constraint until the end of the simulation.
    In contrast, after the abrupt change occurs, $y(t)$ temporarily violates the constraint during reidentification in the exact and overparameterized cases $\hat{n}\ge n=4.$ 
    Additionally, after $t=338$ s, $y(t)$ is more damped than the underparameterized cases, and the output constraint is enforced at all times.
    Since the last step command is not achievable, OFMPCOI moves $y(t)$ away from the command and toward the constraint.
    (b) The standard deviation of the measurement noise is  $\sigma = 0.1.$
    Note that, due to the higher sensor-noise level, the oscillation amplitude and the output-constraint violation for $\hat{n}<n=4$ are more pronounced in (b) than in (a).
    Furthermore, note that, in both (a) and (b), for $\hat{n}=2,$ RLS is able to capture the two integrators of the plant, whereas, for $\hat{n}=3,$ the three poles of the model are unable to capture the two integrators and the two undamped poles of the plant, causing the identification and the transient reponse to be poorer.
    }
    \label{fig:ex3LTV}
\end{figure*}

\begin{example}\label{ex:ex3LTV}
\textit{Abrupt change from an asymptotically stable plant to an unstable plant with measurement noise and output constraint.}
Consider the SISO, asymptotically stable, continuous-time plant
\begin{align}
    G(s) = \dfrac{4}{s^2+0.1s+0.3}, \label{eq:ex3LTVAS}
\end{align}
which abruptly changes at time $t_\rmc=300$ s to the SISO, unstable, continuous-time plant \eqref{eq:ex4} with $\zeta=0.01.$
Note that the order of \eqref{eq:ex4} is $n = 2$ and the order of \eqref{eq:ex3LTVAS} is $n=4.$
A realization of \eqref{eq:ex3LTVAS} is given by
\begin{align}
    \dot{x} & = 
    \left[
    \begin{matrix}
        -0.1 & -0.6 \\
        0.5 & 0
    \end{matrix}
    \right]x + 
    \left[
    \begin{matrix}
        4 \\
        0
    \end{matrix}
    \right]u, \label{eq:ex3LTVss1} \\
    y & =
    \left[
    \begin{matrix}
        0 & 2
    \end{matrix}
    \right]x, \label{eq:ex3LTVss2}
\end{align}
and a realization of \eqref{eq:ex4} is given by \eqref{eq:ex4ss1}, \eqref{eq:ex4ss2}.

The data are sampled with sample period $T_\rms = 1$ s.
Let $u_{\mathrm{min}} = -2,$ $u_{\mathrm{max}} = 2,$ $\Delta u_{\mathrm{min}} = -1,$ $\Delta u_{\mathrm{max}} = 1,$ $\ell=20,$ $\bar{Q}=10I_{\ell-1},$ $\bar{P}=20,$ $R=I_\ell,$ $P_0=10^3I_{2\hat{n}},$ and $\theta_0 = [0_{1\times (2\hat{n}-1)} \ 1]^\rmT.$
The initial plant \eqref{eq:ex3LTVAS} is initialized with $x(0) = [2.5 \ -\mspace{-5mu}1.4]^\rmT$, and the modified plant \eqref{eq:ex4} is initialized with $x(0) = [-0.039 \ 1.508 \ 0 \ 0]^\rmT.$
The output constraint is given by \eqref{eq:constraint}, where $\SC = [1 \ -\mspace{-5mu}1]^\rmT$ and $\SD = [-20 \ -\mspace{-5mu}20]^\rmT.$
The noisy measurements $y_{\rmn,k}$ are given by \eqref{eq:ynoisy}.
For this example, OFMPCOI uses \eqref{eq:QPwithslack}--\eqref{eq:slackconstraint} with $S=10^6 I_2.$  
Figure \ref{fig:ex3LTV} shows the response of OFMPCOI for two different levels of sensor noise and various values of $\hat{n}$ using the command profile \eqref{eq:commandex2LTV} and VRF \eqref{eq:vrf} with $\eta=0.5,$ $\tau_\rmn = 25,$ and $\tau_\rmd = 125.$
\end{example}

\clearpage

\begin{figure*}[t!]
    \centering
    \begin{subfigure}[t]{0.49\textwidth}
        \centering
        \includegraphics[trim = 10 10 10 40, width=\textwidth]{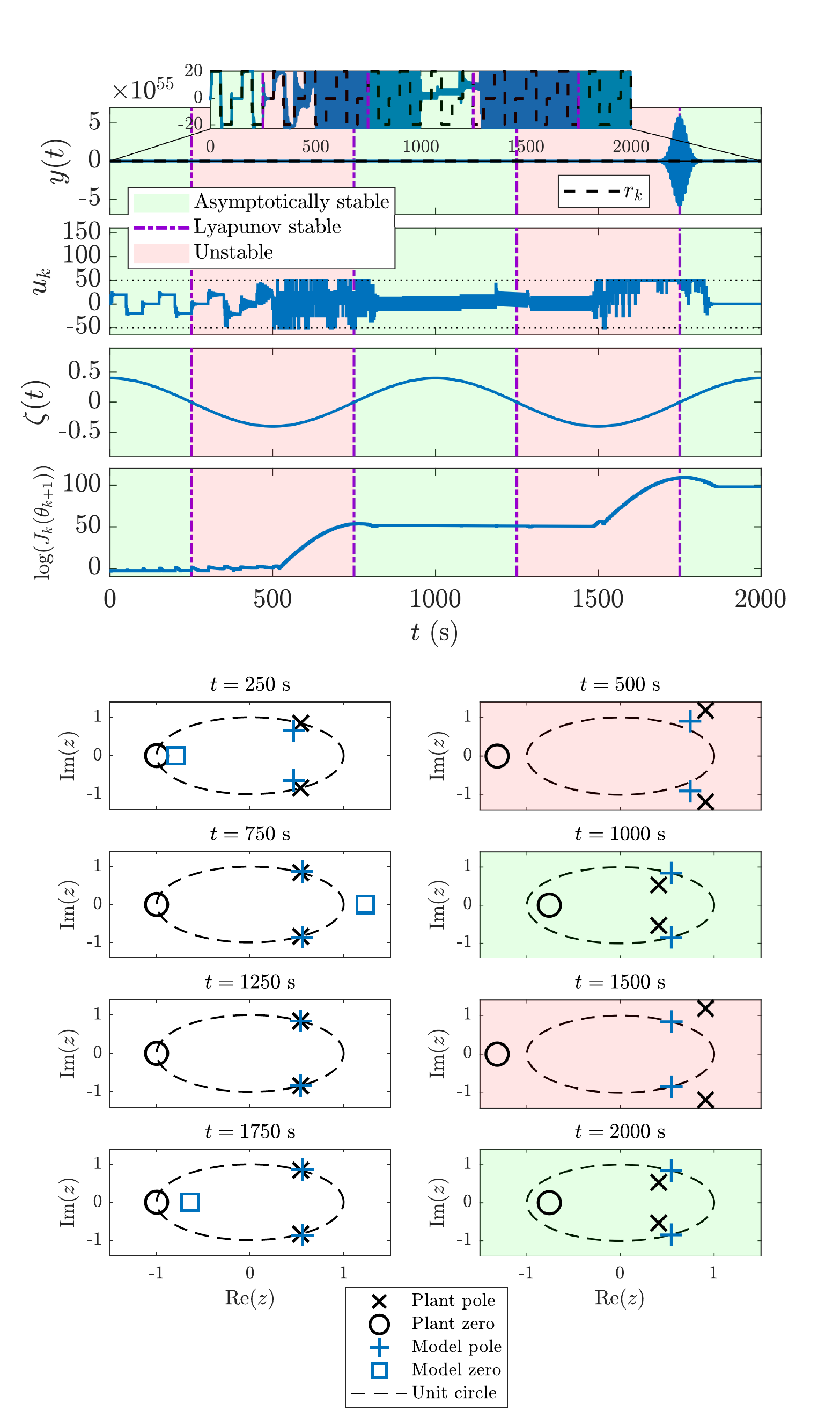}
        \caption{}
        \label{fig:ex1smootha}
    \end{subfigure}%
    ~ 
    \begin{subfigure}[t]{0.49\textwidth}
        \centering
        \includegraphics[trim = 10 10 10 40, width=\textwidth]{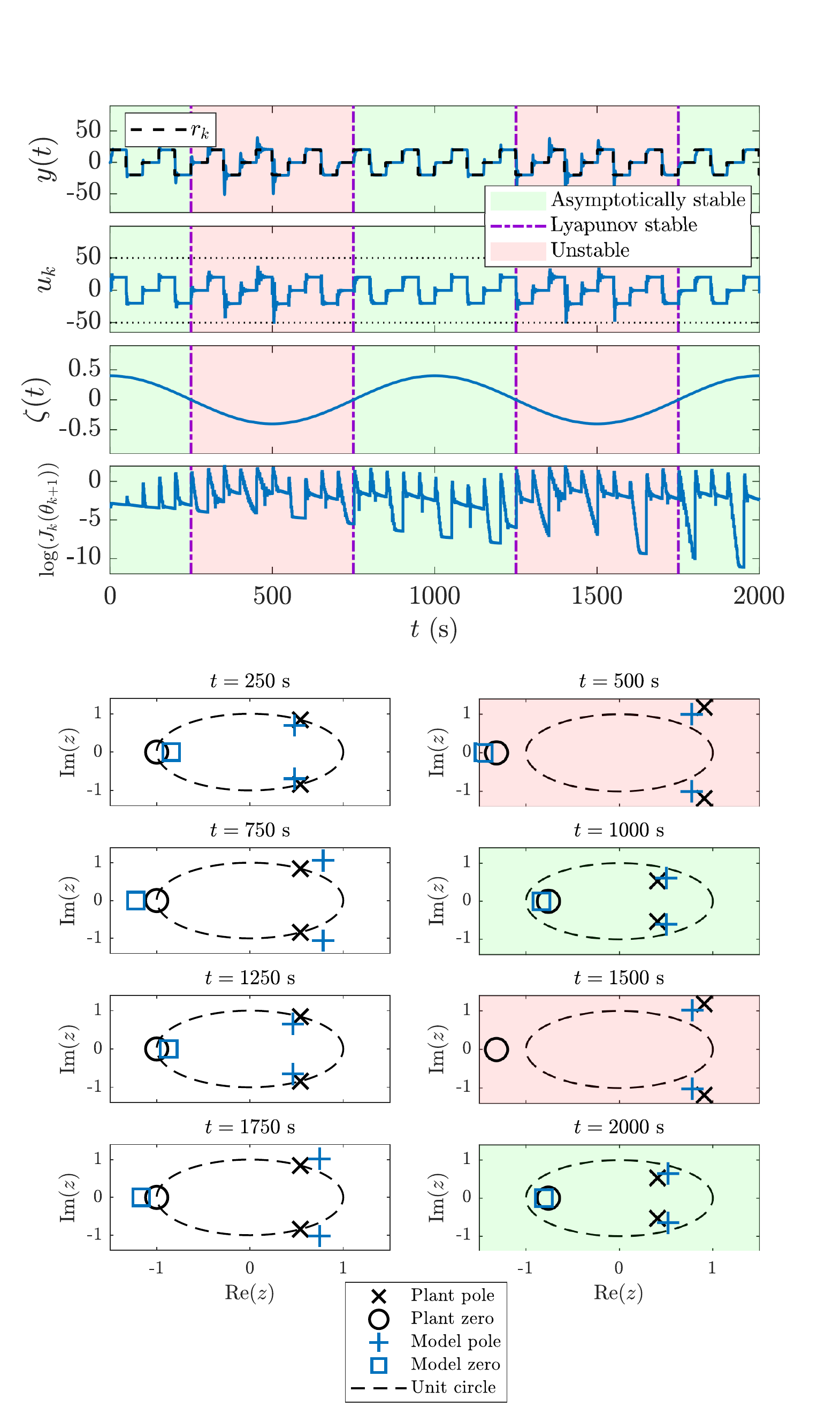}
        \caption{}
        \label{fig:ex1smoothb}
    \end{subfigure}
    \caption{Example \ref{ex:ex1smooth}.
    Command following for the SISO, continuous-time plant \eqref{eq:ex4LTV1}, \eqref{eq:ex4LTV2}, where the damping ratio $\zeta$ gradually changes over time according to \eqref{eq:zetadyn}.
    For constant values of $\zeta$, the plant is asymptotically stable for $\zeta>0,$ Lyapunov stable for $\zeta=0,$ and unstable for $\zeta<0,$ corresponding to the green area, purple vertical-dashed line, and red area, respectively.
    OFMPCOI uses \eqref{eq:QPnoslack}--\eqref{eq:DeltaUcon}  with $\hat{n}=n=2$ and the periodic step command  \eqref{eq:commandex4LTV}.
    (a) CRF with $\lambda=1.$
    Note that $y(t)$ approaches the step commands until approximately $t=360$ s.
    Afterwards, however, since $\lambda=1,$ RLS accumulates all past instantaneous costs, making reidentification more difficult and causing $y(t)$ to diverge during the unstable time intervals, that is, for $250$ s $\le t \le 750$ s and for $1250$ s $\le t \le 1750$ s.
    Although $y(t)$ diverges during the unstable time intervals, $y(t)$ approaches the step commands when the plant reverts to the asymptotically stable case.
    Note that, after $t=750$ s, the identified poles remain on the unit circle.
    Furthermore, $J_k(\theta_{k+1})$ sharply increases approximately at $t=500$ s and $t=1500$ s.
    (b) CRF with $\lambda=0.99.$
    Compared to (a), $y(t)$ approaches the step commands at all times.
    Additionally, note that, since $\lambda<1,$ the values of $J_k(\theta_{k+1})$ are lower than in (a), facilitating reidentification, as seen in the eight bottom-most plots.
    }
    \label{fig:ex1smoothab}
\end{figure*}

\begin{figure*}[t!]
    \centering
    \begin{subfigure}[t]{0.49\textwidth}
        \centering
        \includegraphics[trim = 10 10 10 40, width=\textwidth]{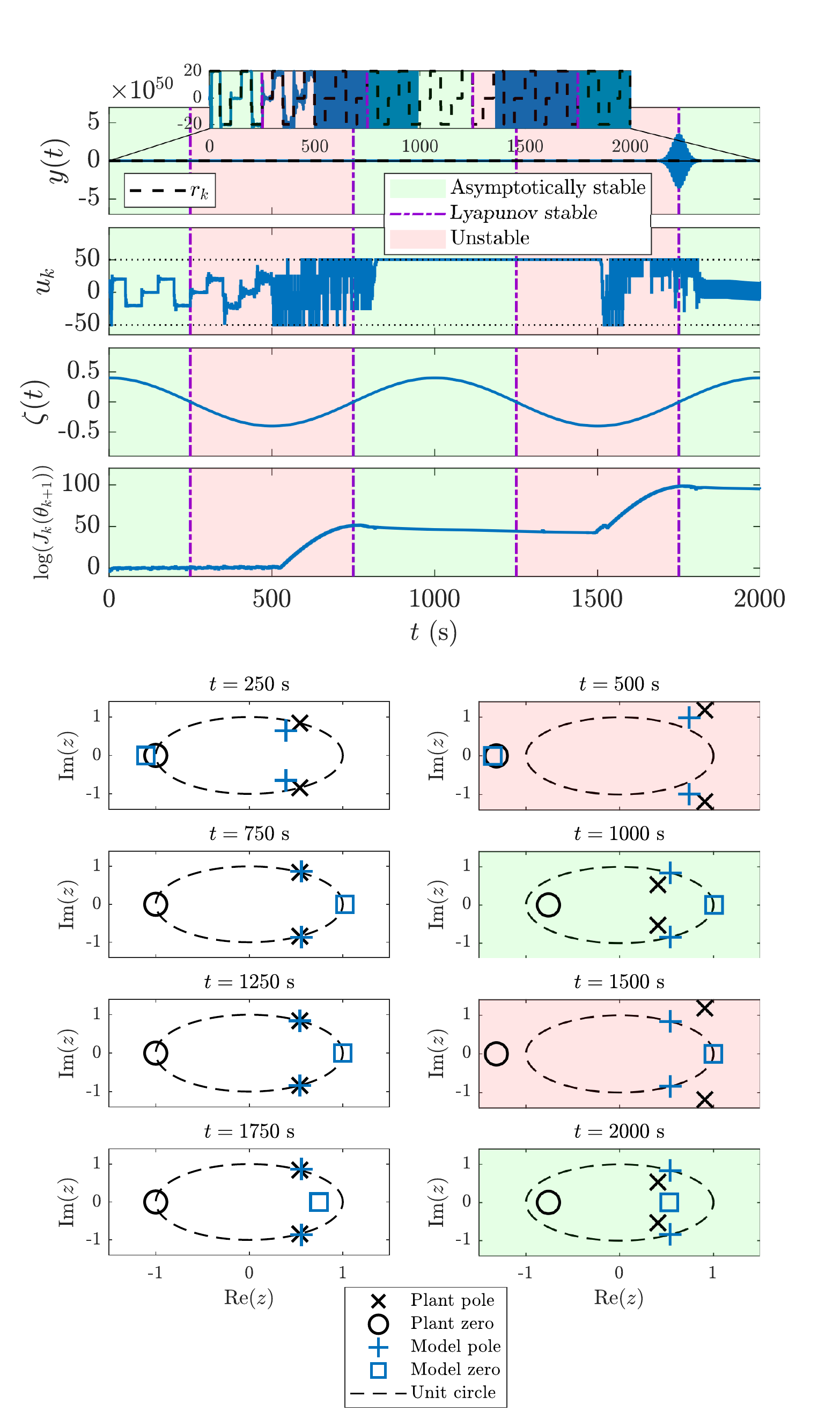}
        \caption{}
        \label{fig:ex1smoothc}
    \end{subfigure}%
    ~ 
    \begin{subfigure}[t]{0.49\textwidth}
        \centering
        \includegraphics[trim = 10 20 10 40, width=\textwidth]{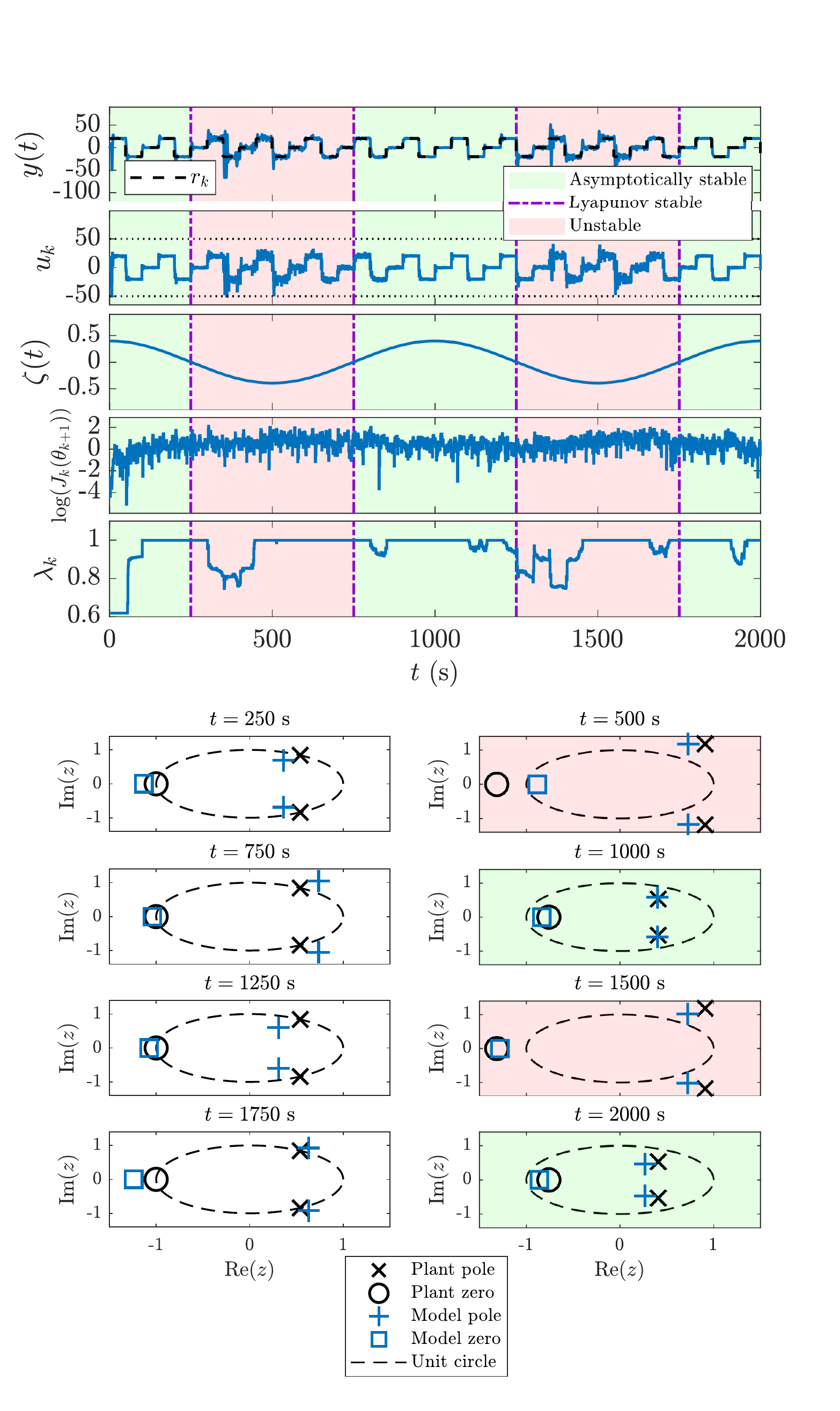}
        \caption{}
        \label{fig:ex1smoothd}
    \end{subfigure}
    \caption{Example \ref{ex:ex1smooth}.
    Command following for the SISO, continuous-time plant \eqref{eq:ex4LTV1}, \eqref{eq:ex4LTV2}, where $\zeta$ gradually changes over time according to \eqref{eq:zetadyn}.
    OFMPCOI uses \eqref{eq:QPnoslack}--\eqref{eq:DeltaUcon} with the noisy measurement $y_{\rmn,k}$ given by \eqref{eq:ynoisy}, where the standard deviation of the measurement noise is $\sigma=1.$
    (a) CRF with $\lambda=0.99.$
    The output $y(t)$ approaches the step commands until approximately $t=500$ s and diverges during the unstable time intervals.
    However, note that $y(t)$ approaches the step commands when the plant reverts to the asymptotically stable case.
    Similarly to the case shown in Figure \ref{fig:ex1smootha}, note that $J_k(\theta_{k+1})$ sharply increases at approximately $t=500$ s and $t=1500$ s, and that the poles remain on the unit circle from $t=750$ s.
    (b) VRF with $\eta=0.5,$ $\tau_\rmn=50,$ and $\tau_\rmd = 250.$
    Compared to (a), $y(t)$ approaches the step commands at all times.
    Note that the values of $J_k(\theta_{k+1})$ are lower than in (a) and that $\lambda_k$ is mainly activated at each transition between the asymptotically stable case and the unstable case, facilitating reidentification, as shown in the eight bottom-most plots.
    }
    \label{fig:ex1smoothcd}
\end{figure*}

\begin{example}\label{ex:ex1smooth}
\textit{Gradual, periodic change between an asymptotically stable plant, a Lyapunov-stable plant, and an unstable plant.}
Consider the SISO, continuous-time plant
\begin{align}
    \dot{x} & =
    \left[
    \begin{matrix}
        -2\zeta & -1 \\
        1 & 0
    \end{matrix}
    \right] x + \left[
    \begin{matrix}
        1 \\ 0
    \end{matrix}
    \right] u, \label{eq:ex4LTV1} \\
    y & = \left[
    \begin{matrix}
        0 & 1
    \end{matrix}
    \right]x, \label{eq:ex4LTV2}
\end{align}
where the damping ratio $\zeta$  gradually changes over time according to
\begin{align}
    \zeta(t) = A_\rmg \cos(\omega_\rmg t), \label{eq:zetadyn}
\end{align}
where $A_\rmg$ and $\omega_\rmg$ are the amplitude and the frequency of the damping-ratio periodic change, respectively.
Note that, in the case where $\zeta$ is constant, \eqref{eq:ex4LTV1}, \eqref{eq:ex4LTV2} is asymptotically stable for $\zeta > 0,$ Lyapunov stable for $\zeta = 0,$ and unstable for $\zeta < 0.$

The data are sampled with sample period $T_\rms = 1$ s.
Let $u_{\mathrm{min}} = -50,$ $u_{\mathrm{max}} = 50,$ $\Delta u_{\mathrm{min}} = -25,$ $\Delta u_{\mathrm{max}} = 25,$ $\ell=20,$ $\bar{Q}=10I_{\ell-1},$ $\bar{P}=20,$ $R=I_\ell,$ $P_0=10^3I_{2\hat{n}},$ $\theta_0 = [0_{1\times (2\hat{n}-1)} \ 1]^\rmT,$ $\hat{n}=n=2,$ $A_\rmg = 0.4,$ and $\omega_\rmg = \pi/500.$
The plant \eqref{eq:ex4LTV1}, \eqref{eq:ex4LTV2} is initialized with $x(0) = [2.5 \ -\mspace{-5mu}1.4]^\rmT,$ and no output constraint is considered in this example.
OFMPCOI uses \eqref{eq:QPnoslack}--\eqref{eq:DeltaUcon} with the measurement $y_k$ and the periodic step command 
\begin{align}
    r_k & =
    \begin{cases}
        20, & 150 \kappa \le k < 50 + 150 \kappa, \\
        -20, & 50  + 150 \kappa \le k < 100  + 150 \kappa, \\
        0, & 100  + 150 \kappa \le k < 150  + 150 \kappa,
    \end{cases} \label{eq:commandex4LTV}
\end{align}
where $\kappa\in\{0,1,2,\ldots\}$.
Figure \ref{fig:ex1smoothab} compares the responses of OFMPCOI using CRF with $\lambda=1$ and $\lambda=0.99.$

Next, instead of using $y_k,$ OFMPCOI uses the noisy measurement $y_{\rmn,k}$ \eqref{eq:ynoisy}, where the standard deviation of the measurement noise is $\sigma=1.$
Figure \ref{fig:ex1smoothcd} shows the response of OFMPCOI using CRF with $\lambda=0.99$ and VRF with $\eta=0.5,$ $\tau_\rmn=50,$ and $\tau_\rmd = 250.$

This example exhibits the manifestations of consistency and exigency.
Due to the slow changing dynamics of the plant and the inertia of the RLS dynamics, there exists an unavoidable bias between the coefficient estimates and the true coefficients, which indicates a lack of consistency within closed-loop identification at each time instant.
However, exigency is manifested by the fact that RLS is able to prioritize at each time instant which information needs to be identified in order to follow the step commands.

\end{example}

\clearpage

\section{CONCLUSIONS}

This article presented a numerical investigation of output-feedback MPC with online identification (OFMPCOI), which uses closed-loop identification to identify a model for constrained receding-horizon control optimization.
The identified model uses an input-output structure with block observable canonical form realization, which facilitates output feedback and avoids the need for full-state feedback and the use of an observer.

The performance of OFMPCOI was investigated numerically in order to investigate the interplay between closed-loop identification and control.
Three issues were considered, namely, persistency, consistency, and exigency,
Persistency is needed for the unambiguous estimation of model parameters, consistency ensures that the parameter estimates are asymptotically unbiased, and exigency prioritizes the identification of model features that are essential to the control objective.
When the persistency of the commands and disturbances is not sufficient for identification, OFMPCOI was shown to provide additional self-generated persistency through the control input.

Since the control input produced by OFMPCOI is correlated with the disturbance and sensor noise, the parameter estimates obtained from closed-loop identification based on recursive least squares (RLS) are not consistent, and thus are asymptotically biased. 
Although the bias was shown to degrade the accuracy of OFMPCOI, this effect was mitigated to some extent by the use of RLS with variable-rate forgetting (VRF).
An alternative approach is to use instrumental variables, which provides consistency for closed-loop identification despite signal correlation albeit at higher computational cost \cite{IVCLID}.

Finally, the interplay between control and identification was further examined in terms of exigency, which is the ability of the closed-loop identification to prioritize the model features needed to achieve the control objective.
Numerical examples showed that exigency is manifested in the continuing refinement of the identified model.

The numerical examples in this article were designed to expose features of OFMPCOI that are candidates for theoretical investigation.
Future research will use these examples to motivate the development of stability and performance guarantees.
A major next step is to extend the method to various classes of nonlinear systems.
Finally, the performance of OFMPCOI on these numerical examples provides further motivation to apply this technique to physical examples.

\section{ACKNOWLEDGMENTS}

The authors appreciate numerous helpful comments and suggestions provided by the reviewers.
This research was supported by ONR under BRC grant N00014-18-1-2211.

\clearpage

\printbibliography

\clearpage

\section{Author Biographies}

\begin{wrapfigure}[5]{L}{0.18\textwidth}
    \vspace{-4ex}
   \includegraphics[width=.17\textwidth]{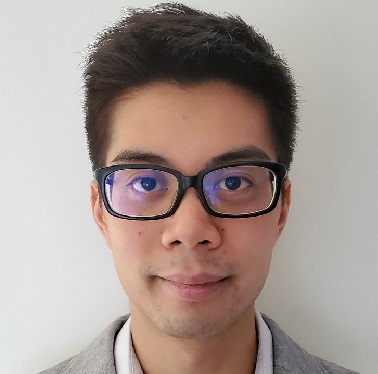}
\end{wrapfigure}
\noindent {\bf Tam W. Nguyen} received the M.Sc. degree in mechatronics engineering and the Ph.D. degree in control systems from the Free University of Brussels, Belgium.
He then served as a postdoctoral researcher in the Aerospace Engineering Department at the University of Michigan, Ann Arbor.
His interests are in constrained control for aerial-robotics applications.

 \vspace{16mm}

\begin{wrapfigure}[5]{L}{0.18\textwidth}
    \vspace{-4ex}
   \includegraphics[width=.17\textwidth]{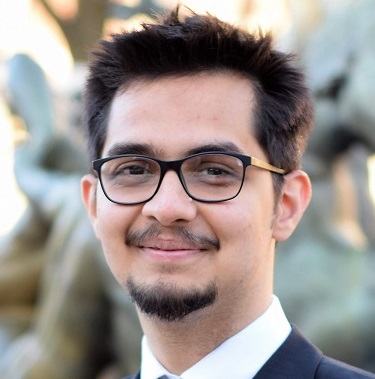}
\end{wrapfigure}
\noindent {\bf Syed Aseem Ul Islam} received the B.Sc. degree in aerospace engineering from the Institute of Space Technology, Islamabad and is currently pursuing the Ph.D. degree in flight dynamics and control from the University of Michigan in Ann Arbor.
His interests are in data-driven adaptive control for aerospace applications.

 \vspace{16mm}

\begin{wrapfigure}[6]{L}{0.18\textwidth}
   \vspace{-4ex}
   \includegraphics[width=.17\textwidth]{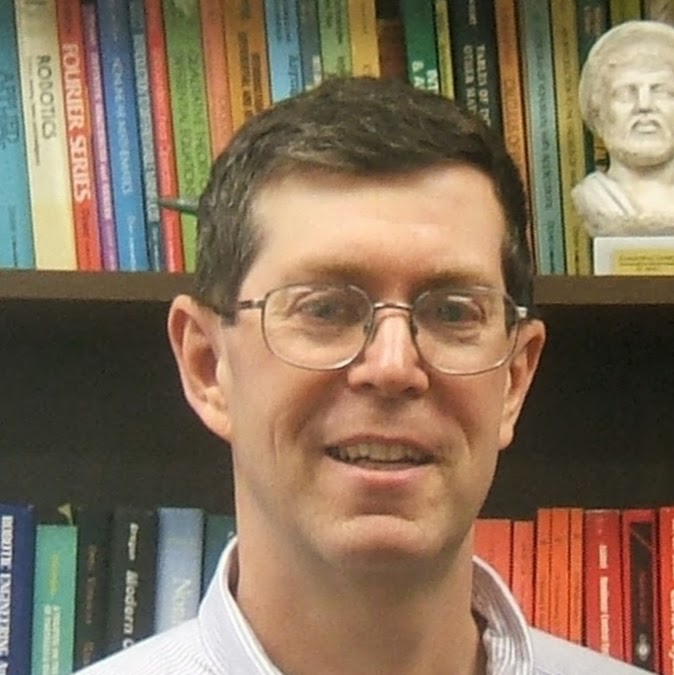}
\end{wrapfigure}
\noindent {\bf Dennis S. Bernstein} received the Sc.B. degree from Brown University and the Ph.D. degree from the University of Michigan in Ann Arbor, Michigan, where he is currently professor in the Aerospace Engineering Department.  His interests are in identification, estimation, and control for aerospace applications.  He is the author of {\it Scalar, Vector, and Matrix Mathematics}, published by Princeton University Press. 

 \vspace{20mm}

 \begin{wrapfigure}[5]{L}{0.18\textwidth}
    \vspace{-4ex}
   \includegraphics[width=.17\textwidth]{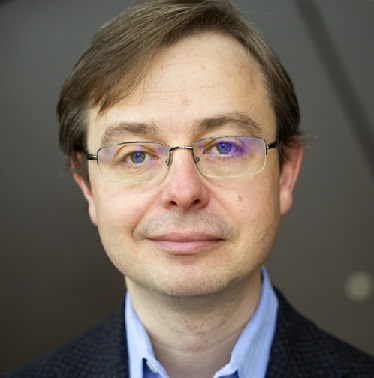}
\end{wrapfigure}
\noindent {\bf Ilya V. Kolmanovsky} is a professor in the department
of aerospace engineering at the University of Michigan. His research interests are in control theory for systems with state and control constraints, and in control applications to aerospace and automotive systems. He received his Ph.D. degree from the University of Michigan in 1995. He is a Fellow of IEEE and is named as an inventor on more than one hundred United States patents.

 \vspace{16mm}

\clearpage

{\bf About This Issue:} {\it 
In ``Output-Feedback  Model Predictive Control
with Online  Identification,''  Tam Nguyen, Aseem Ul Islam, Dennis Bernstein, and Ilya Kolmanovsky explore the properties of an output-feedback model predictive control algorithm that uses online system identification to update the plant model.
Their article presents a numerical investigation of three fundamental issues, namely, persistency (the ability of the control to self-generate control signals that enhance identification), consistency (model bias arising from identification within closed-loop operation), and exigency (the extent to which online identification emphasizes model characteristics that are critical to meeting performance objectives).

}

\clearpage

{\bf Summary:} {\it 
Model predictive control (MPC) is a widely used modern control technique with numerous successful application in diverse areas.
Much of this success is due to the ability of MPC to enforce state and control constraints, which are crucial in many applications of control.
In order to avoid the need for an observer, output-feedback model predictive control with online identification (OFMPCOI) uses the block observable canonical form whose state consists of past values of the control inputs and measured outputs.
Online identification is performed using recursive least squares (RLS) with variable-rate forgetting.
The article describes the algorithmic details of OFMPCOI and  numerically investigates its performance through a collection of numerical examples that highlight various control challenges, such as model order uncertainty, sensor noise, prediction horizon, stabilization, magnitude and move-size saturation, and stabilization.
The numerical examples are used to probe the performance of OFMPCOI in terms of persistency, consistency, and exigency.
Since OFMPCOI does not employ a separate control perturbation to enhance persistency, the focus is on self-generated persistency during transient operation.
For closed-loop identification using RLS, sensor noise gives rise to bias in the identified model, and the goal is to determine the effect of the lack of consistency.
Finally, the numerical examples reveal exigency, which is the extent to which the online identification emphasizes model characteristics that are most relevant to meeting performance objectives.
}

\end{document}